\documentclass[%
 aip,
 amsmath,amssymb,
 reprint,%
]{revtex4-1}

\usepackage[utf8]{inputenc}
\usepackage[T1]{fontenc}

\usepackage[unicode=true]{hyperref}

\usepackage{graphicx}
\graphicspath{{figures/}}

\usepackage{xcolor}

\usepackage{mathtools}
\usepackage{physics}

\usepackage{mathptmx}
\usepackage{etoolbox}

\usepackage{aligned-overset}

\newcommand{\hodge}{{\star}}
\DeclareMathOperator{\sym}{sym}

\newcommand{\bR}{\mathbb{R}}

\newcommand{\cD}{\mathcal{D}}

\newcommand{\cL}{\mathcal{L}}

\newcommand{\cO}{\mathcal{O}}

\newcommand{\cX}{\mathcal{X}}

\newcommand{\cZ}{\mathcal{Z}}

\newcommand{\QuantitySet}{\mathbf{Q}}

\usepackage[capitalize,nameinlink]{cleveref}

\makeatletter
\def\@email#1#2{%
 \endgroup
 \patchcmd{\titleblock@produce}
  {\frontmatter@RRAPformat}
  {\frontmatter@RRAPformat{\produce@RRAP{*#1\href{mailto:#2}{#2}}}\frontmatter@RRAPformat}
  {}{}
}%
\makeatother

\begin{document}

\title{%
  Energy-based, geometric, and compositional formulation of fluid and plasma models
}

\author{Markus Lohmayer}
\email{markus.lohmayer@fau.de}
\affiliation{
Friedrich-Alexander Universität Erlangen-Nürnberg,
Erlangen, Germany
}
\author{Michael Kraus}
\affiliation{
Max-Planck-Institut für Plasmaphysik,
Garching, Germany
}
\author{Sigrid Leyendecker}
\affiliation{
Friedrich-Alexander Universität Erlangen-Nürnberg,
Erlangen, Germany
}

\date{\today}

\begin{abstract}
%
Fluid dynamics plays a crucial role
in various multiphysics applications,
including
energy systems, electronics cooling, and biomedical engineering.
%
%
Developing models
for complex coupled systems
can be challenging and time-consuming.
In particular,
ensuring the consistent integration of models
from diverse physical domains
requires meticulous attention.
%
%
Considering the example of
(electro-)magneto hydrodynamics
(on a fixed spatial domain
and
with linear polarization and magnetization),
this article
demonstrates how
relatively complex models
can be composed from
simpler parts
by means of
a formal language
for multiphysics modeling.
%
%
The Exergetic Port-Hamiltonian Systems (EPHS) modeling language features
a simple graphical syntax
for expressing
the energy-based interconnection of subsystems.
This reduces cognitive load
and
facilitates communication,
especially in multidisciplinary environments.
As the example demonstrates,
existing models can be easily integrated
as subsystems of new models.
%
Specifically,
an ideal fluid model
is used as a subsystem of
a Navier-Stokes-Fourier fluid model,
which in turn is reused as a subsystem of
an (electro-)magneto hydrodynamics model.
The energy-based, compositional approach
simplifies understanding complex models,
and
it makes it easy to
encapsulate, reuse, and replace (parts of) models.
Moreover,
structural properties of EPHS models
guarantee fundamental properties of
thermodynamic systems, such as
conservation of energy,
non-negative entropy production,
and Onsager reciprocal relations.

\end{abstract}

\maketitle

\section{Introduction}%

We first introduce
the EPHS modeling language,
which provides an energy-based framework for composable physical models.
Then,
we discuss why we care about
stating our models in a coordinate-invariant formalism.
Next,
we discuss some related work.
Finally,
we present an outline of the following sections.

\subsection{EPHS modeling language}%

Exergetic Port-Hamiltonian Systems (EPHS)
offer
a compositional and thermodynamically-consistent language for
the specification of dynamic, multiphysical models~%
\onlinecite{2024LohmayerLynchLeyendecker}.
This includes
models from classical mechanics
and electromagnetism
as well as
irreversible processes
with local thermodynamic equilibrium.
An EPHS model is defined by
a power-preserving interconnection of primitive subsystems.
The primitive systems represent fundamental physical behaviors,
namely
storage as well as reversible and irreversible exchange of energy.
The interconnection is expressed
in a simple graphical syntax.
Since expressions in the syntax are composable,
models can be defined hierarchically
to manage complexity
and
to make their parts reusable.
Due to structural properties,
models automatically conform with
the first and second law of thermodynamics,
Onsager reciprocal relations,
and possibly further conservation laws
such as conservation of mass.

The primary objective of the EPHS language is
to simplify understanding, communication and development
of models across various physical domains.
Moreover,
we expect that
the structured approach to model specification
can be leveraged to automatically generate
structure-preserving numerical schemes
for simulation, optimization, and control
of multiphysical systems.
Further,
the framework can probably facilitate
the integration of
first-principles-based modeling
with scientific machine learning.

The application of EPHS to spatially-lumped models~%
\cite{2024LohmayerLynchLeyendecker}
such as multibody systems~%
\cite{2024LohmayerCapobiancoLeyendecker}
suggests that
the language can be effectively used in practice, provided that
a suitable software implementation
and
structure-preserving time discretization
are developed.
This work explores another direction,
namely the formulation of spatially-distributed models
in the EPHS language.
At the same time,
we demonstrate the effectiveness of
the compositional approach
by constructing two plasma models,
which directly reuse
a Navier-Stokes-Fourier (NSF) fluid model
and
a Maxwell electromagnetism model.
The NSF model directly reuses
an ideal compressible fluid model.
This sheds light on
the energetic, thermodynamic and geometric nature of the considered models.

\subsection{Coordinate-invariant formulation}%

Although less commonly used than the classically taught \emph{vector calculus} (VC) formalism,
\emph{exterior calculus} (EC) provides
a powerful language for expressing physical laws
in a coordinate-independent
and geometrically intuitive manner.
While VC implicitly relies on
the Euclidean structure of the spatial domain,
EC is based on
more general spaces, called Riemannian manifolds.
It thereby separates different aspects
that are intermingled in VC,
such as differentiation,
parameterization of the spatial domain with coordinate charts
and the use of its (not necessarily Euclidean) metric.
At the same time,
EC unifies the derivative operators
gradient, curl and divergence from VC
into one operator,
called the exterior derivative.
This is by virtue of a graded algebra:
the exterior algebra of
a $3$-dimensional vector space
contains so-called \mbox{$k$-vectors},
which can be thought of as
oriented $k$-dimensional volumes
for $k = 0, \, \ldots, \, 3$.
Rather than working with scalar and vector fields,
EC relies on the more nuanced concept of
differential \mbox{$k$-forms}.
A \mbox{$k$-form} on a manifold essentially provides,
at each point,
a `scale' to measure some quantity associated to \mbox{$k$-vectors}.
A \mbox{$0$-form} is a scalar field
that measures a quantity associated to points.
A \mbox{$1$-form} can be integrated along a curve,
a \mbox{$2$-form} can be integrated over a surface and
a \mbox{$3$-form} can be integrated over a volume,
giving the total of the measured quantity in each case.
The generalized Stokes theorem of EC unifies
the fundamental theorem of calculus ($k=1$),
the Stokes theorem ($k=2$),
and the divergence theorem ($k=3$).
Integration is defined intrinsically on manifolds,
meaning without a dependence on
specific coordinates or the Riemannian metric.
Balance equations can consequently be expressed
in a way that essentially relies only on
the topology of the spatial domain,
while
the metric appears only in constitutive relations.

Thermodynamic models
as well as
computational models
are commonly based on
a finite reticulation of the spatial domain.
It hence can seem a bit odd that
physical modeling traditionally relies on differential calculus,
with discretization coming as an unavoidable afterthought,
see e.g.~\onlinecite{2014Tonti}.
While we do not break with this tradition,
we expect that the models presented here
can be naturally transformed into finite-dimensional EPHS
based on discrete analogues of exterior calculus,
see e.g.~\onlinecite{2003Hirani,2010ArnoldFalkWinther}.
Discrete exterior calculus formulations of fluid models
with variational structure
are explored for instance in~%
\onlinecite{2011PavlovMullenTongKansoMarsdenDesbrun,2022CoueraudGay}.

\subsection{Related work}%

In~\onlinecite{2024LohmayerLynchLeyendecker},
we discuss how
the EPHS language
leverages ideas from
four research fields:
(1)~graphical and energy-based modeling
of physical systems with bond graphs~%
\cite{2012KarnoppMargolisRosenberg,2010Borutzky},
(2)~the metriplectic and GENERIC framework
for nonequilibrium thermodynamics~%
\cite{1986Morrison,2005Oettinger,2018PavelkaKlikaGrmela},
(3)~port-Hamiltonian systems theory~%
\cite{2009DuindamMacchelliStramigioliBruyninckx,2014SchaftJeltsema},
and
(4)~applied category theory research on
the formalization of graphical languages
as well as compositional dynamical systems~%
\cite{2019FongSpivak,2023Myers}.
Here,
our discussion focuses on fluid models
with Hamiltonian, port-Hamiltonian and metriplectic structure.
While some familiarity
with (port-)Hamiltonian and metriplectic or GENERIC systems
is assumed here,
this is not the case for the main text.

\subsubsection{Hamiltonian fluid mechanics}

The discovery of
the geometry underlying
classical fluid mechanics
was pioneered by Vladimir Arnold
who found that
Euler's equation for
a freely rotating rigid body
and
Euler's equation for
an incompressible ideal fluid
possess a similar
noncanonical Hamiltonian structure,
see~\onlinecite{1966Arnold,1969Arnold}.
This so-called Lie-Poisson structure
can be derived via symmetry reduction.
For a freely rotating rigid body,
the canonical configuration space is
the Lie group of $3 \times 3$ rotation matrices
$\mathrm{SO}(3)$.
The invariance (or symmetry) of
the Hamiltonian function
(kinetic energy)
under superimposed rotations
(left translation in group theory terms)
leads to
a reduced system
with a Lie-Poisson structure
on the linear dual of
the left Lie algebra $\mathfrak{so}(3)$
associated with $\mathrm{SO}(3)$.
This gives
an evolution equation for
the momentum of the body
expressed in a body-attached frame.
For an incompressible ideal fluid,
the canonical Hamiltonian description
tracks fluid particles
(Lagrangian viewpoint in continuum mechanics terms).
The corresponding canonical configuration space is
the group of volume-preserving diffeomorphisms
on the spatial domain
occupied by the fluid.
For this to be a group
(and for the system to be autonomous),
the assumption is made that
the spatial domain has no boundary
or that
the diffeomorphisms leave the boundary invariant,
implying that the fluid velocity is tangential to the boundary.
The invariance of the Hamiltonian function
under particle relabling
(right translation in group theory terms)
similarly leads to
a reduced system
with a Lie-Poisson structure
on the linear dual of
the right Lie algebra of the diffeomorphism group.
This gives
an evolution equation for
the fluid momentum
(or alternatively for the fluid velocity)
expressed in an inertial reference frame
(Eulerian viewpoint in continuum mechanics terms).
For more details concerning
Hamiltonian mechanics on Lie groups
and symmetry reduction, see e.g.~%
\onlinecite{1970MarsdenAbraham,1970EbinMarsden,1999MarsdenRatiu}.

Using a semidirect product of Lie groups,
the geometric approach
can be extended to
arbitrary rigid-body motions
on the one hand
and to
compressible fluids
on the other.
For rigid body dynamics,
the semidirect product
$\mathrm{SE}(3) = \mathrm{SO}(3) \ltimes \mathbb{R}^3$
also keeps track of translations,
which are acted upon by rotations.
For compressible fluid dynamics,
a semidirect product
additionally keeps track of
mass and entropy (densities),
which are acted upon by diffeomorphisms (advection).
More details can be found for instance in~%
\onlinecite{1984MarsdenRatiuWeinstein2,1984MarsdenRatiuWeinstein1}.

\subsubsection{Port-Hamiltonian fluid dynamics}

While
the assumption that
diffeomorphisms leave the boundary invariant
is fundamental at the Lie group (configuration) level,
it plays no important role
at the Lie algebra (velocity) level.
Based on this observation,
the Lie-Poisson structure
can be generalized to
a Stokes-Dirac structure~\cite{2002SchaftMaschke},
leading to
a port-Hamiltonian model
of (incompressible or compressible) ideal fluid dynamics
that explicitly takes into account
energy exchange across the boundary of the spatial domain.
This was first explored in~%
\onlinecite{2001SchaftMaschke}
and discussed in more detail in~%
\onlinecite{%
2021RashadCalifanoSchullerStramigioli1,%
2021RashadCalifanoSchullerStramigioli2}.
In~\cref{sec:ideal-fluid},
the ideal fluid model
is expressed
using the EPHS language.

In~\onlinecite{2021CalifanoRashadSchullerStramigioli},
the port-Hamiltonian ideal fluid model is extended
to account for the loss of mechanical energy due to viscosity.
The extension is based on an in-domain port,
which is closed with a resistive relation
that removes energy.
The resulting Navier-Stokes model
neglects the thermal energy domain,
as its Hamiltonian storage function
represents only the mechanical energy
(at the macroscopic level).
The thermal energy domain is neglected
based on either the incompressibility or barotropicity assumption.
The latter assumes
a compressible fluid whose
internal energy,
and hence also pressure,
depend solely on the mass density.

In~\onlinecite{2022LohmayerLeyendecker2},
it is briefly shown how
to extend this to
a thermodynamically-consistent port-Hamiltonian Navier-Stokes-Fourier system.
Its internal energy function
not only depends on the mass density
but also on the fluid's entropy.
Moreover,
a subsystem representing
the irreversible process of thermal conduction is added.
The storage function of the port-Hamiltonian system
is recognized as the exergy content of the fluid,
i.e.~the amount of energy
that is theoretically available to do work
with respect to a fixed reference environment,
see~\onlinecite{2021LohmayerKotyczkaLeyendecker}.
This enables
the thermodynamically-consistent combination of
reversible and irreversible dynamics,
as known from the metriplectic~\cite{1986Morrison}
or GENERIC~\cite{1997GrmelaOettinger} formalism.
In~\cref{sec:nsf},
the Navier-Stokes-Fourier fluid model
is stated in a more detailed and improved manner.

\subsubsection{Hamiltonian and metriplectic plasma models}

Plasma,
sometimes known as the fourth state of matter,
refers to an ionized gas,
i.e.~a gas consisting of free electrons and ions.
%
%
A plasma hence is electrically conductive,
and it can excite and interact with electromagnetic waves.
There is a vast zoo of plasma models.

Kinetic theories describe the electron and the ion species
separately in terms of a particle distribution function.
In the non-dissipative case,
collisionless plasmas can be described by
the Vlasov-Maxwell system
with Hamiltonian structure~%
\cite{1982MarsdenWeinstein}.
In the dissipative case,
collisional plasmas can be described by
the Vlasov-Maxwell-Landau system
having a metriplectic structure~%
\cite{2022HirvijokiBurbyBrizard}.

Assuming local thermodynamic equilibrium,
two-fluid theory
is concerned with a continuum description,
considering electrons and ions as
separate fluid components
with their own mass density and velocity.
In the non-dissipative case,
a Hamiltonian two fluid-model is presented in~%
\onlinecite{1982SpencerKaufmann}.
%

Simplifying to a single fluid component,
an electro-magneto hydrodynamic (EMHD) model can be formulated,
which has a parameter representing
the effective charge per unit of mass.
For high-frequency electromagnetic waves,
this parameter can be chosen as the electron mass.
Considering that
electrons have a much smaller mass than ions,
the EMHD model then focuses on the motion of electrons
and
considers ions as a stationary background.
%
An ideal EMHD model with Hamiltonian structure
is shown in~\onlinecite{2018PavelkaKlikaGrmela}.
In~\cref{sec:emhd},
a dissipative extension of this model is stated
using the EPHS language and exterior calculus.

Assuming high electric conductivity and quasi-neutrality,
the influence of the electric field is significant only
on small time and length scales.
Magneto hydrodynamics (MHD) models hence
neglect the displacement current term in Maxwell's equations
and
consider ions and electrons as a single conducting fluid.
MHD models are commonly used in astrophysics and fusion research, when
the characteristic time scales of fluid motion are long
compared to the timescales over which
electric fields change significantly.
In the non-dissipative case,
the Hamiltonian structure of MHD
is shown in~\onlinecite{1980MorrisonGreene}
and
a coordinate-invariant formulation using differential forms
is presented in~\onlinecite{2020GilbertVanneste}.
In the dissipative case,
a metriplectic model considering
viscosity and electric resistivity
is presented in~\onlinecite{2012MaterassiTassi}.
In~\cref{sec:mhd},
a similar model is stated using the EPHS language and exterior calculus.
An extended MHD model
considering thermodynamic cross effects
with metriplectic structure
is derived in~\onlinecite{2020CoquinotMorrison}.

\subsection{Outline}%

To get familiar with the EPHS framework,
\cref{sec:ephs} considers
an ideal barotropic fluid model
on a one-dimensional spatial domain.
\Cref{sec:ideal-fluid}
states the EPHS model
of an ideal compressible fluid
on a three-dimensional spatial domain,
modeled as
a Riemannian manifold
with boundary.
\Cref{sec:nsf}
extends the ideal fluid model
to a Navier-Stokes-Fourier system
by adding
thermal conduction
and
viscosity models.
\Cref{sec:em}
states a Maxwell model
describing the propagation of electromagnetic waves.
\Cref{sec:emhd}
combines
the Navier-Stokes-Fourier system
and
the Maxwell system
into an electro-magneto hydrodynamics (EMHD) model
that describes
the motion of an electrically-charged fluid
interacting with
internally generated
and externally applied
electric and magnetic fields.
\Cref{sec:mhd}
presents a magneto hydrodynamics (MHD) model
that neglects the electric field,
assuming that it evolves much faster than the magnetic field.
\Cref{sec:conclusion}
concludes with
a discussion.

\Cref{sec:geometry}
summarizes
the exterior calculus formalism
used in the main text.
%
%
\Cref{sec:metriplectic}
summarizes how
EPHS models are related to
the metriplectic or GENERIC formalism.

\section{Ideal barotropic fluid model in 1D}%
\label{sec:ephs}

Here,
we introduce the EPHS framework,
considering a spatially-distributed system,
namely an ideal barotropic fluid
on a one-dimensional spatial \mbox{domain}
$\cZ = [0, \, L] \subset \bR$
with
$L \in \bR_{>0}$
and
coordinate $z \in \cZ$.

\subsection{Syntax}

\begin{figure}
  \centering
	\includegraphics[width=5.8cm]{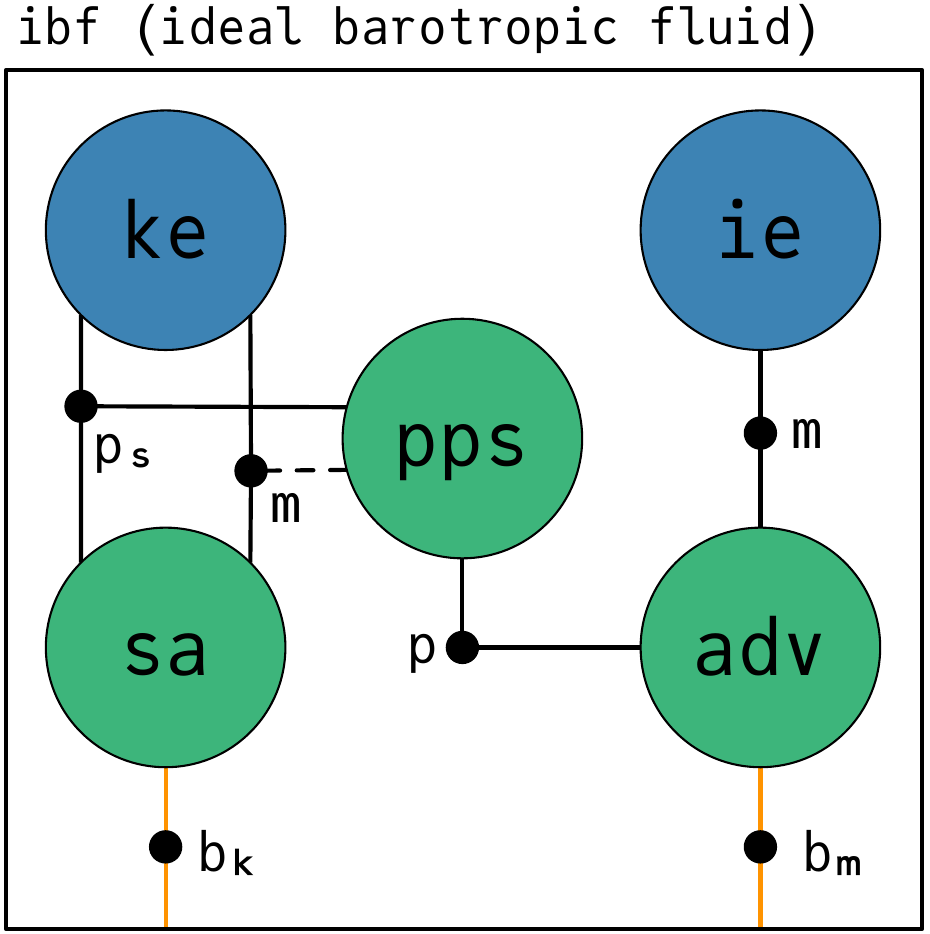}
  \caption{%
    Interconnection pattern
    for an ideal barotropic fluid model.
    Box $\mathtt{ke}$ represents
    storage of kinetic energy,
    which is exchanged both in terms of
    (specific) momentum via port $\mathtt{ke.p_s}$
    and
    mass via port $\mathtt{ke.m}$.
    Box $\mathtt{pps}$ represents
    the reversible transformation between
    two different representations of momentum, namely
    specific momentum (or velocity)
    exchanged via port $\mathtt{pps.p_s}$
    and momentum density
    exchanged via port $\mathtt{pps.p}$.
    As the transformation depends on the mass density,
    the box has a state port $\mathtt{pps.m}$,
    drawn as a dashed line.
    Box $\mathtt{sa}$ represents
    self-advection of kinetic energy
    and
    the boundary port $\mathtt{b_k}$ accounts for
    advection across
    the boundary $\partial \cZ$.
    %
    Box $\mathtt{ie}$ represents
    storage of internal energy,
    which is exchanged in terms of
    mass via port $\mathtt{ie.m}$.
    Box $\mathtt{adv}$ represents
    advection of
    internal energy in terms of mass
    and
    the boundary port $\mathtt{b_m}$ accounts for
    advection across the boundary $\partial \cZ$.
  }%
  \label{fig:ibf}
\end{figure}

A primary feature of the EPHS language is
its graphical syntax.
Expressions in the syntax are called
\emph{interconnection patterns},
since their purpose is to specify how
increasingly complex systems can be formed
by interconnecting simpler subsystems.
The pattern shown in~\cref{fig:ibf}
is used to specify the considered fluid model
in terms of five subsystems.
The round inner boxes represent
(the interfaces of) the subsystems,
while
the rectangular outer box represents
the interface of the resulting \emph{composite system}.
An \emph{interface} is a collection of \emph{ports},
which are drawn as lines emanating from
the respective box.
For instance,
box $\mathtt{ke}$
represents an interface with two ports
named $\mathtt{p_s}$ and $\mathtt{m}$.
Similarly,
the outer box represents an interface with two ports
named $\mathtt{b_k}$ and $\mathtt{b_m}$.
Ports are connected to \emph{junctions},
which are drawn as black dots.
To distinguish ports
with the same name
belonging to different subsystem interfaces,
the name of the box is used as a prefix.
For instance,
the three ports
$\mathtt{ke.m}$,
$\mathtt{sa.m}$ and
$\mathtt{pps.m}$
are connected to the same junction.
In the graphical representation,
the name of the port
is normally written next to the box to which it belongs
but
whenever ports connected to a junction
have the same name,
we write it only once at the junction.
This is the case with all port names in~\cref{fig:ibf}.

The language is termed compositional
since expressions in the syntax
can be hierarchically nested.
When the outer interface of one expression
equals,
up to a given renaming of ports,
an inner interface of another expression,
the two expressions can be uniquely composed,
giving a single expression~%
\onlinecite{2024LohmayerLynchLeyendecker}.
Consequently,
complex interconnection patters and models
can be easily avoided
by refactoring them into
a hierarchy of simple and reusable parts.

\subsection{Semantics}

The meaning of a composite system
is determined by
the meaning of its interconnection pattern
and the meaning of the subsystems
that fill the inner boxes.
The meaning of the interconnection pattern
arises from its junctions,
where the connected ports can exchange
energy
and
state information.
Since subsystems may themselves be composite systems,
we note that
the assignment of meaning to interconnection patterns
is compatible with their composition.
According to~\onlinecite{2024LohmayerLynchLeyendecker},
the semantics of composable physical systems
are defined mathematically
in terms of
certain types of relations and their composition.
However, for simplicity of exposition,
we focus our attention on
the equations that define these relations.
Instead of talking about composition of relations,
we thus speak about
combining equations and
eliminating common port variables.

Since a hierarchy of interconnection patterns
can be uniquely flattened into a single pattern,
a composite system
is ultimately given by
an interconnection of primitive systems,
which are also called components.
The different kinds of components
are characterized by
a particular structure of their defining equations.
Together with the structure of
the equations that give meaning to interconnection patterns,
this guarantees, in particular, that any model respects
the first and the second law of thermodynamics.

The color filling of the inner boxes
of an interconnection pattern
is merely an annotation,
indicating what kind of system shall fill each box.
The following list provides an overview:
\begin{itemize}
  \item{%
    Blue boxes are filled with
    \emph{storage components},
    which are primitive systems
    representing storage of energy.
  }
  \item{%
    Green boxes are filled with
    \emph{reversible components},
    which represent
    a reversible coupling between energy domains,
    a reversible transformation,
    or a constraint.
    Reversible components conserve energy, entropy, and exergy.
  }
  \item{%
    Red boxes are filled with
    \emph{irreversible components},
    which model an irreversible process.
    Irreversible components conserve energy,
    produce entropy, and destroy exergy.
  }
  \item{%
    Yellow boxes are filled with
    nested \emph{composite systems}.
    A yellow box
    and
    the outer box of the pattern of the nested system
    are identified.
  }
\end{itemize}

\subsection{Ports and interfaces}

According to the definition
of spatially-lumped systems
in~\onlinecite{2024LohmayerLynchLeyendecker},
each port has two attributes.
One attribute is Boolean and
distinguishes between \emph{state ports} and \emph{power ports}.
The other attribute assigns
a physical quantity,
along with an underlying state space.
At a junction,
only ports with the same quantity (and state space) can be connected.
A state port,
drawn as a dashed line,
exchanges state information
through its \emph{state variable}.
This dynamic variable represents
the physical quantity given by the respective attribute,
and it takes values in its underlying state space.
A power port,
drawn as a solid line,
also allows for energy exchange
via two additional port variables
called \emph{flow} and \emph{effort}.
These dynamic variables respectively take values in
the tangent and cotangent bundles
over the port's state space.
The flow variable gives
a rate of change of the underlying state variable
and
the effort variable is defined such that
the duality pairing of both variables gives
the (exergetic) power
that is exchanged via the port.

Regarding spatially-distributed systems,
we have to add some remarks:

1.~%
State spaces depend on a spatial domain.
For instance,
the quantity associated to port $\mathtt{pps.m}$ is
$(C^\infty(\cZ), \, \mathtt{mass}) \in \QuantitySet$,
indicating that
the port shares information about
the fluid's mass density
with
the underlying state space being
the space of smooth functions on $\cZ = [0, \, L]$.
The set $\QuantitySet$ contains
the possible quantities
(parametrized over spatial domains).
We write
$\mathtt{pps.m.x} \in C^\infty(\cZ)$
to refer to the state variable,
which, like any port variable, implicitly depends on time.

2.~%
Considering for instance
the power port $\mathtt{ke.m}$,
the port variables take values in
the direct sum
$\mathrm{T} \cX \oplus \mathrm{T}^* \cX$
of the tangent and the cotangent bundle
over the port's state space
$\cX = C^\infty(\cZ)$.
Based on the vector space structure
of this infinite-dimensional state space,
we identify
$
\mathrm{T} \cX \oplus \mathrm{T}^* \cX
\cong
\cX \times \cX \times \cX^*
\ni (
  \mathtt{ke.m.x}, \,
  \mathtt{ke.m.f}, \,
  \mathtt{ke.m.e}
)
$.
Glossing over functional analytic aspects,
we furthermore identify
the dual space $\cX^*$ with $\cX$ itself.

3.~%
The duality pairing
involves an integral over the spatial domain.
For instance,
port $\mathtt{ke.m}$
shares information about mass
through its state variable
$\mathtt{ke.m.x} \in C^\infty(\cZ)$,
and
it can exchange energy by exchanging mass.
Its flow variable
$\mathtt{ke.m.f} \in C^\infty(\cZ)$
equals the rate at which mass is locally added to the system via the port
and
the corresponding effort variable
$\mathtt{ke.m.e} \in C^\infty(\cZ)$
equals the amount of kinetic energy per unit of mass.
The duality pairing of the effort and flow variables is
\begin{equation*}
  \langle \mathtt{ke.m.e} \mid \mathtt{ke.m.f} \rangle
  \: = \:
  \int_0^L \mathtt{ke.m.e} \cdot \mathtt{ke.m.f} \, \, \dd z
  \,.
\end{equation*}
The net power supplied to the storage component
is then given by
$
\langle \mathtt{ke.p_s.e} \mid \mathtt{ke.p_s.f} \rangle
\, + \,
\langle \mathtt{ke.m.e} \mid \mathtt{ke.m.f} \rangle
$,
and it equals
the rate at which kinetic energy is stored
in the domain $\cZ$.

4.~%
Two power ports may be fundamentally incompatible,
despite having the same physical quantity.
For instance,
ports $\mathtt{sa.m}$ and $\mathtt{adv.m}$
both have the quantity
$(C^\infty(\cZ), \, \mathtt{mass})$,
but for $\mathtt{sa.m}$,
mass carries kinetic energy,
whereas for $\mathtt{adv.m}$,
it carries internal energy.
To safeguard against ill-formed connections
at the level of syntax,
we distinguish three kinds of power ports
using the symbols
$\{ \mathsf{k}, \, \mathsf{p}, \, \mathsf{i} \}$.
Here,
$\mathsf{k}$ stands for
kinetic or magnetic energy,
$\mathsf{p}$ stands for
potential or electric energy, and
$\mathsf{i}$ stands for internal energy.
We hence write
$((C^\infty(\cZ), \, \mathtt{mass}), \, \mathsf{k})$
for the attributes of port $\mathtt{sa.m}$
and
$((C^\infty(\cZ), \, \mathtt{mass}), \, \mathsf{i})$
for the attributes of port $\mathtt{adv.m}$.

5.~%
While state and power ports
model some coupling within the spatial domain,
a \emph{boundary port},
drawn as an orange line,
models energy exchange
across the boundary of the domain.
%
%
A reversible or irreversible component
that involves a differential operator
has one or more boundary ports.
To obtain the power balance equation
for the component,
integration by parts is used,
and the resulting boundary terms
are split
into the flow and effort variables
of the boundary ports.
These ports can then be used to
impose boundary conditions
or to
join instances of the same component
defined on neighboring domains.
For instance,
in~\cref{fig:ibf},
the boundary ports
of the reversible components are exposed,
allowing for such connections to be made at a higher level,
where the fluid model is seen as a subsystem.
The flow variables
$\mathtt{sa.b_k.f}$
and
$\mathtt{adv.b_m.f}$
are equal to
the mass flow rate
into the system
across $\partial \cZ$.
The effort variables
$\mathtt{sa.b_k.e}$
and
$\mathtt{adv.b_m.e}$
respectively are equal to
the amount of kinetic and internal energy per unit of mass
(except that the latter includes a shift by
a chemical potential of the reference environment
to give the amount of exergy).
To assign meaning to interconnection patterns,
the attributes of a boundary port
have to determine the spaces
in which the flow and effort variables take values.
To safeguard against ill-formed connections
at the level of syntax,
a hash value could be included as an attribute
that identifies
the differential operator,
and the splitting of the boundary term
that leads to the boundary port.
We write $\mathbf{B}$ for
the set of possible attributes of boundary ports,
but we omit the details for now.

Formally,
an interface
$I = (N, \, \tau)$
is given by a finite set of ports $N$
and
a function
$
  \tau \colon N \to
  \QuantitySet
  \sqcup
  \QuantitySet \times \{ \mathsf{k}, \, \mathsf{p}, \, \mathsf{i} \}
  \sqcup
  \mathbf{B}
$
that assigns their attributes.
For instance,
the interface
$I_\text{ke} = (N_\text{ke}, \, \tau_\text{ke})$
defining the box $\mathtt{ke}$
is given by
the set of ports
$N_\text{ke} = \{ \mathtt{p_s}, \, \mathtt{m} \}$
and the function $\tau_\text{ke}$
determined by
\begin{equation}
  \begin{alignedat}{2}
    &\tau_\text{ke}(\mathtt{p_s})
    \: &&= \:
    ((C^\infty(\cZ), \, \mathtt{specific\_momentum}), \, \mathsf{k})
    \\
    &\tau_\text{ke}(\mathtt{m})
    \: &&= \:
    ((C^\infty(\cZ), \, \mathtt{mass}), \, \mathsf{k})
    \,.
  \end{alignedat}
  \label{eq:tau_ke}
\end{equation}

\subsection{Semantics of interconnection patterns}

In general,
the semantics of an interconnection pattern
is given by
three equations per junction.
\begin{subequations}
  The first one is called
  \emph{equality of state},
  as it demands equality of the state variables
  of all connected ports.
  For instance, we have
  \begin{equation}
    \mathtt{ke.m.x}
    \: = \:
    \mathtt{sa.m.x}
    \: = \:
    \mathtt{pps.m.x}
    \,.
  \end{equation}
  Similarly,
  \emph{equality of effort}
  requires the effort variables
  of all connected power ports to be equal.
  We hence have
  \begin{equation}
    \mathtt{ke.m.e}
    \: = \:
    \mathtt{sa.m.e}
    \,.
  \end{equation}
  Finally,
  \emph{equality of net flow}
  demands that
  the sum of the flow variables of all connected inner power ports
  is equal to
  the sum of the flow variables of all connected outer power ports.
  For the present example,
  this gives
  \begin{equation}
    \mathtt{ke.m.f}
    \: + \:
    \mathtt{sa.m.f}
    \: = \: 0
    \,,
  \end{equation}
  where the right hand side is zero
  because no connected power port belongs to the outer box.
  For a junction where boundary ports are connected,
  equality of effort and equality of net flow apply
  in the same way.
  \label{eq:junction}
\end{subequations}

\subsection{Components}

In this subsection,
we discuss each of the five primitive subsystems.

The storage component
$(I_\text{ke}, \, E_\text{ke})$
filling box $\mathtt{ke}$
is defined by
its interface $I_\text{ke}$,
see~\cref{eq:tau_ke},
and
its energy function
$E_\text{ke} \colon \cX_\text{ke} \to \bR$.
Here,
$\cX_\text{ke} = C^\infty(\cZ) \times C^\infty(\cZ)$
is given by the Cartesian product of
the state spaces of the ports in $I_\text{ke}$.
The kinetic energy is given by
\begin{equation*}
  E_\text{ke}(\upsilon, \, \rho)
  \: = \:
  \int_\cZ \frac{1}{2} \cdot \rho \cdot \upsilon^2 \: \dd z
  \,,
\end{equation*}
where
$\upsilon = \mathtt{ke.p_s.x}$
and
$\rho = \mathtt{ke.m.x}$
are introduced for notational convenience.
At a storage component,
the flow variables are
the rates of change of the respective state variables
and
the effort variables are
the partial functional derivatives of the exergy function.
For purely mechanical and electromagnetic forms of energy,
the exergy function is equal to the energy function.
We thus have
\begin{subequations}
  \begin{alignat}{2}
    &\mathtt{ke.p_s.f}
    \: &&= \:
    \dot{\upsilon}
    \\
    &\mathtt{ke.p_s.e}
    \: &&= \:
    \rho \cdot \upsilon
    \\
    &\mathtt{ke.m.f}
    \: &&= \:
    \dot{\rho}
    \\
    &\mathtt{ke.m.e}
    \: &&= \:
    \upsilon^2 / 2
    \,.
  \end{alignat}
  \label{eq:ke_1d}
\end{subequations}
The stored power is consequently given by
\begin{equation*}
  \dot{E}_\text{ke}
  \: = \:
  \int_\cZ \left(
  \mathtt{ke.p_s.e} \cdot \mathtt{ke.p_s.f}
  \, + \,
  \mathtt{ke.m.e} \cdot \mathtt{ke.m.f}
  \right) \dd z
  \,.
\end{equation*}

The storage component
$(I_\text{ie}, \, E_\text{ie})$
filling box $\mathtt{ie}$
has the interface
$I_\text{ie} = (N_\text{ie}, \tau_\text{ie})$
with
$N_\text{ie} = \{ \mathtt{m} \}$
and
$
\tau_\text{ie}(\mathtt{m}) =
((C^\infty(\cZ), \, \mathtt{mass}), \, \mathsf{i})
$.
Its energy function
$E_\text{ie} \colon C^\infty(\cZ) \to \bR$
has the form
\begin{equation*}
  E_\text{ie}(\rho)
  \: = \:
  \int_\cZ U(\rho) \, \dd z
  \,,
\end{equation*}
where
$U \colon \bR \to \bR$
defines the internal energy of the barotropic fluid
in terms of its mass density
$\rho = \mathtt{ie.m.x}$.
We note that
internal energy refers to
energy stored at more microscopic scales,
which are not resolved by
the thermodynamic or macroscopic model.
The derivative
$\mu = \frac{\partial U}{\partial \rho}$
is the chemical potential
and
the pressure is given by
$\pi = \mu \cdot \rho - U(\rho)$.
This is seen by
considering a control volume
that is small enough
such that the contained fluid is in thermodynamic equilibrium.
Let $\bar{U}(m, \, v)$ be the internal energy of the contained fluid,
given in terms of its mass $m$ and volume $v$.
In equilibrium thermodynamics,
the chemical potential is defined by
$\mu = \frac{\partial \bar{U}(m, \, v)}{\partial m}$
and the pressure is defined by
$\pi = -\frac{\partial \bar{U}(m, \, v)}{\partial v}$.
Because mass, volume and energy are extensive quantities,
$\bar{U}$ is a homogeneous function of degree one, meaning that
\begin{equation*}
  \forall \, c \in \bR_{>0} \, \colon \,
  c \cdot \bar{U}(m, \, v)
  \: = \:
  \bar{U}(c \cdot m, \, c \cdot v)
  \,.
\end{equation*}
For $c = 1 \mathrm{m}^3 / v$,
we get
\begin{equation*}
  \frac{\bar{U}(m, \, v)}{v}
  \: = \:
  \frac{\bar{U}(\rho \cdot 1 \mathrm{m}^3, \, 1 \mathrm{m}^3)}{1 \mathrm{m}^3}
  \: \eqqcolon \:
  U(\rho)
  \,,
\end{equation*}
where
$\rho = m / v$.
Based on this,
we obtain
\begin{equation*}
  \mu
  \: = \:
  \frac{\partial \bar{U}(m, \, v)}{\partial m}
  \: = \:
  \frac{\partial \bigl( v \cdot U(\frac{m}{v}) \bigr)}{\partial m}
  \: = \:
  \frac{\partial U(\rho)}{\partial \rho}
\end{equation*}
and
\begin{equation*}
  \begin{split}
    \pi
    \: &= \:
    -\frac{\partial \bar{U}(m, \, v)}{\partial v}
    \: = \:
    -\frac{\partial \bigl( v \cdot U(\frac{m}{v}) \bigr)}{\partial v}
    \\
    \: &= \:
    -\Bigl(
      U(\rho) +
      v \cdot \frac{\partial U(\rho)}{\partial \rho} \cdot
      \Bigl( -\frac{m}{v^2} \Bigr)
    \Bigr)
    \: = \:
    \mu \cdot \rho - U(\rho)
    \,.
  \end{split}
\end{equation*}
Assuming that the exergy reference environment
contains a mass species
with chemical potential
$\textcolor{violet}{\mu_0}$,
the exergy storage function is,
modulo an added constant,
given by
\begin{equation*}
  H_\text{ie}(\rho)
  \: = \:
  \int_\cZ \bigl(
    U(\rho) - \textcolor{violet}{\mu_0} \cdot \rho
  \bigr) \, \dd z
  \,.
\end{equation*}%
The flow and effort variables are then given by
\begin{subequations}
  \begin{align}
    \mathtt{ie.m.f}
    \: &= \:
    \dot{\rho}
    \\
    \mathtt{ie.m.e}
    \: &= \:
    \mu - \textcolor{violet}{\mu_0}
    \,.
  \end{align}%
  \label{eq:ie_1d}%
\end{subequations}
\Cref{sec:metriplectic} provides more insight
regarding the use of exergy as a storage function
for the consistent combination of
reversible and irreversible dynamics.
We note that
the intensive quantities,
which define the reference environment
and are written in \textcolor{violet}{violet},
eventually cancel out in the evolution equations.
Here,
the presence of the mass species
in the environment
and the resulting shift of the effort by
$\textcolor{violet}{\mu_0}$
are not substantial.

The reversible component
$(I_\text{pps}, \, \cD_\text{pps})$
filling box $\mathtt{pps}$
has the interface
$I_\text{pps} = (N_\text{pps}, \tau_\text{pps})$
with
$N_\text{pps} = \{ \mathtt{p_s}, \, \mathtt{p}, \, \mathtt{m} \}$
and
\begin{equation*}
  \begin{alignedat}{2}
    &\tau_\text{pps}(\mathtt{p_s})
    \: &&= \:
    ((C^\infty(\cZ), \, \mathtt{specific\_momentum}), \, \mathsf{k})
    \\
    &\tau_\text{pps}(\mathtt{p})
    \: &&= \:
    ((C^\infty(\cZ), \, \mathtt{momentum}), \, \mathsf{k})
    \\
    &\tau_\text{pps}(\mathtt{m})
    \: &&= \:
    (C^\infty(\cZ), \, \mathtt{mass})
    \,.
  \end{alignedat}
\end{equation*}
The reversible transformation between
the two representations of momentum
is expressed by
the Dirac structure $\cD_\text{pps}$
given by
\begin{equation}
  \begin{bmatrix}
      \mathtt{pps.p_s.f} \\
      \mathtt{pps.p.e}
  \end{bmatrix}
  \: = \:
  \begin{bmatrix}
      0 & -1/\rho \\
      1/\rho & 0
  \end{bmatrix}
  \,
  \begin{bmatrix}
      \mathtt{pps.p_s.e} \\
      \mathtt{pps.p.f}
  \end{bmatrix}
  \,,
  \label{eq:pps_1d}
\end{equation}
where
$\rho = \mathtt{pps.m.x}$.
Due to the skew-symmetry of the matrix in~\cref{eq:pps_1d},
the net power at the component vanishes, i.e.
\begin{equation*}
  \int_\cZ \left(
    \mathtt{pps.p_s.e} \cdot \mathtt{pps.p_s.f} +
    \mathtt{pps.p.e} \cdot \mathtt{pps.p.f}
  \right) \dd z
  \: = \: 0
  \,.
\end{equation*}

The reversible component
$(I_\text{sa}, \, \cD_\text{sa})$
filling box $\mathtt{sa}$
has the interface
$I_\text{sa} = (N_\text{sa}, \tau_\text{sa})$
with
$N_\text{sa} = \{ \mathtt{p_s}, \, \mathtt{m}, \, \mathtt{b_k} \}$
and
\begin{equation*}
  \begin{alignedat}{2}
    &\tau_\text{sa}(\mathtt{p_s})
    \: &&= \:
    ((C^\infty(\cZ), \, \mathtt{specific\_momentum}), \, \mathsf{k})
    \\
    &\tau_\text{sa}(\mathtt{m})
    \: &&= \:
    ((C^\infty(\cZ), \, \mathtt{mass}), \, \mathsf{k})
    \,.
  \end{alignedat}
\end{equation*}
In this article,
we don't specify the attributes of boundary ports.
%
%
Self-advection of the fluid's kinetic energy
is expressed by
the Stokes-Dirac structure $\cD_\text{sa}$
given by
\begin{subequations}
  \begin{align}
    \Biggl[
      \begin{array}{c}
        \mathtt{sa.p_s.f} \\
        \mathtt{sa.m.f}
      \end{array}
    \Biggr]
    \: &= \:
    \Biggl[
      \begin{array}{cc}
        0 & \frac{\partial}{\partial z}(\_) \\
        \frac{\partial}{\partial z}(\_) & 0
      \end{array}
    \Biggr]
    \,
    \Biggl[
      \begin{array}{c}
        \mathtt{sa.p_s.e} \\
        \mathtt{sa.m.e}
      \end{array}
    \Biggr]
    \label{eq:sa_1d_domain}
    \\
    \Biggl[
      \begin{array}{c}
        \mathtt{sa.b_k.f} \\
        \mathtt{sa.b_k.e}
      \end{array}
    \Biggr]
    \: &= \:
    \Biggl[
      \begin{array}{cc}
        -(\_)\vert_{\partial \cZ} & 0 \\
        0 & (\_)\vert_{\partial \cZ}
      \end{array}
    \Biggr]
    \,
    \Biggl[
      \begin{array}{c}
        \mathtt{sa.p_s.e} \\
        \mathtt{sa.m.e}
      \end{array}
    \Biggr]
    \,,
  \end{align}%
  \label{eq:sa_1d}%
\end{subequations}
where
$\_$ denotes a placeholder
for the respective component of the vector of effort variables,
$\frac{\partial}{\partial z}$
is the spatial derivative,
and
$\vert_{\partial \cZ}$
denotes restriction to the boundary.
The matrix in~\cref{eq:sa_1d_domain}
is formally skew-symmetric,
since the adjoint of
$\frac{\partial}{\partial z}(\_)$ is
$-\frac{\partial}{\partial z}(\_)$.
In other words,
the minus sign required for skew-symmetry
cancels with
the minus sign resulting from integration by parts.
The net power at all three ports is zero,
since
\begin{equation*}
  \begin{split}
    &\int_\cZ \bigl(
      \mathtt{p_s.e} \cdot \mathtt{p_s.f}
      \: + \:
      \mathtt{m.e} \cdot \mathtt{m.f}
    \bigr) \dd z
    \: = \:
    \int_\cZ \frac{\partial}{\partial z} \bigl(
      \mathtt{p_s.e} \cdot \mathtt{m.e}
    \bigr) \, \dd z
    \\
    &= \:
    \bigl( \mathtt{m.e} \cdot \mathtt{p_s.e} \bigr)\vert_{z=L}
    -
    \bigl( \mathtt{m.e} \cdot \mathtt{p_s.e} \bigr)\vert_{z=0}
    \: = \:
    -\int_{\partial \cZ} \mathtt{b_k.e} \cdot \mathtt{b_k.f}
    \,.
  \end{split}
\end{equation*}
The `integral' over $\partial \cZ$ is simply
the sum of the integrand evaluated at the two points,
taking into account their opposite orientation, i.e.
\begin{equation}
  \int_{\partial \cZ} (\_) \, \dd z
  \: = \:
  (\_)\vert_{z=L} - (\_)\vert_{z=0}
  \,.
  \label{eq:stokes_1d_rhs}
\end{equation}
If the boundary port $\mathtt{b_k}$
is not connected with another port,
equality of net flow gives
$\mathtt{b_k.f} = 0$.
This corresponds to an isolated boundary condition,
since $\mathtt{b_k.f}$ is
the incoming mass flux at the boundary.

The reversible component
$(I_\text{adv}, \, \cD_\text{adv})$
filling box $\mathtt{adv}$
has the interface
$I_\text{adv} = (N_\text{adv}, \tau_\text{adv})$
with
$N_\text{adv} = \{ \mathtt{p}, \, \mathtt{m}, \, \mathtt{b_m} \}$
and
\begin{equation*}
  \begin{alignedat}{2}
    &\tau_\text{adv}(\mathtt{p})
    \: &&= \:
    ((C^\infty(\cZ), \, \mathtt{momentum}), \, \mathsf{p})
    \\
    &\tau_\text{adv}(\mathtt{m})
    \: &&= \:
    ((C^\infty(\cZ), \, \mathtt{mass}), \, \mathsf{p})
    \,.
  \end{alignedat}
\end{equation*}
Advection of internal energy
is expressed by
the Stokes-Dirac structure $\cD_\text{adv}$
given by
\begin{subequations}
  \begin{align}
    \Biggl[
      \begin{array}{c}
        \mathtt{adv.p.f} \\
        \mathtt{adv.m.f}
      \end{array}
    \Biggr]
    \: &= \:
    \Biggl[
      \begin{array}{cc}
        0 & \rho \cdot \frac{\partial}{\partial z}(\_) \\
        \frac{\partial}{\partial z}(\rho \cdot \_) & 0
      \end{array}
    \Biggr]
    \,
    \Biggl[
      \begin{array}{c}
        \mathtt{adv.p.e} \\
        \mathtt{adv.m.e}
      \end{array}
    \Biggr]
    \label{eq:adv_1d_domain}
    \\
    \Biggl[
      \begin{array}{c}
        \mathtt{adv.b_m.f} \\
        \mathtt{adv.b_m.e}
      \end{array}
    \Biggr]
    \: &= \:
    \Biggl[
      \begin{array}{cc}
        -(\rho \cdot \_)\vert_{\partial \cZ} & 0 \\
        0 & (\_)\vert_{\partial \cZ}
      \end{array}
    \Biggr]
    \,
    \Biggl[
      \begin{array}{c}
        \mathtt{adv.p.e} \\
        \mathtt{adv.m.e}
      \end{array}
    \Biggr]
    \,,
  \end{align}
  \label{eq:adv_1d}
\end{subequations}
where
$\rho = \mathtt{adv.m.x}$.
Since,
for any $\varphi, \, \psi \in C^\infty(\cZ)$,
\begin{equation*}
  \begin{split}
    \int_\cZ
      \varphi \cdot
      \frac{\partial}{\partial z}(\rho \cdot \psi) \:
    \dd z
    \: = \:
    &-\int_\cZ
      \frac{\partial}{\partial z}(\varphi) \cdot
      \rho \cdot \psi \:
    \dd z
    \\
    &+ \: \int_{\partial \cZ}
      \varphi \cdot \rho \cdot \psi \:
    \dd z
    \,,
  \end{split}
\end{equation*}
the adjoint of
$\frac{\partial}{\partial z}(\rho \cdot \_)$ is
$-\rho \cdot \frac{\partial}{\partial z}(\_)$.
Thus,
the matrix in~\cref{eq:adv_1d_domain} is
formally skew-symmetric.
Again,
we split the boundary term
arising from integration by parts
into flow and effort variables of the boundary port
such that
$\mathtt{b_m.f} = 0$
is a physically possible
isolated boundary condition.
Since
$\mathtt{adv.m.e} = \mathtt{pps.p.e}$
is the velocity,
$\mathtt{b_m.f}$
is again the incoming mass flow rate.
It can be easily checked that
the net power at all three ports is zero.

\subsection{Composite system}

Finally,
we collect the equations
that define the semantics of the composite system.
Combining~%
\cref{eq:ke_1d,eq:ie_1d,eq:pps_1d,eq:sa_1d,eq:adv_1d}
with
the equations associated to
the interconnection pattern in~\cref{fig:ibf}
(see~\cref{eq:junction})
and
eliminating port variables gives
the following system of equations
for the composite system.

The equation for balance of mass
is given by
\begin{equation*}
  \begin{split}
    \dot{\rho}
    \: &= \:
    \mathtt{ke.m.f}
    \: = \:
    -\mathtt{sa.m.f}
    \: = \:
    -\frac{\partial}{\partial z}(\mathtt{sa.p_s.e})
    \\
    \: &= \:
    -\frac{\partial}{\partial z}(\mathtt{ke.p_s.e})
    \: = \:
    -\frac{\partial}{\partial z}(\rho \cdot \upsilon)
    \,.
  \end{split}
\end{equation*}
This also follows
if we start with
$\dot{\rho} = \mathtt{ie.m.f}$,
since mass is a state variable for
both kinetic energy and internal energy.
%

The equation for
balance of (specific) momentum
reads
\begin{equation*}
  \begin{split}
    \dot{\upsilon}
    \: &= \:
    \mathtt{ke.p_s.f}
    \: = \:
    -\mathtt{sa.p_s.f} - \mathtt{pps.p_s.f}
    \\
    \: &= \:
    -\frac{\partial}{\partial z}(\mathtt{sa.m.e}) +
    \frac{1}{\rho} \cdot \mathtt{pps.p.f}
    \\
    \: &= \:
    -\frac{\partial}{\partial z}(\mathtt{ke.m.e}) -
    \frac{1}{\rho} \cdot \mathtt{adv.p.f}
    \\
    \: &= \:
    -\frac{\partial}{\partial z} \Bigl( \frac{1}{2} \cdot \upsilon^2 \Bigr) -
    \frac{1}{\rho} \cdot \rho \cdot
    \frac{\partial}{\partial z}(\mu - \textcolor{violet}{\mu_0})
    \\
    \: &= \:
    -\upsilon \cdot \frac{\partial \upsilon}{\partial z} -
    \frac{1}{\rho} \cdot \frac{\partial \pi}{\partial z}
    \,.
  \end{split}
\end{equation*}
The last equality follows from
\begin{equation*}
  \frac{\partial \pi}{\partial z} =
  \frac{\partial}{\partial z}\bigl( \mu \cdot \rho - U(\rho) \bigr) =
  \frac{\partial \mu}{\partial z} \cdot \rho +
  \mu \cdot \frac{\partial \rho}{\partial z} -
  \frac{\partial U(\rho)}{\partial \rho} \cdot \frac{\partial \rho}{\partial z}
  \,,
\end{equation*}
where the last two terms cancel.
The balance equation can be rewritten as
\begin{equation*}
  \rho \cdot \left(
    \dot{\upsilon}
    \, + \,
    \upsilon \cdot \frac{\partial \upsilon}{\partial z}
  \right)
  \: = \:
  -\frac{\partial \pi}{\partial z}
  \,,
\end{equation*}
which can be recognized as
Newton's second law of motion applied to a fluid parcel,
since the term in parentheses is
the material derivative of the velocity.

Regarding boundary ports,
the incoming mass flux at the boundary is given by
$
-(\rho \cdot \upsilon)\vert_{\partial \cZ} =
\mathtt{b_k.f} = \mathtt{b_m.f}
$.
The effort variable
$
\mathtt{b_k.e} =
(\upsilon^2 / 2)\vert_{\partial \cZ}
$
is the kinetic energy per unit of incoming mass.
The chemical potential
$
\mu\vert_{\partial \cZ} =
\textcolor{violet}{\mu_0} + \mathtt{b_m.e}
$
is the internal energy per unit of mass
and
$
\mathtt{b_m.e} =
\mu\vert_{\partial \cZ} - \textcolor{violet}{\mu_0}
$
is the corresponding exergy content.

\section{Ideal fluid model}%
\label{sec:ideal-fluid}

\begin{figure}
  \centering
	\includegraphics[width=5.8cm]{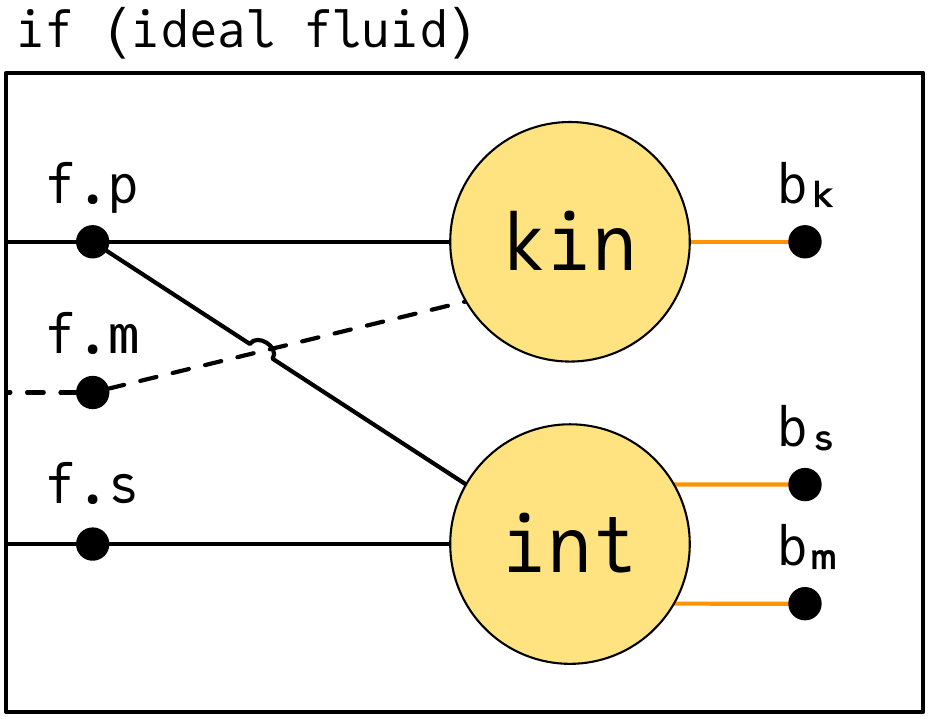}
  \caption{%
    Interconnection pattern
    for an ideal fluid model.
    Box $\mathtt{kin}$ represents
    the kinetic energy subsystem
    and
    box $\mathtt{int}$ represents
    the internal energy subsystem.
    The boundary ports
    $\mathtt{b_k}$,
    $\mathtt{b_m}$ and
    $\mathtt{b_s}$,
    which account for
    advection of kinetic and internal energy
    across $\partial \cZ$,
    are not exposed,
    leading to a model
    with impermeable boundary.
  }%
  \label{fig:if_expanded}
\end{figure}

\begin{figure}
  \centering
	\includegraphics[width=5.8cm]{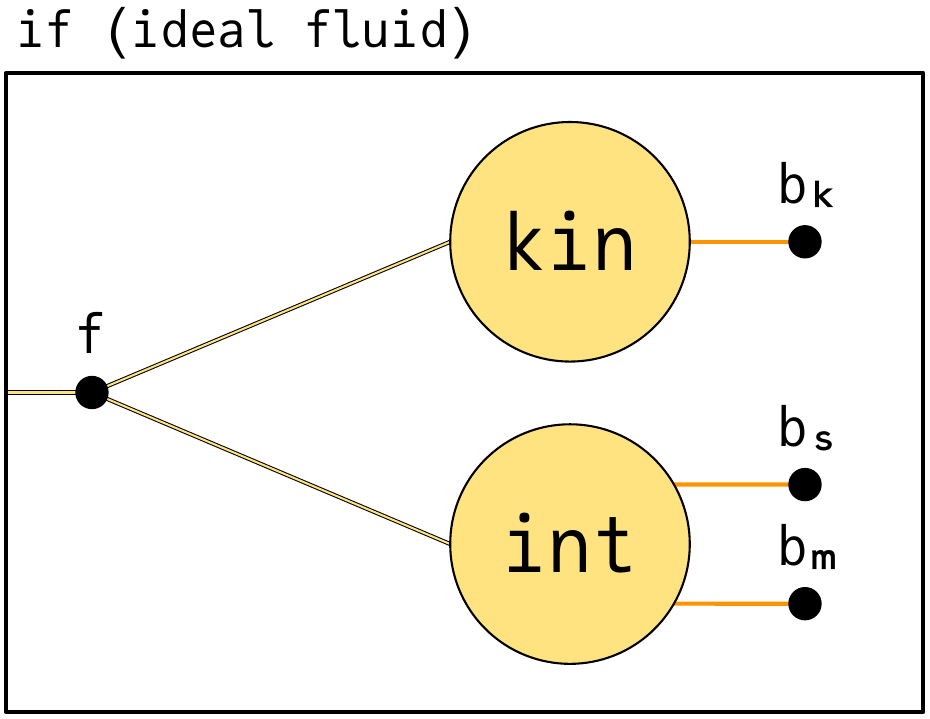}
  \caption{%
    The pattern from~\cref{fig:if_expanded}
    is shown in a simplified form,
    where multiple ports with a common prefix are drawn as
    a single line, called a multiport.
  }%
  \label{fig:if}
\end{figure}

In this section,
we implement
an EPHS model of an ideal compressible fluid,
whose spatial domain is
a three-dimensional Riemannian manifold
$(\cZ, \, g)$
with boundary $\partial \cZ$,
see~\cref{sec:geometry}.
We note that the fluid models in this and the next section
can also be used if
the spatial domain is one- or two-dimensional.
Regarding the equations,
nothing changes,
except that in the two-dimensional case
a few minus signs need to be added in certain places,
see~\onlinecite{2022LohmayerLeyendecker2}.
Similar to
the port-Hamiltonian ideal fluid model
in~\onlinecite{%
2021RashadCalifanoSchullerStramigioli1,%
2021RashadCalifanoSchullerStramigioli2},
we view the model as being decomposed into
a kinetic energy system
and
an internal energy system,
see~\cref{fig:if_expanded}.
The interconnection pattern
is redrawn in~\cref{fig:if},
using a syntactic sugar called multiports.
This means that
multiple ports with a common prefix
are drawn as a single connection.
The simplified graphical representation
implies that
ports with the same name are connected.
We note that this does not change
the mathematical content of an interconnection pattern,
as defined in~\onlinecite{2024LohmayerLynchLeyendecker}.

\subsection{Kinetic energy system}%

\begin{figure}
  \centering
	\includegraphics[width=5.8cm]{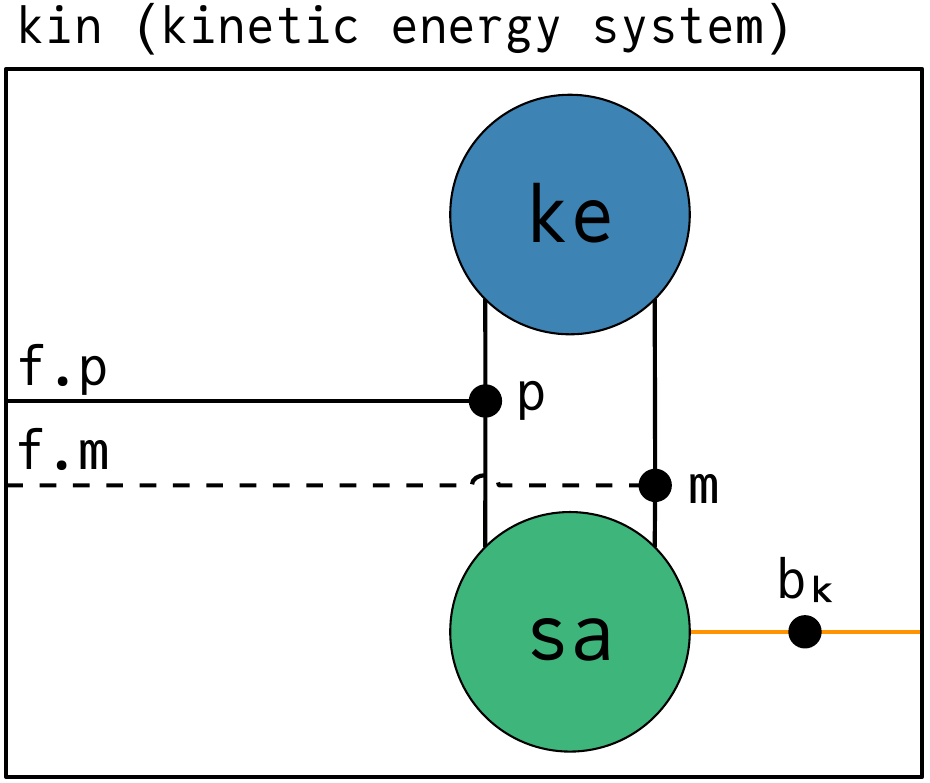}
  \caption{%
    Interconnection pattern
    for the canonical definition of
    the kinetic energy system.
    Box $\mathtt{ke}$ represents
    storage of kinetic energy,
    which is exchanged in terms of
    momentum via port $\mathtt{ke.p}$
    and
    mass via port $\mathtt{ke.m}$.
    Box $\mathtt{sa}$ represents
    self-advection of kinetic energy
    and
    the boundary port $\mathtt{b_k}$
    accounts for
    advection across $\partial \cZ$.
    The outer port $\mathtt{f.p}$
    allows for
    exchange of kinetic energy
    with other systems on the same domain $\cZ$
    and
    the outer state port $\mathtt{f.m}$
    shares
    information about the fluid mass
    with other systems on $\cZ$.
  }%
  \label{fig:kin}
\end{figure}

\begin{figure}
  \centering
	\includegraphics[width=5.8cm]{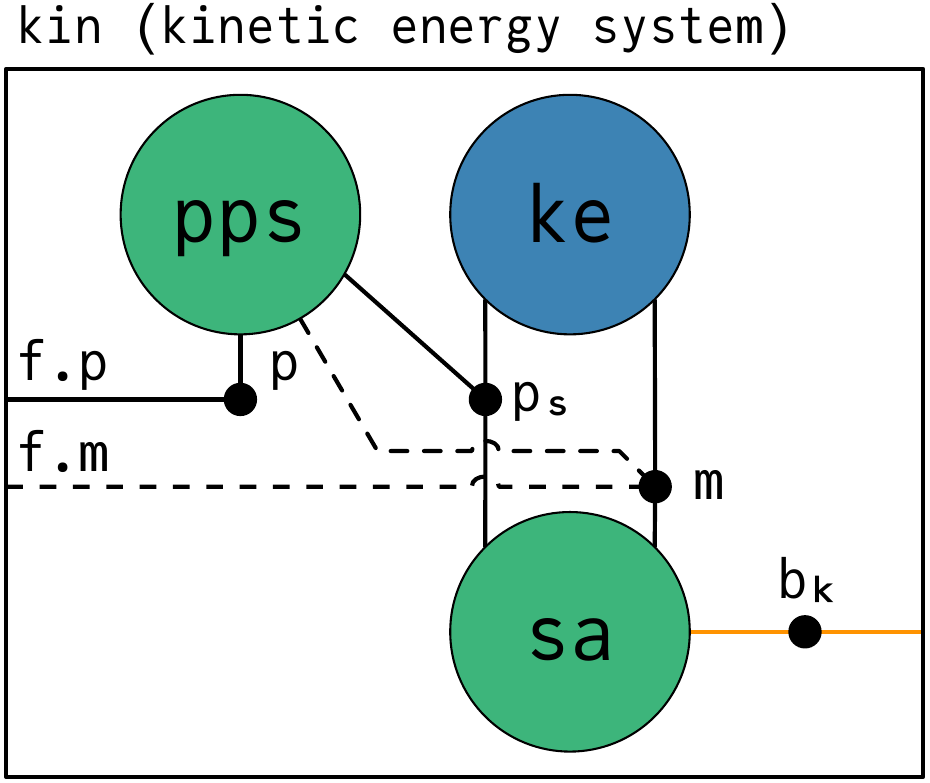}
  \caption{%
    Interconnection pattern
    of the kinetic energy system.
    In contrast to
    the pattern
    in~\cref{fig:kin},
    kinetic energy
    is expressed in terms of
    specific momentum (or velocity),
    rather than momentum density.
    To retain the same outer interface,
    the additional box $\mathtt{pps}$ represents the
    reversible transformation between
    the two alternative state variables.
  }%
  \label{fig:kin_pps}
\end{figure}

The canonical definition of
the kinetic energy system
uses
the interconnection pattern
in~\cref{fig:kin}
with
a storage component
filling box $\mathtt{ke}$
that models storage of kinetic energy
and
a reversible component
filling box $\mathtt{sa}$
that models self-advection of kinetic energy.
We however work with the pattern
shown in~\cref{fig:kin_pps}.
The additional reversible component
filling box $\mathtt{pps}$
performs a variable transformation between
momentum density
and
specific momentum,
which is equivalent to velocity.
Using velocity as a state variable
is common in fluid mechanics,
and it simplifies
the expression for the Stokes-Dirac structure
that defines
the reversible component in box $\mathtt{sa}$,
as shown in~\onlinecite{2021RashadCalifanoSchullerStramigioli1}.

Next,
we discuss each of the three primitive systems
filling the inner boxes
of the pattern in~\cref{fig:kin_pps},
and then
we collect the equations that
define the semantics of the composite system.

\subsubsection{Storage of kinetic energy}

The fluid mass
$\tilde{m} \in \tilde{\Omega}^3(\cZ)$
is naturally expressed as
a twisted $3$-form,
as its integral $\int_\cZ \tilde{m}$
gives the total mass.
The mass density is
then given by
the Hodge dual
$\rho = \hodge \tilde{m} \in \Omega^0(\cZ)$.
The velocity
$u \in \Gamma(\mathrm{T} \cZ)$
is naturally expressed as
a vector field on $\cZ$,
since an integral curve of $u$
describes the trajectory of a fluid particle.
A force
$\xi \in \Gamma(\mathrm{T}^* \cZ)$
acting on the fluid
is understood as
a covector field on $\cZ$,
such that
the associated mechanical power
is given by the duality pairing of force and velocity
$\langle \xi \mid u \rangle$.
According to Newton's second law,
the time derivative of the momentum $p$
is equal to the force $\xi$ acting on the fluid,
i.e.~$\dot{p} = \xi$.
Hence,
we also have
$p \in \Gamma(\mathrm{T}^* \cZ)$.
The specific momentum $\upsilon$
is then given by the covector field
$
\upsilon = p / \rho
\in \Gamma(\mathrm{T}^* \cZ) \cong \Omega^1(\cZ)
$,
and it holds that
$u = \upsilon^\sharp$.

We want to mention that
momentum and forces can also be modeled as
$\mathrm{T}^* \cZ$-valued twisted $3$-forms.
Applying the Hodge star to the form part
of the momentum
then gives the specific momentum
as a $\mathrm{T}^* \cZ$-valued $0$-form.
Although this better reflects
the geometric nature of these quantities,
we avoid bundle-valued forms,
as they add complexity,
especially in terms of the notation.
However,
for modeling shear viscosity in~\cref{sec:nsf},
bundle-valued forms become a necessity.

The storage component
$(I_\text{ke}, \, E_\text{ke})$
filling box $\mathtt{ke}$
is defined by
its interface
$I_\text{ke} = (\{ \mathtt{p_s}, \, \mathtt{m} \}, \, \tau_\text{ke})$
with
\begin{alignat*}{2}
  &\tau_\text{ke}(\mathtt{p_s})
  \: &&= \:
  ((\Omega^1(\cZ), \, \mathtt{specific\_momentum}), \, \mathsf{k})
  \\
  &\tau_\text{ke}(\mathtt{m})
  \: &&= \:
  ((\tilde{\Omega}^3(\cZ), \, \mathtt{mass}), \, \mathsf{k})
\end{alignat*}
and
its energy function
$E_\text{ke} \colon \Omega^1(\cZ) \times \tilde{\Omega}^3(\cZ) \to \bR$
given by
\begin{equation*}
  \begin{split}
    E_\text{ke}(\upsilon, \, \tilde{m})
		\: &= \:
    \int_\cZ \frac{1}{2} \cdot \tilde{m} \cdot g(\upsilon^\sharp, \, \upsilon^\sharp)
		\: = \:
    \int_\cZ \frac{1}{2} \cdot \tilde{m} \cdot \iota_{\upsilon^\sharp} \upsilon
		\\
    \overset{\eqref{eq:interior_product}}&{=} \:
    \int_\cZ \frac{1}{2} \cdot \tilde{m} \cdot
    \hodge (\upsilon \wedge \hodge \upsilon)
		\: = \:
    \int_\cZ \frac{1}{2} \cdot \hodge \tilde{m} \cdot
    (\upsilon \wedge \hodge \upsilon)
    \,,
  \end{split}
\end{equation*}
where
$\upsilon = \mathtt{p_s.x}$
and
$\tilde{m} = \mathtt{m.x}$.

The effort variables
are defined as the
partial functional derivatives
of the exergy storage function
$H_\text{ke} = E_\text{ke}$.
For instance,
$\mathtt{p_s.e} = \delta_\upsilon H_\text{ke}$
is defined by
\begin{equation*}
  H_\text{ke}(\upsilon + \epsilon \cdot \delta \upsilon, \, \tilde{m})
  \: = \:
  H_\text{ke}(\upsilon, \, \tilde{m})
  \, + \,
  \epsilon \cdot \int_\cZ \delta_\upsilon H_\text{ke} \wedge \delta \upsilon
  \, + \,
  \cO(\epsilon^2)
\end{equation*}
for $\epsilon \rightarrow 0$
and any $\delta \upsilon \in \Omega^1(\cZ)$.
%
%
The flow and effort variables are hence given by
\begin{subequations}
  \begin{alignat}{2}
    &\mathtt{p_s.f}
    \: &&= \:
    \dot{\upsilon}
    \: \in \: \Omega^1(\cZ)
    \\
    &\mathtt{p_s.e}
    \: &&= \:
    \hodge \tilde{m} \cdot \hodge \upsilon
    \: \in \: \tilde{\Omega}^2(\cZ)
    \\
    &\mathtt{m.f}
    \: &&= \:
    \dot{\tilde{m}}
    \: \in \: \tilde{\Omega}^3(\cZ)
    \\
    &\mathtt{m.e}
    \: &&= \:
    \hodge (\upsilon \wedge \hodge \upsilon) / 2
    \: \in \: \Omega^0(\cZ)
    \,.
  \end{alignat}
  \label{eq:ke}
\end{subequations}
The stored power is consequently given by
\begin{equation*}
  \dot{E}_\text{ke}
  \: = \:
  \int_\cZ \left(
  \mathtt{p_s.e} \wedge \mathtt{p_s.f}
  \, + \,
  \mathtt{m.e} \wedge \mathtt{m.f}
  \right)
  \,.
\end{equation*}

\subsubsection{Transformation between momentum and specific momentum}

The reversible component
filling box $\mathtt{pps}$
transforms between
momentum density at port $\mathtt{p}$
and
specific momentum at port $\mathtt{p_s}$.
More precisely,
it relates
the flow variable
$\mathtt{p.f} \in \Omega^1(\cZ)$
representing an external force acting on the fluid
to
the flow variable
$\mathtt{p_s.f} \in \Omega^1(\cZ)$
representing the same force given per unit of mass,
and dually,
it relates
the effort variable
$
\mathtt{p_s.e} =
\hodge \tilde{m} \cdot \hodge \upsilon \in \tilde{\Omega}^2(\cZ)
$
representing the mass flux
to
the effort variable
$
\mathtt{p.e} =
\hodge \upsilon \in \tilde{\Omega}^2(\cZ)
$
representing the velocity or volume flux
(i.e.~fluid velocity \emph{through} area).

The reversible component
$(I_\text{pps}, \, \cD_\text{pps})$
filling box $\mathtt{pps}$
is defined by its interface
$
I_\text{pps} =
(\{ \mathtt{p_s}, \, \mathtt{p}, \, \mathtt{m} \}, \tau_\text{pps})
$
with
\begin{equation*}
  \begin{alignedat}{2}
    &\tau_\text{pps}(\mathtt{p_s})
    \: &&= \:
    ((\Omega^1(\cZ), \, \mathtt{specific\_momentum}), \, \mathsf{k})
    \\
    &\tau_\text{pps}(\mathtt{p})
    \: &&= \:
    ((\Omega^1(\cZ), \, \mathtt{momentum}), \, \mathsf{k})
    \\
    &\tau_\text{pps}(\mathtt{m})
    \: &&= \:
    (\tilde{\Omega}^3(\cZ), \, \mathtt{mass})
  \end{alignedat}
\end{equation*}
and
its Dirac structure $\cD_\text{pps}$
given by
\begin{equation}
  \begin{bmatrix}
      \mathtt{p_s.f} \\
      \mathtt{p.e}
  \end{bmatrix}
  \: = \:
  \begin{bmatrix}
      0 & -1 / \hodge \tilde{m} \\
      1 / \hodge \tilde{m} & 0
  \end{bmatrix}
  \,
  \begin{bmatrix}
      \mathtt{p_s.e} \\
      \mathtt{p.f}
  \end{bmatrix}
  \,,
  \label{eq:pps}
\end{equation}
where
$\tilde{m} = \mathtt{m.x}$.

Due to the skew-symmetry of the matrix in~\cref{eq:pps},
the net power at the component vanishes, i.e.
\begin{equation*}
  \int_\cZ \left(
    \mathtt{p_s.e} \wedge \mathtt{p_s.f} +
    \mathtt{p.e} \wedge \mathtt{p.f}
  \right) \dd z
  \: = \: 0
  \,.
\end{equation*}

\subsubsection{Self-advection of kinetic energy}

The reversible component
filling box $\mathtt{sa}$ models
the self-advection of kinetic energy.
It encapsulates
a generalization of
the well-known Lie-Poisson structure for ideal fluids,
which includes boundary ports,
see e.g.~\onlinecite{2021RashadCalifanoSchullerStramigioli1}.

The reversible component
$(I_\text{sa}, \, \cD_\text{sa})$
filling box $\mathtt{sa}$
is defined by
its interface
$I_\text{sa} = (\{ \mathtt{p_s}, \, \mathtt{m}, \, \mathtt{b_k} \}, \tau_\text{sa})$
with
\begin{equation*}
  \begin{alignedat}{2}
    &\tau_\text{sa}(\mathtt{p_s})
    \: &&= \:
    ((\Omega^1(\cZ), \, \mathtt{specific\_momentum}), \, \mathsf{k})
    \\
    &\tau_\text{sa}(\mathtt{m})
    \: &&= \:
    ((\tilde{\Omega}^3(\cZ), \, \mathtt{mass}), \, \mathsf{k})
  \end{alignedat}
\end{equation*}
and
the Stokes-Dirac structure $\cD_\text{sa}$
given by
\begin{subequations}
  \begin{align}
    \Biggl[
      \begin{array}{c}
        \mathtt{p_s.f} \\
        \mathtt{m.f}
      \end{array}
    \Biggr]
    \: &= \:
    \Biggl[
      \begin{array}{cc}
        S(\_) & \dd(\_) \\
        \dd(\_) & 0
      \end{array}
    \Biggr]
    \,
    \Biggl[
      \begin{array}{c}
        \mathtt{p_s.e} \\
        \mathtt{m.e}
      \end{array}
    \Biggr]
    \label{eq:sa_domain}
    \\
    \Biggl[
      \begin{array}{c}
        \mathtt{b_k.f} \\
        \mathtt{b_k.e}
      \end{array}
    \Biggr]
    \: &= \:
    \Biggl[
      \begin{array}{cc}
        -i^*(\_) & 0 \\
        0 & i^*(\_)
      \end{array}
    \Biggr]
    \,
    \Biggl[
      \begin{array}{c}
        \mathtt{p_s.e} \\
        \mathtt{m.e}
      \end{array}
    \Biggr]
    \,,
    \label{eq:sa_boundary}
  \end{align}%
  \label{eq:sa}%
\end{subequations}
where
\begin{equation*}
  S(\_)
  \: = \:
  \frac{1}{\hodge \tilde{m}} \cdot
  \iota_{(\sharp \circ \hodge)(\_)}\dd \upsilon
  \: \overset{\eqref{eq:interior_product}}{=} \:
  -\frac{1}{\hodge \tilde{m}} \cdot
  \hodge \bigl( \hodge (\_) \wedge \hodge \dd \upsilon \bigr)
\end{equation*}
and
$i^*$ denotes the pullback along the inclusion
$i \colon \partial \cZ \to \cZ$.

The name Stokes-Dirac structure refers to the fact that
Stokes theorem, see~\cref{eq:stokes},
is used to define the boundary ports in~\cref{eq:sa_boundary}
such that
the formally skew-symmetric operator matrix in~\cref{eq:sa_domain}
defines a power-preserving relation
among all port variables:
\begin{equation*}
  \begin{split}
    &\int_\cZ \Bigl(
      \mathtt{p_s.e} \wedge \mathtt{p_s.f}
      +
      \mathtt{m.e} \wedge \mathtt{m.f}
    \Bigr)
    \, =
    \\
    &\int_\cZ
      \mathtt{p_s.e} \wedge \Bigl(
        -\frac{1}{\hodge \tilde{m}} \cdot \hodge \bigl( \hodge(\mathtt{p_s.e}) \wedge \hodge \dd \upsilon \bigr)
      \Bigr)
    \, +
    \\
    &\int_\cZ \Bigl(
      \mathtt{p_s.e} \wedge \dd(\mathtt{m.e})
      \, + \,
      \mathtt{m.e} \wedge \dd(\mathtt{p_s.e})
    \Bigr)
    \, =
    \\
    &\int_\cZ \dd(\mathtt{p_s.e} \wedge \mathtt{m.e})
    \, = \,
    \int_{\partial \cZ} i^*(\mathtt{p_s.e} \wedge \mathtt{m.e})
    \, =
    \\
    &\int_{\partial \cZ} i^*(\mathtt{p_s.e}) \wedge i^*(\mathtt{m.e})
    \, = \,
    -\int_{\partial \cZ} \mathtt{b_k.e} \wedge \mathtt{b_k.f}
    \,.
  \end{split}
\end{equation*}
The term related to $S$ on the second line vanishes since
\begin{equation*}
  \mathtt{p_s.e} \wedge \hodge \bigl( \hodge(\mathtt{p_s.e}) \wedge \hodge \dd \upsilon \bigr)
  \, = \,
  \hodge(\mathtt{p_s.e}) \wedge \hodge(\mathtt{p_s.e}) \wedge \hodge \dd \upsilon
  \, = \,
  0
  \,.
\end{equation*}

\subsubsection{Interconnected kinetic energy system}

Combining~%
\cref{eq:ke,eq:pps,eq:sa}
with
the equations for
the interconnection pattern in~\cref{fig:kin_pps}
and
eliminating port variables gives
the following equations
for the composite system:
\begin{subequations}
	\begin{align}
		\dot{\upsilon}
		\: &= \:
    \hodge(\upsilon \wedge \hodge \dd \upsilon)
    -\dd \bigl( \hodge (\upsilon \wedge \hodge \upsilon) / 2 \bigr)
    +\frac{1}{\hodge \tilde{m}} \cdot \mathtt{f.p.f}%
		\\
    \dot{\tilde{m}}
		\: &= \:
    -\dd (\hodge \tilde{m} \cdot \hodge \upsilon)
		\\
		\mathtt{f.p.e}
		\: &= \:
		\hodge \upsilon%
		\\
		\mathtt{b_k.f}
		\: &= \:
    -i^*(\hodge \tilde{m} \cdot \hodge \upsilon)
		\\
		\mathtt{b_k.e}
		\: &= \:
		i^* \bigl( \hodge (\upsilon \wedge \hodge \upsilon) / 2 \bigr)
    \,.
	\end{align}%
	\label{eq:kin}%
\end{subequations}

The term
$\dd(\hodge \tilde{m} \cdot \hodge \upsilon)$
is equal to
the Lie derivative
$\cL_u \tilde{m}$,
since $\tilde{m}$ is advected by
the flow of the time-dependent vector field $u = \upsilon^\sharp$.
The flow variable $\mathtt{f.p.f}$ represents
an external force acting on the fluid particles
and
the effort variable $\mathtt{f.p.e}$ is
the velocity
(or volume flux through a surface).
The flow variable $\mathtt{b_k.f}$ is
the incoming mass flux through $\partial \cZ$
and
the effort variable $\mathtt{b_k.e}$ is
the kinetic energy per unit of mass.

\subsection{Internal energy system}%

\begin{figure}
  \centering
	\includegraphics[width=5.8cm]{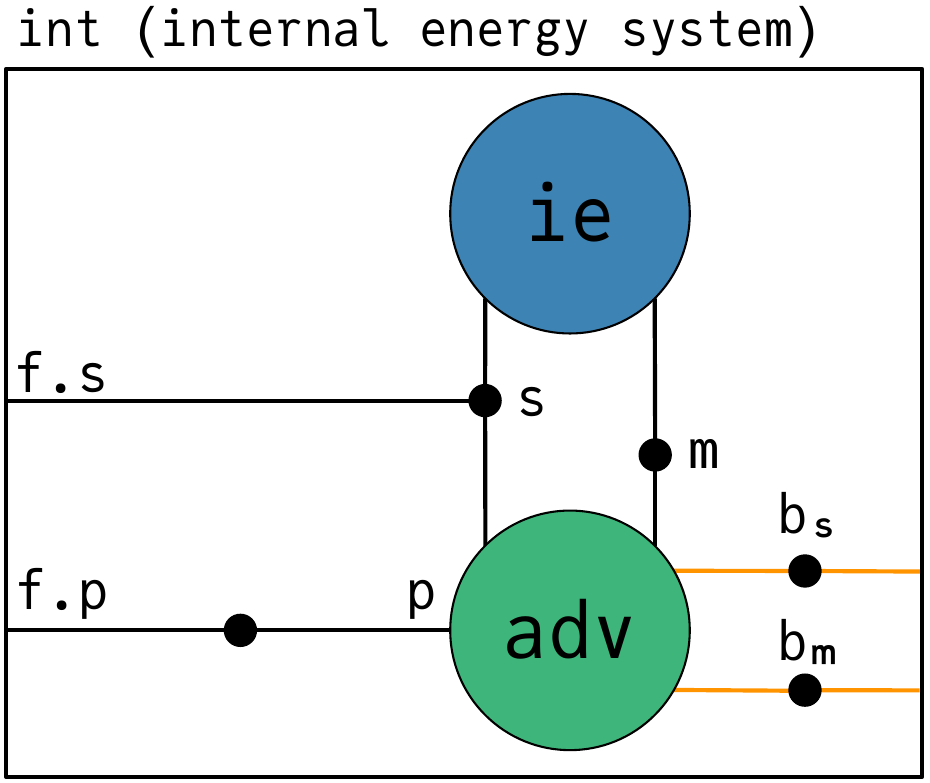}
  \caption{%
    Interconnection pattern
    of the internal energy system.
    Box $\mathtt{ie}$ represents
    storage of internal energy,
    which is exchanged in terms of
    entropy via port $\mathtt{ie.s}$
    and
    mass via port $\mathtt{ie.m}$.
    Box $\mathtt{adv}$ represents
    advection of internal energy
    and
    the boundary ports
    $\mathtt{b_s}$ and $\mathtt{b_m}$
    account for
    advection
    across $\partial \cZ$.
    The outer ports $\mathtt{f.s}$ and $\mathtt{f.p}$
    respectively allow for
    exchange of thermal and kinetic energy
    with other systems on the same domain $\cZ$.
  }%
  \label{fig:int}
\end{figure}

The internal energy system is defined by
the interconnection pattern
shown in~\cref{fig:int} with
a storage component
filling box $\mathtt{ie}$
that models storage of internal energy
and
a reversible component
filling box $\mathtt{adv}$
that models advection of internal energy.

Next,
we discuss the two primitive systems
filling the inner boxes of the pattern in~\cref{fig:int},
and then we collect the equations
that define the semantics of the composite system.

\subsubsection{Storage of internal energy}

The entropy of the fluid
reflects incomplete information
about its state
at more microscopic scales,
which are not resolved by the macroscopic model.
Based on the assumption of local equilibrium,
thermodynamic properties of the fluid
are defined by a potential
$U \colon \bR \times \bR \to \bR$,
which yields the internal energy density
$U(\hodge \tilde{s}, \, \hodge \tilde{m})$
as a function of
the entropy density $\hodge \tilde{s}$ and
the mass density $\hodge \tilde{m}$.
Specifically,
the intensive quantities
temperature $\theta$,
chemical potential $\mu$, and
pressure $\pi$
are given pointwise throughout $\cZ$ by
\begin{subequations}
  \begin{align}
    \theta
    \: &= \:
    \frac{\partial U(\hodge \tilde{s}, \, \hodge \tilde{m})}{\partial(\hodge \tilde{s})}
    \\
    \mu
    \: &= \:
    \frac{\partial U(\hodge \tilde{s}, \, \hodge \tilde{m})}{\partial(\hodge \tilde{m})}
    \\
    \pi
    \: &= \:
    \theta \cdot \hodge \tilde{s}
    \, + \,
    \mu \cdot \hodge \tilde{m}
    \, - \,
    U(\hodge \tilde{s}, \, \hodge \tilde{m})
    \label{eq:pi}
    \,.
  \end{align}
\end{subequations}
The last equation follows from a similar argument
as given for the barotropic case in~\cref{sec:ephs}.

The storage component
$(I_\text{ie}, \, E_\text{ie})$
filling box $\mathtt{ie}$
is defined by
its interface
$I_\text{ie} = (\{ \mathtt{s}, \, \mathtt{m} \}, \, \tau_\text{ie})$
with
\begin{alignat*}{2}
  &\tau_\text{ie}(\mathtt{s})
  \: &&= \:
  ((\tilde{\Omega}^3(\cZ), \, \mathtt{entropy}), \, \mathsf{i})
  \\
  &\tau_\text{ie}(\mathtt{m})
  \: &&= \:
  ((\tilde{\Omega}^3(\cZ), \, \mathtt{mass}), \, \mathsf{i})
\end{alignat*}
and
its energy function
$E_\text{ie} \colon \bR \to \bR$
given by
\begin{equation*}
  E_\text{ie}(\tilde{s}, \, \tilde{m})
  \: = \:
  \int_\cZ \hodge U(\hodge \tilde{s}, \, \hodge \tilde{m})
  \,,
\end{equation*}
where
$\tilde{s} = \mathtt{s.x}$
and
$\tilde{m} = \mathtt{m.x}$.

With
$\textcolor{violet}{\theta_0}$
and
$\textcolor{violet}{\mu_0}$
denoting the constant temperature and chemical potential
of the reference environment,
the exergy storage function
is, modulo an added constant, given by
\begin{equation}
  H_\text{ie}(\tilde{s}, \, \tilde{m})
  \: = \:
  \int_\cZ \bigl(
    \hodge U(\hodge \tilde{s}, \, \hodge \tilde{m})
    - \textcolor{violet}{\theta_0} \cdot \tilde{s}
    - \textcolor{violet}{\mu_0} \cdot \tilde{m}
  \bigr)
  \,,
\end{equation}
see also~\cref{sec:metriplectic}.
The flow and effort variables are thus given by
\begin{subequations}
  \begin{alignat}{2}
    \mathtt{s.f}
    \: &= \:
    \dot{\tilde{s}}
    \\
    \mathtt{s.e}
    \: &= \:
    \theta - \textcolor{violet}{\theta_0}
    \\
    \mathtt{m.f}
    \: &= \:
    \dot{\tilde{m}}
    \\
    \mathtt{m.e}
    \: &= \:
    \mu - \textcolor{violet}{\mu_0}
    \,.
  \end{alignat}
  \label{eq:ie}
\end{subequations}

\subsubsection{Advection of internal energy}

The reversible component
filling box $\mathtt{adv}$ models
advection of internal energy.
The effort variable $\mathtt{adv.p.e}$
provides the fluid velocity from the kinetic energy system,
which appears in the Lie derivative of entropy and mass.
The flow variable $\mathtt{adv.p.f}$
is the force resulting from local variations in pressure.

The reversible component
$(I_\text{adv}, \, \cD_\text{adv})$
filling box $\mathtt{adv}$
is defined by
its interface
$
I_\text{adv} =
(\{ \mathtt{p}, \, \mathtt{s}, \, \mathtt{m}, \, \mathtt{b_s}, \, \mathtt{b_m} \}, \tau_\text{adv})
$
with
\begin{equation*}
  \begin{alignedat}{2}
    &\tau_\text{adv}(\mathtt{p})
    \: &&= \:
    ((\Omega^1(\cZ), \, \mathtt{momentum}), \, \mathsf{k})
    \\
    &\tau_\text{adv}(\mathtt{s})
    \: &&= \:
    ((\tilde{\Omega}^3(\cZ), \, \mathtt{entropy}), \, \mathsf{i})
    \\
    &\tau_\text{adv}(\mathtt{m})
    \: &&= \:
    ((\tilde{\Omega}^3(\cZ), \, \mathtt{mass}), \, \mathsf{i})
  \end{alignedat}
\end{equation*}
and its Stokes-Dirac structure $\cD_\text{adv}$
given by
\begin{subequations}
	\begin{align}
		\left[
			\begin{array}{c}
				\mathtt{p.f} \\
				\mathtt{s.f} \\
				\mathtt{m.f}
			\end{array}
		\right]
		&=
		\left[
			\begin{array}{ccc}
        0 & \hodge \tilde{s} \cdot \dd(\_) & \hodge \tilde{m} \cdot \dd(\_) \\
        \dd ( \hodge \tilde{s} \cdot \_) & 0 & 0 \\
        \dd ( \hodge \tilde{m} \cdot \_) & 0 & 0
			\end{array}
		\right]
		\left[
			\begin{array}{c}
				\mathtt{p.e} \\
				\mathtt{s.e} \\
				\mathtt{m.e}
			\end{array}
		\right]%
		\label{eq:adv_domain}
		\\
		\left[
			\begin{array}{c}
				\mathtt{b_s.f} \\
				\mathtt{b_s.e} \\
				\mathtt{b_m.f} \\
				\mathtt{b_m.e}
			\end{array}
		\right]
		&=
		\left[
			\begin{array}{ccc}
                -i^*( \hodge \tilde{s} \cdot \_) & 0 & 0 \\
				0 & i^*(\_) & 0 \\
                -i^*( \hodge \tilde{m} \cdot \_) & 0 & 0 \\
				0 & 0 & i^*(\_)
			\end{array}
		\right]
		\,
		\left[
			\begin{array}{c}
				\mathtt{p.e} \\
				\mathtt{s.e} \\
				\mathtt{m.e}
			\end{array}
		\right]
    \,,%
		\label{eq:adv_boundary}%
	\end{align}%
	\label{eq:adv}
\end{subequations}
where
$\tilde{s} = \mathtt{s.x}$
and
$\tilde{m} = \mathtt{m.x}$.

With the boundary ports
defined in~\cref{eq:adv_boundary},
the formally skew-symmetric operator matrix in~\cref{eq:adv_domain}
defines a power-preserving relation
among the port variables:
\begin{equation*}
  \begin{split}
    &\int_\cZ \Bigl(
      \mathtt{p.e} \wedge \mathtt{p.f}
      +
      \mathtt{s.e} \wedge \mathtt{s.f}
      +
      \mathtt{m.e} \wedge \mathtt{m.f}
    \Bigr)
    \, =
    \\
    &\int_\cZ
      \Bigl(
        \hodge \tilde{s} \cdot \mathtt{p.e} \wedge \dd(\mathtt{s.e})
        +
        \hodge \tilde{m} \cdot \mathtt{p.e} \wedge \dd(\mathtt{m.e})
        +
        \\
        &\quad\quad\;\,
        \mathtt{s.e} \wedge \dd(\hodge \tilde{s} \cdot \mathtt{p.e})
        +
        \mathtt{m.e} \wedge \dd(\hodge \tilde{m} \cdot \mathtt{p.e})
      \Bigr)
    \, =
    \\
    &\int_\cZ \Bigl(
      \dd \bigl( \mathtt{s.e} \wedge (\hodge \tilde{s} \cdot \mathtt{p.e}) \bigr)
      +
      \dd \bigl( \mathtt{m.e} \wedge (\hodge \tilde{m} \cdot \mathtt{p.e}) \bigr)
    \Bigr)
    \, =
    \\
    &\int_{\partial \cZ} \Bigl(
      i^*(\mathtt{s.e}) \wedge i^*(\hodge \tilde{s} \cdot \mathtt{p.e})
      +
      i^*(\mathtt{m.e}) \wedge i^*(\hodge \tilde{m} \cdot \mathtt{p.e})
    \Bigr)
    \, =
    \\
    -&\int_{\partial \cZ} \Bigl(
      \mathtt{b_s.e} \wedge \mathtt{b_s.f}
      +
      \mathtt{b_m.e} \wedge \mathtt{b_m.f}
    \Bigr)
    \,.
  \end{split}
\end{equation*}
It can easily be verified
that~\cref{eq:adv_domain}
satisfies the conditions for
conservation of entropy and mass,
as
$
\dd(\textcolor{violet}{\theta_0}) =
\dd(\textcolor{violet}{\mu_0}) = 0
$,
see~\onlinecite{2024LohmayerLynchLeyendecker}.

\subsubsection{Interconnected internal energy system}

Combining~%
\cref{eq:ie,eq:adv}
with
the equations for
the interconnection pattern in~\cref{fig:int}
and
eliminating port variables gives
the following equations
for the composite system:
\begin{subequations}
	\begin{align}
    \dot{\tilde{s}}
		\: &= \:
    -\dd ( \hodge \tilde{s} \cdot \mathtt{f.p.e} ) + \mathtt{f.s.f}
		\label{eq:int_s}
		\\
    \dot{\tilde{m}}
		\: &= \:
    -\dd ( \hodge \tilde{m} \cdot \mathtt{f.p.e} )
		\label{eq:int_m}
		\\
		\mathtt{f.p.f}
		\: &= \:
		\dd \pi%
		\label{eq:int_p}%
		\\
		\mathtt{f.s.e}
		\: &= \:
    \theta - \textcolor{violet}{\theta_0}%
		\\
		\mathtt{b_s.f}
		\: &= \:
    -i^* ( \hodge \tilde{s} \cdot \mathtt{f.p.e} )%
		\\
		\mathtt{b_s.e}
		\: &= \:
    i^*(\theta - \textcolor{violet}{\theta_0})
		\\
		\mathtt{b_m.f}
		\: &= \:
    -i^* ( \hodge \tilde{m} \cdot \mathtt{f.p.e} )%
		\\
		\mathtt{b_m.e}
		\: &= \:
    i^*(\mu - \textcolor{violet}{\mu_0})
    \,.
	\end{align}%
	\label{eq:int}%
\end{subequations}

When interconnecting this model with
the kinetic energy system,
we have $\mathtt{f.p.e} = \hodge \upsilon$.
Hence,
\cref{eq:int_s,eq:int_m} contain
the Lie derivatives
$\cL_u \tilde{s} = \dd(\hodge \tilde{s} \cdot \hodge \upsilon)$
and
$\cL_u \tilde{m} = \dd(\hodge \tilde{m} \cdot \hodge \upsilon)$,
since $\tilde{s}$ and $\tilde{m}$ are advected by
the flow of the time-dependent vector field $u = \upsilon^\sharp$.
\Cref{eq:int_p} follows from
\begin{equation*}
	\begin{split}
		\dd \pi
    \: &\overset{\eqref{eq:pi}}{=} \:
    \hodge \tilde{s} \cdot \dd \theta  +
    \theta \cdot \dd \hodge \tilde{s} +
    \hodge \tilde{m} \cdot \dd \mu  +
    \mu \cdot \dd \hodge \tilde{m}
    \\
    & \quad
    -\pdv{U(\hodge \tilde{s}, \, \hodge \tilde{m})}{(\hodge \tilde{s})} \cdot \dd \hodge \tilde{s}
    -\pdv{U(\hodge \tilde{s}, \, \hodge \tilde{m})}{(\hodge \tilde{m})} \cdot \dd \hodge \tilde{m}
		\\
		\: &= \:
    \hodge \tilde{s} \cdot \dd \theta +
    \hodge \tilde{m} \cdot \dd \mu
    \,.
	\end{split}%
\end{equation*}
The flow variable $\mathtt{f.s.f}$ represents
an external entropy source term
and
the effort variable $\mathtt{f.s.e}$ is
the temperature
(relative to the reference environment).
The flow variable $\mathtt{b_s.f}$ is the entropy influx
across $\partial \cZ$
and
the effort variable $\mathtt{b_s.e}$ is
the temperature at the boundary
(relative to the environment).
Similarly,
the flow variable $\mathtt{b_m.f}$ is the mass influx
across $\partial \cZ$
and
the effort variable $\mathtt{b_m.e}$ is
the chemical potential at the boundary
(relative to the environment).

\subsection{Interconnected ideal fluid model}%

Combining~%
\cref{eq:kin,eq:int}
with
the equations for
the interconnection pattern in~\cref{fig:if_expanded}
and
eliminating port variables gives
the following system of equations on $\cZ$:
\begin{subequations}
  \begin{align}
    \begin{split}
      \dot{\upsilon}
      \: &= \:
      +\hodge(\upsilon \wedge \hodge \dd \upsilon)
      -\dd \bigl( \hodge (\upsilon \wedge \hodge \upsilon) / 2 \bigr)
      \\
      &\quad\;
      -\frac{1}{\hodge \tilde{m}} \cdot \dd \pi
      \, + \,
      \frac{1}{\hodge \tilde{m}} \cdot \mathtt{f.p.f}%
    \end{split}
    \\
    \dot{\tilde{m}}
    \: &= \:
    -\dd (\hodge \tilde{m} \cdot \hodge \upsilon)
    \\
    \dot{\tilde{s}}
    \: &= \:
    -\dd (\hodge \tilde{s} \cdot \hodge \upsilon)
    \, + \,
    \mathtt{f.s.f}
    \label{eq:if_s}
    \\
    \mathtt{f.p.e}
    \: &= \:
    \hodge \upsilon%
    \\
    \mathtt{f.s.e}
    \: &= \:
    \theta - \textcolor{violet}{\theta_0}
    \,.%
    \label{eq:if_fse}
  \end{align}%
  \label{eq:if}%
\end{subequations}
Moreover,
the boundary conditions
$
\mathtt{kin.b_k.f} =
\mathtt{int.b_m.f} =
-i^*(\hodge \tilde{m} \cdot \hodge \upsilon) =
0
$
and
$
\mathtt{int.b_s.f} =
-i^*(\hodge \tilde{s} \cdot \hodge \upsilon) =
0
$
are implied.

\section{Navier-Stokes-Fourier fluid model}%
\label{sec:nsf}

\begin{figure}
  \centering
	\includegraphics[width=5.8cm]{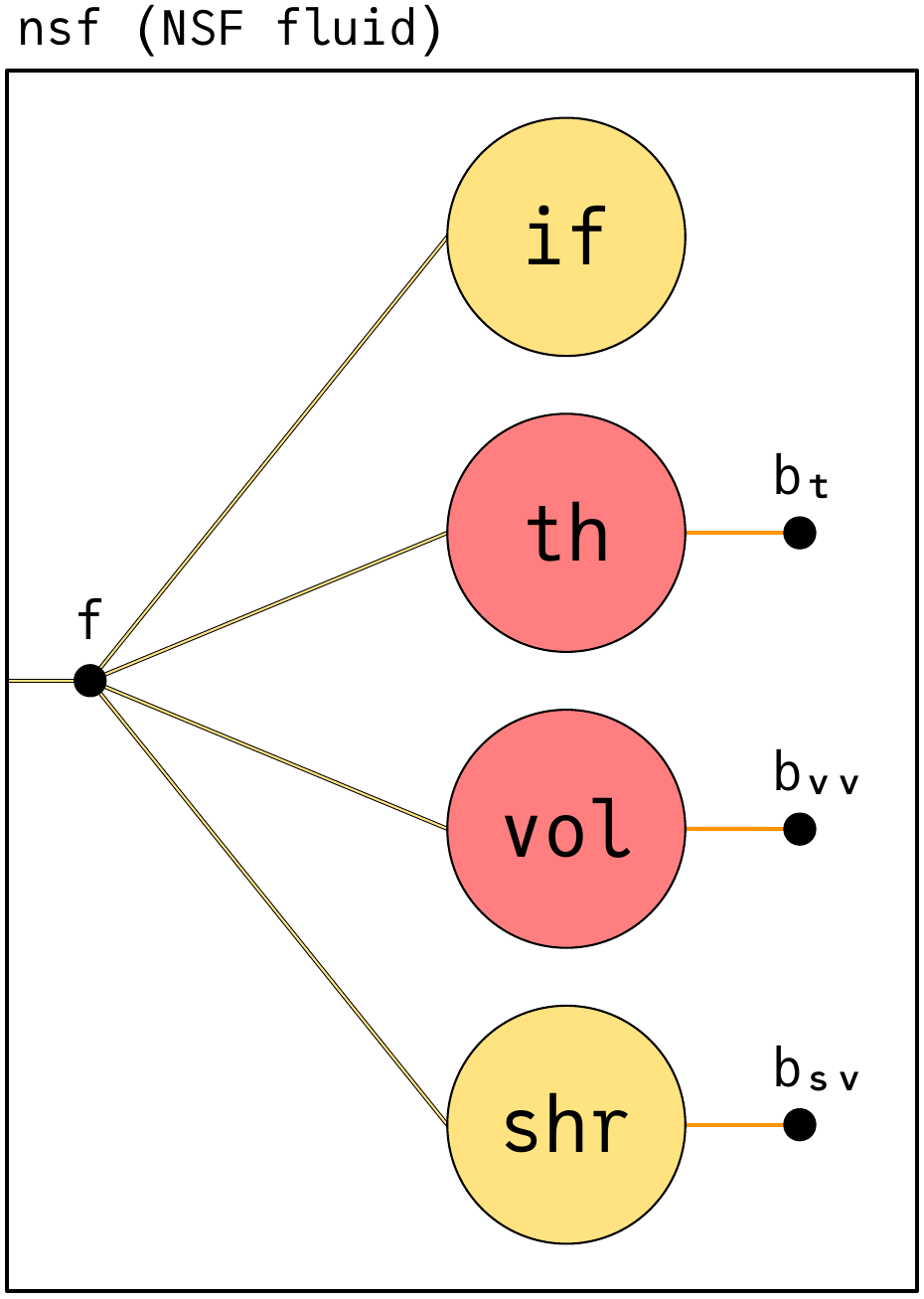}
  \caption{%
    Interconnection pattern
    of the Navier-Stokes-Fourier model.
    Box $\mathtt{if}$ represents
    the ideal fluid model
    from~\cref{sec:ideal-fluid}.
    Box $\mathtt{th}$ represents
    thermal conduction.
    Boxes $\mathtt{vol}$ and $\mathtt{shr}$ represent
    volume and shear viscosity, respectively.
    The boundary ports
    $\mathtt{b_t}$,
    $\mathtt{b_{vv}}$ and
    $\mathtt{b_{sv}}$,
    which account for
    thermal conduction and
    viscous transfer of momentum
    across $\partial \cZ$
    are not exposed,
    leading to an isolated model.
  }%
  \label{fig:nsf}
\end{figure}

In this section,
we extend the ideal fluid model
to a Navier-Stokes-Fourier (NSF) model,
which takes into account
thermal conduction,
volume viscosity,
and shear viscosity,
see~\cref{fig:nsf}.
The simplified graphical representation
of the interconnection pattern
using the multiport $\mathtt{f}$
hides to some extent that
the NSF model has
the same outer interface as the ideal fluid model.
The interface has
a power port $\mathtt{f.p}$
to exchange kinetic energy in terms of momentum,
a power port $\mathtt{f.s}$
to exchange thermal energy in terms of entropy, and
a state port $\mathtt{f.m}$
to exchange information about the mass density.

Next,
we discuss the three additional systems,
and then we collect the equations
for the composite system.

\subsection{Thermal conduction}%

Thermal conduction
is a relaxation process
that counteracts
a non-uniform temperature
$\theta \in \Omega^0(\cZ)$
with a diffusive heat flux
$\tilde{\phi} \in {\tilde{\Omega}}^2(\cZ)$.
Considering this process separately,
the first law states that
the change of internal energy
$
\tilde{u} = \hodge U(\hodge \tilde{s}, \, \hodge \tilde{m})
\in {\tilde{\Omega}}^3(\cZ)
$
is due to
the heat flux
$\tilde{\phi}$
across
the boundary
$\partial \cZ$, i.e.
\begin{equation*}
  \int_\cZ \dot{\tilde{u}}
  \: = \:
  \int_{\partial \cZ} i^*(-\tilde{\phi})
  \: \overset{\eqref{eq:stokes}}{=} \:
  \int_\cZ \dd (-\tilde{\phi})
  \,.
\end{equation*}
Concerning the minus sign,
we note that
$\tilde{\phi}$ represents
the outgoing heat flux.
While this runs counter to
the convention in thermodynamics,
where incoming energy has a positive sign,
it aligns with
the induced Stokes orientation on $\partial \cZ$,
which is defined based on
an outward-pointing vector field.
Since the same applies to
any submanifold of $\cZ$ (control volume),
the local energy balance equation
$\dot{\tilde{u}} = \dd(-\tilde{\phi})$
holds.

According to Fourier's law,
the heat flux is
proportional to
the local change in temperature
and
goes from hot to cold.
Specifically,
it states that
$\tilde{\phi} = -\kappa_t \cdot \hodge \, \dd \theta$,
where
the thermal conductivity parameter $\kappa_t \geq 0$
is constant or a function of $\theta$.

Using the chain rule to
expand the time derivative of $\tilde{u}$
with respect to $\tilde{s}$
gives
\begin{equation*}
  \dot{\tilde{u}}
  \: = \:
  \pdv{U(\hodge \tilde{s}, \, \hodge \tilde{m})}{(\hodge \tilde{s})} \cdot \dot{\tilde{s}}
  \: = \:
  \theta \cdot \dot{\tilde{s}}
  \,.
\end{equation*}
The rate of entropy change
due to thermal conduction
is hence given by
\begin{equation}
  \dot{\tilde{s}}
  \: = \:
  \frac{1}{\theta} \cdot \dot{\tilde{u}}
  \: = \:
  \frac{1}{\theta} \cdot \dd(-\tilde{\phi})
  \: = \:
  \frac{1}{\theta} \cdot \dd( \kappa_t \cdot \hodge \, \dd \theta )
  \,.
  \label{eq:th_entropy}
\end{equation}
This dynamics is encapsulated by
the following component.

The irreversible component
$(I_\text{th}, \, \cO_\text{th})$
filling box $\mathtt{th}$
is defined by
its interface
$
I_\text{th} =
(\{ \mathtt{f.s}, \, \mathtt{b_t} \}, \tau_\text{th})
$
with
\begin{equation*}
  \begin{alignedat}{2}
    &\tau_\text{th}(\mathtt{f.s})
    \: &&= \:
    ((\tilde{\Omega}^3(\cZ), \, \mathtt{entropy}), \, \mathsf{i})
  \end{alignedat}
\end{equation*}
and
the Stokes-Onsager structure $\cO_\text{th}$
given by
\begin{subequations}
	\begin{align}
		\left[
			\mathtt{f.s.f}
		\right]
		\: &= \:
		\left[
			-\frac{1}{\theta} \cdot
			\dd (
				\hodge \biggl(
					\kappa_t \cdot
					\theta^2 \cdot
          \frac{1}{\textcolor{violet}{\theta_0}} \cdot
					\dd (\frac{1}{\theta} \cdot \_ )
				\biggr)
			)
		\right]
		\,
		\left[
			\mathtt{f.s.e}
		\right]
		\label{eq:th_domain}
		\raisetag{1.1em}
    \\
		\begin{split}
      \mathtt{b_t.f}
      \: &= \:
      i^*\biggl(
        \frac{1}{\theta} \cdot
        \hodge \biggl(
          \kappa_t \cdot
          \theta^2 \cdot
          \frac{1}{\textcolor{violet}{\theta_0}} \cdot
          \dd (\frac{1}{\theta} \cdot \mathtt{f.s.e} )
        \biggr)
      \biggr)
      \\
      \mathtt{b_t.e}
      \: &= \:
      i^*(\mathtt{f.s.e})
      \,,%
    \end{split}
		\raisetag{1.1em}
	\end{align}%
	\label{eq:th}%
\end{subequations}
where
$\theta = \textcolor{violet}{\theta_0} + \mathtt{f.s.e}$.

\Cref{eq:th_domain} simplifies to
$
\mathtt{f.s.f} =
-\frac{1}{\theta} \cdot \dd( \kappa_t \cdot \hodge \dd \theta)
$,
since
$\mathtt{f.s.e} = \theta - \textcolor{violet}{\theta_0}$,
see also~\cref{eq:if_fse},
and
$\dd(\theta^{-1}) = -\theta^{-2} \cdot \dd \theta$.
Ignoring the advection term in~\cref{eq:if_s}
and using balance of net flow
gives
$
\dot{\tilde{s}} =
\mathtt{if.f.s.f} = -\mathtt{th.f.s.f} =
\frac{1}{\theta} \cdot \dd( \kappa_t \cdot \hodge \dd \theta)
$,
which agrees with~\cref{eq:th_entropy}.
Connecting this component with
a storage component
representing the thermal capacity of a solid
hence results in a simple `heat equation'.
Concerning the boundary $\partial \cZ$,
the flow variable
$
\mathtt{b_t.f} =
i^*( \frac{1}{\theta} \cdot (-\tilde{\phi}) ) =
i^*( \frac{1}{\theta} \cdot \kappa_t \cdot \hodge \dd \theta )
\in \tilde{\Omega}^2(\partial \cZ)
$
is the entropy influx across $\partial \cZ$
and
the effort variable
$
\mathtt{b_t.e} =
i^*( \theta - \textcolor{violet}{\theta_0} )
\in \Omega^0(\partial \cZ)
$
is the temperature at the boundary
(relative to the exergy reference environment).

According to~\onlinecite{2024LohmayerLynchLeyendecker},
an Onsager structure is a relation
among flow variables $f$ and effort variables $e$
that is of the form
$f = \frac{1}{\textcolor{violet}{\theta_0}} \cdot M(e) \, e$,
where
$M$ is a smooth function that yields,
for each value of the effort variable $e$,
a symmetric non-negative definite linear operator $M(e)$,
see also~\cref{sec:metriplectic}.
Onsager reciprocal relations hold
due to the symmetry.
The second law holds due to
the non-negative definiteness.
The first law is ensured through
the kernel of $M(e)$ for each $e$.
Specifically,
conservation of energy requires that
$M(e) \, e^\prime = 0$,
where $e^\prime$ is $e$ without any shifts by
the constant \textcolor{violet}{intensive properties}
of the reference environment.

For spatially-distributed systems,
the symmetry of $M(e)$ is made apparent
by applying integration by parts
to obtain a weak form,
where the boundary conditions are implicit.
Rather than stating the weak form
with an arbitrary test function $\psi \in \Omega^0(\cZ)$,
we fix $\psi = \mathtt{f.s.e}$,
which directly leads to the following power balance equation:
\begin{equation*}
  \begin{split}
    &\int_\cZ \mathtt{f.s.e} \wedge \mathtt{f.s.f}
    \, + \,
    \int_{\partial \cZ} \mathtt{b_t.e} \wedge \mathtt{b_t.f}
    \: = \:
    \\
    \textcolor{violet}{\theta_0} \cdot
    &\int_\cZ
    \frac{1}{\textcolor{violet}{\theta_0}} \cdot
    \dd (\frac{1}{\theta} \cdot \mathtt{f.s.e} )
    \wedge
    \hodge \biggl(
      \kappa_t \cdot
      \theta^2 \cdot
      \frac{1}{\textcolor{violet}{\theta_0}} \cdot
      \dd (\frac{1}{\theta} \cdot \mathtt{f.s.e} )
    \biggr)
    \: = \:
    \\
    \textcolor{violet}{\theta_0} \,
    &\int_\cZ
    \frac{1}{\theta^2} \cdot \dd \theta \wedge \hodge (\kappa_t \cdot \dd \theta)
    \, \geq \, 0
  \end{split}
\end{equation*}
Replacing the first occurrence of the effort variable $\mathtt{f.s.e}$
on the first and second line
by an arbitrary test function $\psi \in \Omega^0(\cZ)$
gives the weak form.
We can identify the term
$
\frac{1}{\textcolor{violet}{\theta_0}} \cdot
\dd (\frac{1}{\theta} \cdot \mathtt{f.s.e} )
= -\dd(\frac{1}{\theta})
$
as the \emph{thermodynamic force}
that drives the irreversible process.
The map
$\hodge( \kappa_t \cdot \theta^2 \cdot \_)$
turns this into
the \emph{thermodynamic flux} $-\tilde{\phi}$.
The integrand is hence given by
the product of force and flux,
which gives
the local entropy production rate.
Multiplying the integral with
the leading factor $\textcolor{violet}{\theta_0}$
gives the total exergy destruction rate
due to thermal conduction.
These rates are non-negative
due to the symmetry and non-negative definiteness.
Since energy is exchanged at the boundary,
the condition ensuring conservation of energy
only applies to the bulk part of the strong form
in~\cref{eq:th_domain}.
With
$e = \mathtt{f.s.e} = \theta - \textcolor{violet}{\theta_0}$
and
$e^\prime = \theta$,
the condition is satisfied since
$\dd(\frac{1}{\theta} \cdot e^\prime) = \dd(1) = 0$.

\subsection{Volume viscosity}%

When an ideal fluid contracts,
kinetic energy is reversibly transformed into internal energy,
and vice versa for expansion.
Volume viscosity is a relaxation process
that counteracts
local changes in volume
with a pressure-like quantity,
which in particular leads to
a damping of acoustic waves.

The Riemannian volume form
$\hodge 1 \in \tilde{\Omega}^3(\cZ)$
is the natural measure for volume.
Its Lie derivative
$\cL_{u} (\hodge 1)$
gives the rate of volume change
for a fluid element moving with the flow.
We have
\begin{equation*}
	\cL_{u} (\hodge 1)
	\: = \:
	\dd( \iota_{\upsilon^\sharp} (\hodge 1) )
	\: \overset{\eqref{eq:interior_product}}{=} \:
	\dd( \hodge ( \upsilon \wedge \hodge \, \hodge 1 ) )
	\: = \:
	\dd \, \hodge \upsilon
  \,.
\end{equation*}
The Hodge dual
$\hodge \, \dd \, \hodge \upsilon \in \Omega^0(\cZ)$
is equal to the divergence of
the velocity vector field $u$.

The pressure-like quantity
that counteracts local changes in volume
is modeled as
the divergence of the velocity
multiplied by
a volume viscosity coefficient
$\mu_v \geq 0$,
which may depend on
velocity and temperature.
Consequently,
the rate at which kinetic energy is dissipated into heat
is given by
the product of
the pressure-like quantity
$\mu_v \cdot \hodge \, \dd \, \hodge \upsilon \in \Omega^0(\cZ)$
and
the rate of volume change
$\dd \, \hodge \upsilon \in \tilde{\Omega}^3(\cZ)$.
The entropy production rate
is thus given by
$
\frac{1}{\theta} \cdot \mu_v \cdot
(\hodge \, \dd \, \hodge \, \upsilon) \cdot
(\dd \, \hodge \, \upsilon)
\in \tilde{\Omega}^3(\cZ)
$.
This is encapsulated
by the following component.

The irreversible component
$(I_\text{vol}, \, \cO_\text{vol})$
filling box $\mathtt{vol}$
is defined by
its interface
$
I_\text{vol} =
(\{ \mathtt{f.p}, \, \mathtt{f.s}, \, \mathtt{b_{vv}} \}, \tau_\text{vol})
$
with
\begin{equation*}
  \begin{alignedat}{2}
    &\tau_\text{vol}(\mathtt{f.p})
    \: &&= \:
    ((\Omega^1(\cZ), \, \mathtt{momentum}), \, \mathsf{p})
    \\
    &\tau_\text{vol}(\mathtt{f.s})
    \: &&= \:
    ((\tilde{\Omega}^3(\cZ), \, \mathtt{entropy}), \, \mathsf{i})
  \end{alignedat}
\end{equation*}
and
the Stokes-Onsager structure $\cO_\text{vol}$
given by
\begin{subequations}
	\begin{align}
    \begin{split}
      \left[
        \begin{array}{c}
          \mathtt{f.p.f} \\
          \mathtt{f.s.f}
        \end{array}
      \right]
      \: &= \:
      \frac{1}{\textcolor{violet}{\theta_0}} \cdot
      \left[
        \begin{array}{cc}
          A(\_) & B(\_) \\
          C(\_) & D(\_)
        \end{array}
      \right]
      \left[
        \begin{array}{c}
          \mathtt{f.p.e} \\
          \mathtt{f.s.e}
        \end{array}
      \right]
      \\
      A(\_)
      \: &= \:
      -\dd \bigl( \mu_v \cdot \hodge \dd(\_) \cdot \theta \bigr) \\
      B(\_)
      \: &= \:
      +\dd \bigl( \mu_v \cdot (\hodge \dd \hodge \upsilon) \cdot (\_) \bigr) \\
      C(\_)
      \: &= \:
      -\mu_v \cdot \hodge \dd(\_) \cdot (\dd \hodge \upsilon) \\
      D(\_)
      \: &= \:
      +\frac{1}{\theta} \cdot \mu_v \cdot (\hodge \dd \hodge \upsilon) \cdot (\dd \hodge \upsilon) \cdot (\_)
    \end{split}
    \label{eq:vol_domain}
    \\[0.7em]
    \begin{split}
      \mathtt{b_{vv}.f}
      \: &= \:
      -i^*(\mathtt{f.p.e})
      \\
      \mathtt{b_{vv}.e}
      \: &= \:
      -\frac{1}{\textcolor{violet}{\theta_0}} \cdot i^* \bigl(
        \mu_v \cdot \hodge \dd (\mathtt{f.p.e}) \cdot \theta
        -\mu_v \cdot (\hodge \, \dd \, \hodge \, \upsilon) \cdot \mathtt{f.s.e}
      \bigr)
      \,,
		\end{split}
		\raisetag{3.8em}
	\end{align}%
	\label{eq:vol}
\end{subequations}
where
$\hodge \upsilon = \mathtt{f.p.e}$
and
$\theta = \textcolor{violet}{\theta_0} + \mathtt{f.s.e}$.

Simplification of~\cref{eq:vol_domain} yields
the viscous force
$
-\mathtt{f.p.f} =
\dd( \mu_v \cdot \hodge \dd \hodge \upsilon )
$
and
the entropy production rate
$
-\mathtt{f.s.f} =
\frac{1}{\theta} \cdot \mu_v \cdot (\hodge \dd \hodge \upsilon) \cdot \dd \hodge \upsilon
$.
%
Moreover,
the flow variable
$\mathtt{b_{vv}.f} = -i^*(\hodge \upsilon)$
is the velocity (or `volume influx') across $\partial \cZ$
and
the effort variable
$
\mathtt{b_{vv}.e} =
i^*(\mu_v \cdot (\hodge \dd \hodge \upsilon))
$
is the pressure-like quantity at the boundary.

To show that~\cref{eq:vol} defines
a non-negative definite symmetric operator,
we again
apply integration by parts
to obtain a weak form
with implicit boundary conditions.
Letting the test functions
be equal to
the effort variables $\mathtt{f.p.e}$ and $\mathtt{f.s.e}$
gives the following power balance equation:
\begin{equation*}
	\begin{split}
		\int_\cZ (
      \mathtt{f.p.e} &\wedge \mathtt{f.p.f}
			+
			\mathtt{f.s.e} \wedge \mathtt{f.s.f}
		)
		+
		\int_{\partial \cZ}
      \mathtt{b_{vv}.e} \wedge
      \mathtt{b_{vv}.f}
		\: = \:
		\\
		\frac{1}{\textcolor{violet}{\theta_0}} \cdot \int_\cZ \Bigl[
			&+\dd(\mathtt{f.p.e}) \wedge \Bigl(
        \mu_v \cdot \hodge \dd(\mathtt{f.p.e}) \cdot \theta
			\Bigr)
			\\
      & -\dd(\mathtt{f.p.e}) \wedge \Bigl(
        \mu_v \cdot (\hodge \, \dd \, \hodge \upsilon) \cdot \mathtt{f.s.e}
			\Bigr)
			\\
      &-\mathtt{f.s.e} \wedge \Bigl(
        \mu_v \cdot \hodge \dd(\mathtt{f.p.e}) \cdot (\dd \, \hodge \upsilon)
			\Bigr)
			\\
      &+\mathtt{f.s.e} \wedge \Bigl(
        \frac{1}{\theta} \cdot \mu_v \cdot (\hodge \, \dd \, \hodge \upsilon) \cdot (\dd \, \hodge \upsilon) \cdot \mathtt{f.s.e}
			\Bigr)
		\Bigr]
		\: = \:
    \\
    \textcolor{violet}{\theta_0} \cdot \int_\cZ
    &\frac{1}{\theta} \cdot \mu_v \cdot (\hodge \dd \hodge \upsilon) \cdot \dd \hodge \upsilon
		\: \geq \: 0
    \,.
	\end{split}
\end{equation*}
Since the terms on the second and fifth line are symmetric and positive
and the terms on the third and fourth line are equal,
the operator is symmetric
and non-negative definiteness follows from the final line.
Again,
the integrand is equal to the local entropy production rate
and
multiplying the integral with
the leading factor $\textcolor{violet}{\theta_0}$
gives the total exergy destruction rate
due to volume viscosity.

\Cref{eq:vol_domain} satisfies the condition
for conservation of energy
because
$A(\hodge \upsilon) + B(\theta) = 0$
and
$B(\hodge \upsilon) + D(\theta) = 0$,
where
$\hodge \upsilon = \mathtt{f.p.e}$
and
$\theta = \textcolor{violet}{\theta_0} + \mathtt{f.s.e}$.

\subsection{Shear viscosity}%

Here,
we model an irreversible process
that counteracts
general deformations of the fluid.
As this includes
expansion and contraction,
there is some overlap with
the model for volume viscosity.
Assuming that $(\cZ, \, g)$ is Euclidean,
the viscous stress tensor can be split into
an isotropic part corresponding to volume viscosity
and
a deviatoric part corresponding to pure shear stress.
As we do not know how to express this splitting
within the coordinate-invariant framework,
we carry on,
acknowledging that
the presented shear viscosity model
essentially adds more volume viscosity.

The covariant derivative,
denoted by $\nabla$,
is used to
determine local changes in
the velocity field
$u = \upsilon^\sharp \in \Gamma(\mathrm{T} \cZ)$.
Pairing the second leg of
the \emph{`velocity gradient tensor'} field
$\nabla u \in \Gamma(\mathrm{T} \cZ) \otimes \Omega^1(\cZ)$
with a vector (field)
gives the change of the velocity field
in the direction of the given vector (field).

The stress state due to shear viscosity
is described by
the \emph{viscous stress tensor} field
\begin{equation*}
  T
  \: = \:
  \mu_s \cdot \hodge_2 \Bigl( \sym \bigl( \flat_1( \nabla u ) \bigr) \Bigr)
  \: \in \:
  \Gamma(\mathrm{T}^* \cZ) \otimes \tilde{\Omega}^{2}(\cZ)
  \,.
\end{equation*}
The flat map $\flat_1$ acts on the first leg of
the $\mathrm{T} \cZ$-valued $1$-form $\nabla u$,
turning it into
a $\mathrm{T}^* \cZ$-valued $1$-form.
The two legs of
$\flat_1 ( \nabla u ) \in \Omega^1(\cZ) \otimes \Omega^1(\cZ)$
are then symmetrized
such that the stress state described by $T$
does not cause fluid elements to spin,
as required for
conservation of angular momentum.
The \emph{`strain rate tensor'} field
$\sym(\flat_1( \nabla u ))$
is equal to
$\frac{1}{2} \cdot \cL_u g$,
see~\onlinecite{2023GilbertVanneste}.
The Lie derivative of the metric $g$
provides a natural measure for deformation
induced by the flow of $u$.
The Hodge star $\hodge_2$ acts on the second leg
of the strain rate tensor,
yielding a $\mathrm{T}^* \cZ$-valued twisted $2$-form.
Multiplication with
a shear viscosity coefficient $\mu_s \geq 0$,
which may depend on velocity and temperature,
finally gives the stress tensor.
Its first leg
represents the traction force
acting across an oriented surface element
that is given by a $2$-vector paired with the second leg.
Moreover,
pairing
the first leg of $T$
with
the velocity field $u$
leaves a twisted $2$-form
that can be integrated over a surface $S$,
yielding the rate of work $P$
done by the stress
on $S$,
see~\onlinecite{2007KansoArroyoTongYavariMarsdenDesbrun}.
Using the duality pairing for
bundle-valued forms
in~\cref{eq:wedge_dot_pairing},
we have
$P = \int_S u \dot{\wedge} T$.

The net force resulting from
the viscous stress $T$
is given by
the exterior covariant derivative
$
\dd_\nabla(T)
\in \Omega^1(\cZ) \otimes \tilde{\Omega}^3(\cZ)
$.
The first leg
represents the net force
on a volume element
that is given by a $3$-vector paired with the second leg.
The rate at which kinetic energy
is dissipated into heat is given by
$\int_\cZ \nabla u \dot{\wedge} T$
and
the local entropy production rate is
hence given by
$\frac{1}{\theta} \cdot \nabla u \dot{\wedge} T$.

\begin{figure}
  \centering
	\includegraphics[width=5.8cm]{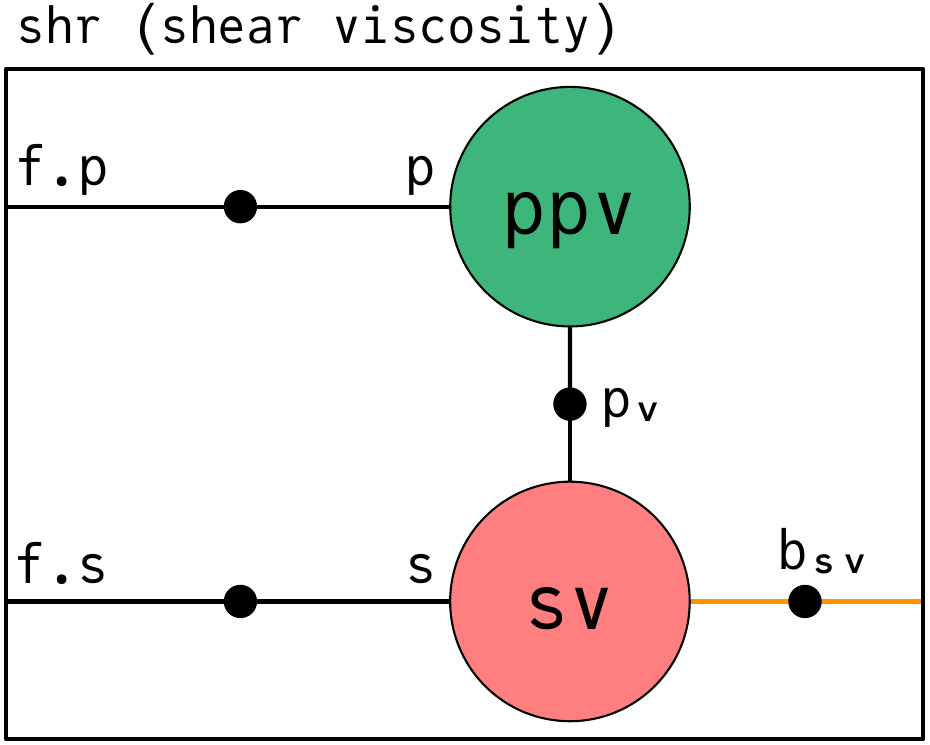}
  \caption{%
    Interconnection pattern of
    the shear viscosity model.
    The outer ports $\mathtt{f.p}$ and $\mathtt{f.s}$
    represent the kinetic and the thermal energy domain of the fluid.
    Box $\mathtt{sv}$ represents
    the irreversible process of shear viscosity
    and
    the boundary port $\mathtt{b_{sv}}$
    accounts for
    viscous transfer of momentum
    across $\partial \cZ$.
    Box $\mathtt{ppv}$ represents
    the reversible transformation between
    the two different geometric representations of momentum.
  }%
  \label{fig:shr}
\end{figure}

The interconnection pattern
of the shear viscosity model
is shown in~\cref{fig:shr}.
First,
we discuss
the reversible component
that realizes the transformation
between momentum given as a $1$-form
and it being given as a $\mathrm{T}^* \cZ$-valued twisted $3$-form.
Then,
we discuss the irreversible component
that encapsulates the irreversible dynamics.
Finally,
we collect the equations
for the composite system.

The reversible component
$(I_\text{ppv}, \, \cD_\text{ppv})$
filling box $\mathtt{ppv}$
is defined by its interface
$
I_\text{ppv} =
(\{ \mathtt{p}, \, \mathtt{p_v} \}, \tau_\text{ppv})
$
with
\begin{equation*}
  \begin{alignedat}{2}
    &\tau_\text{ppv}(\mathtt{p})
    \: &&= \:
    ((\Omega^1(\cZ), \, \mathtt{momentum}), \, \mathsf{k})
    \\
    &\tau_\text{ppv}(\mathtt{p_v})
    \: &&= \:
    ((\Gamma(\mathrm{T}^* \cZ) \otimes \tilde{\Omega}^3(\cZ), \, \mathtt{momentum}), \, \mathsf{k})
  \end{alignedat}
\end{equation*}
and
its Dirac structure $\cD_\text{ppv}$
given by
\begin{equation}
  \begin{bmatrix}
			\mathtt{p.f} \\
			\mathtt{p_v.e} \\
  \end{bmatrix}
	\: = \:
  \begin{bmatrix}
			0 & -\hodge_2(\_) \\
			\sharp( \hodge( \_)) & 0
  \end{bmatrix}
	\,
  \begin{bmatrix}
			\mathtt{p.e} \\
			\mathtt{p_v.f}
  \end{bmatrix}
  \,.%
	\label{eq:ppv}
\end{equation}

To show that~\cref{eq:ppv}
defines a power-preserving relation,
without loss of generality,
let
$\mathtt{p.e} = \hodge \upsilon$
and
$\mathtt{p_v.f} = \alpha \otimes (\hodge 1)$
for some
$\upsilon \in \Omega^1(\cZ)$
and
$\alpha \in \Gamma(\mathrm{T}^* \cZ)$.
We then have
\begin{equation*}
  \begin{split}
    &\int_\cZ \mathtt{p.e} \wedge \mathtt{p.f}
    \, + \,
    \int_\cZ \mathtt{p_v.e} \dot{\wedge} \mathtt{p_v.f}
    \: = \:
    \\
    -&\int_\cZ \hodge \upsilon \wedge \alpha
    \, + \,
    \int_\cZ \upsilon^\sharp \dot{\wedge} \alpha \otimes (\hodge 1)
    \: = \:
    0
    \,.
  \end{split}
\end{equation*}
The last equality follows from
\begin{equation*}
  \upsilon^\sharp \dot{\wedge} \alpha \otimes (\hodge 1)
  \: = \:
  \hodge \iota_{\upsilon^\sharp} \alpha
  \: \overset{\eqref{eq:interior_product}}{=} \:
  \upsilon \wedge \hodge \alpha
  \: = \:
  \hodge \upsilon \wedge \alpha
  \,.
\end{equation*}

The irreversible component
$(I_\text{sv}, \, \cO_\text{sv})$
filling box $\mathtt{sv}$
is defined by
its interface
$
I_\text{sv} =
(\{ \mathtt{p_v}, \, \mathtt{s}, \, \mathtt{b_{sv}} \}, \tau_\text{sv})
$
with
\begin{equation*}
  \begin{alignedat}{2}
    &\tau_\text{sv}(\mathtt{p_v})
    \: &&= \:
    ((\Gamma(\mathrm{T}^* \cZ) \otimes \tilde{\Omega}^3(\cZ), \, \mathtt{momentum}), \, \mathsf{k})
    \\
    &\tau_\text{sv}(\mathtt{s})
    \: &&= \:
    ((\tilde{\Omega}^3(\cZ), \, \mathtt{entropy}), \, \mathsf{i})
  \end{alignedat}
\end{equation*}
and
the Stokes-Onsager structure $\cO_\text{sv}$
given by
\begin{subequations}
	\begin{align}
		\begin{split}
			\raisetag{2em}
			\left[
				\begin{array}{c}
					\mathtt{p_v.f} \\
					\mathtt{s.f}
				\end{array}
			\right]
			\: &= \:
      \frac{1}{\textcolor{violet}{\theta_0}} \cdot
			\left[
				\begin{array}{cc}
					A(\_) & B(\_) \\
					C(\_) & D(\_)
				\end{array}
			\right] \,
			\left[
				\begin{array}{c}
					\mathtt{p_v.e} \\
					\mathtt{s.e}
				\end{array}
			\right]
			\\
			A(\_)
            \: &= \:
      -\dd_\nabla \bigl( \mu_s \cdot \hodge_2(\sym(\flat_1(\nabla(\_)))) \cdot \theta \bigr) \\
			B(\_)
            \: &= \:
      \dd_\nabla \bigl( \mu_s \cdot \hodge_2(\sym(\flat_1(\nabla u ))) \cdot (\_) \bigr) \\
			C(\_)
            \: &= \:
      -\mu_s \cdot \nabla(\_) \dot{\wedge} \hodge_2(\sym(\flat_1(\nabla u))) \\
			D(\_)
        \: &= \:
      \frac{1}{\theta} \cdot \mu_s \cdot \nabla u \dot{\wedge} \hodge_2(\sym(\flat_1(\nabla u ))) \cdot (\_)
		\end{split}
		\label{eq:sv_domain}
		\\
		\begin{split}
			\raisetag{1.2em}
      \mathtt{b_{sv}.f}
			\: &= \:
			-i_2^*(\mathtt{p_v.e})
			\\
      \mathtt{b_{sv}.e}
			\: &= \:
      -\frac{1}{\textcolor{violet}{\theta_0}} \,
      i_2^* \bigl(
				\mu_s \cdot \hodge_2( \sym( \flat_1( \nabla(\mathtt{p_v.e})))) \cdot \theta
        \\ & \qquad \qquad
			  -\mu_s \cdot \hodge_2( \sym( \flat_1( \nabla u ))) \cdot \mathtt{s.e}
			\bigr)
		\end{split}
		\label{eq:sv_boundary}
	\end{align}%
	\label{eq:sv}
\end{subequations}
where
$u = \mathtt{p_v.e}$
and
$\theta = \textcolor{violet}{\theta_0} + \mathtt{s.e}$.

Simplification of~\cref{eq:sv_domain} yields
the net viscous force
$
-\mathtt{p_v.f} =
\dd_\nabla(T)
$
as well as
the entropy production rate
$
-\mathtt{s.f} =
\frac{1}{\theta} \cdot \nabla u \dot{\wedge} T
$.
%
Further,
the flow variable
$
\mathtt{b_{sv}.f} = -i_2^*(u)
\in \Gamma(\mathrm{T} \cZ) \otimes \Omega^0(\partial \cZ)
$
is the velocity at the boundary
expressed as a $2$-point tensor
and
the effort variable
$
\mathtt{b_{sv}.e} = i_2^*(T)
\in \Gamma(\mathrm{T}^* \cZ) \otimes \tilde{\Omega}^2(\partial \cZ)
$
is the viscous stress tensor
pulled back to the boundary as a $2$-point tensor,
see also~\onlinecite{2021CalifanoRashadSchullerStramigioli}.

To show that~\cref{eq:sv} defines
a non-negative definite symmetric operator,
we again
apply integration by parts
to obtain a weak form
with implicit boundary conditions.
Letting the test functions
be equal to
the effort variables $\mathtt{p_v.e}$ and $\mathtt{s.e}$
gives the following power balance equation:
\begin{equation*}
	\begin{split}
		\int_\cZ \mathtt{p_v.e} &\dot{\wedge} \mathtt{p_v.f}
		\, + \,
		\int_\cZ \mathtt{s.e} \wedge \mathtt{s.f}
		+
		\int_{\partial \cZ}
      \mathtt{b_{sv}.e} \dot{\wedge}
      \mathtt{b_{sv}.f}
		\: = \:
		\\
		\frac{1}{\textcolor{violet}{\theta_0}} \cdot \int_\cZ \Bigl[
      &+\nabla(\mathtt{p_v.e}) \dot{\wedge} \Bigl(
        \mu_s \cdot \hodge_2( \sym( \flat_1( \nabla( \mathtt{p_v.e} )))) \cdot \theta
			\Bigr)
			\\
      & -\nabla(\mathtt{p_v.e}) \dot{\wedge} \Bigl(
        \mu_s \cdot \hodge_2( \sym( \flat_1( \nabla u ))) \cdot \mathtt{s.e}
			\Bigr)
			\\
      &-\mathtt{s.e} \cdot \Bigl(
        \mu_s \cdot \nabla( \mathtt{p_v.e} ) \dot{\wedge} \hodge_2( \sym( \flat_1( \nabla u )))
			\Bigr)
			\\
      &+\mathtt{s.e} \cdot \Bigl(
        \frac{1}{\theta} \cdot
        \mu_s \cdot \nabla u \dot{\wedge} \hodge_2( \sym( \flat_1( \nabla u ))) \cdot \mathtt{s.e}
			\Bigr)
		\Bigr]
		\: = \:
    \\
    \textcolor{violet}{\theta_0} \cdot \int_\cZ
    &\frac{1}{\theta} \cdot \nabla u \dot{\wedge} T
		\: \geq \: 0
    \,.
	\end{split}
\end{equation*}
Again,
the integrand is equal to the local entropy production rate
and
multiplying the integral with
the leading factor $\textcolor{violet}{\theta_0}$
gives the total exergy destruction rate
due to shear viscosity.

\Cref{eq:sv_domain} satisfies the condition
for conservation of energy
because
$A(u) + B(\theta) = 0$
and
$B(u) + D(\theta) = 0$,
where
$u = \mathtt{p_v.e}$
and
$\theta = \textcolor{violet}{\theta_0} + \mathtt{s.e}$.


Combining~%
\cref{eq:ppv,eq:sv}
with
the equations for
the interconnection pattern in~\cref{fig:shr}
and
eliminating port variables gives
the following equations
for the composite system:
\begin{subequations}
	\begin{align}
    \mathtt{f.p.f}
    \: &= \:
    -\hodge_2 \, \dd_\nabla T
    \\
    \mathtt{f.s.f}
    \: &= \:
    -\frac{1}{\theta} \cdot \nabla u \dot{\wedge} T
    \\
    \mathtt{b_{sv}.f}
    \: &= \:
    -i_2^*(u)
    \\
    \mathtt{b_{sv}.e}
    \: &= \:
    i_2^*(T)
  \end{align}
  \label{eq:shr}%
\end{subequations}%
Here,
$u = \sharp( \hodge ( \mathtt{f.p.e} ) )$,
$\theta = \textcolor{violet}{\theta_0} + \mathtt{f.s.e}$
and
$T = \mu_s \cdot \hodge_2 \bigl( \sym \bigl( \flat_1 \bigl( \nabla u \bigr) \bigr) \bigr)$.

\subsection{Interconnected Navier-Stokes-Fourier model}%

Combining~%
\cref{eq:if,eq:th,eq:vol,eq:shr}
with
the equations for
the interconnection pattern in~\cref{fig:nsf}
and
eliminating port variables gives
the following system of equations on $\cZ$:
\begin{subequations}
	\begin{align}
		\begin{split}
			\dot{\upsilon}
			\: &= \:
      +\hodge(\upsilon \wedge \hodge \dd \upsilon)
			-\dd (\hodge (\upsilon \wedge \hodge \upsilon) / 2)
      -\frac{1}{\hodge m} \cdot \dd \pi
      \\
      &\quad\:\:
      + \frac{1}{\hodge \tilde{m}} \cdot \dd( \mu_v \cdot \hodge \, \dd \, \hodge \, \upsilon ) \\
      &\quad\:\:
      + \frac{1}{\hodge \tilde{m}} \cdot \hodge_2 \, \dd_\nabla \bigl( \mu_s \cdot \hodge_2(\sym(\nabla \upsilon )) \bigr) \\
      &\quad\:\:
      + \frac{1}{\hodge \tilde{m}} \cdot \mathtt{f.p.f}
			\label{eq:nsf_momentum_balance}
		\end{split}
		\\
    \dot{\tilde{m}}
		\: &= \:
    -\dd (\hodge \tilde{m} \cdot \hodge \upsilon)
		\label{eq:nsf_mass_balance}
		\\
		\begin{split}
      \dot{\tilde{s}}
      \: &= \: -\dd ( \hodge \tilde{s} \cdot \hodge \upsilon ) \\
			&\quad\:\:
      + \frac{1}{\theta} \cdot \dd( \kappa_t \cdot \hodge \dd \theta ) \\
			&\quad\:\:
      + \frac{1}{\theta} \cdot \mu_v \cdot (\hodge \, \dd \, \hodge \, \upsilon) \cdot (\dd \, \hodge \, \upsilon) \\
			&\quad\:\:
      + \frac{1}{\theta} \cdot \nabla \upsilon^\sharp  \dot{\wedge} \bigl( \mu_s \cdot \hodge_2(\sym(\nabla \upsilon )) \bigr) \\
      &\quad\:\:
      + \mathtt{f.s.f}
			\label{eq:nsf_entropy_balance}
		\end{split}%
    \\
    \mathtt{f.p.e}
    \: &= \:
    \hodge \upsilon
    \\
    \mathtt{f.s.e}
    \: &= \:
    \theta - \textcolor{violet}{\theta_0}
	\end{align}%
	\label{eq:nsf}%
\end{subequations}
Moreover,
the boundary conditions
$
\mathtt{if.kin.b_k.f} =
\mathtt{if.int.b_m.f} =
-i^*(\hodge \tilde{m} \cdot \hodge \upsilon) =
0
$ (no advective mass flux),
$
\mathtt{if.int.b_s.f} =
-i^*(\hodge \tilde{s} \cdot \hodge \upsilon) =
0
$ (no advective entropy flux),
$
\mathtt{th.b_t.f} =
-i^*(\frac{1}{\theta} \cdot (- \tilde{\phi})) =
0
$ (no conductive heat flux),
$
\mathtt{vol.b_{vv}.f} =
-i^*(\hodge \upsilon) =
0
$ (no velocity or `volume flux' through the boundary), and
$
\mathtt{shr.b_{sv}.f} =
-i_2^*(u) =
0
$ (vanishing velocity in vicinity of boundary)
are implied.
%

\section{Maxwell model}%
\label{sec:em}

\begin{figure}
	\centering
	\includegraphics[width=5.8cm]{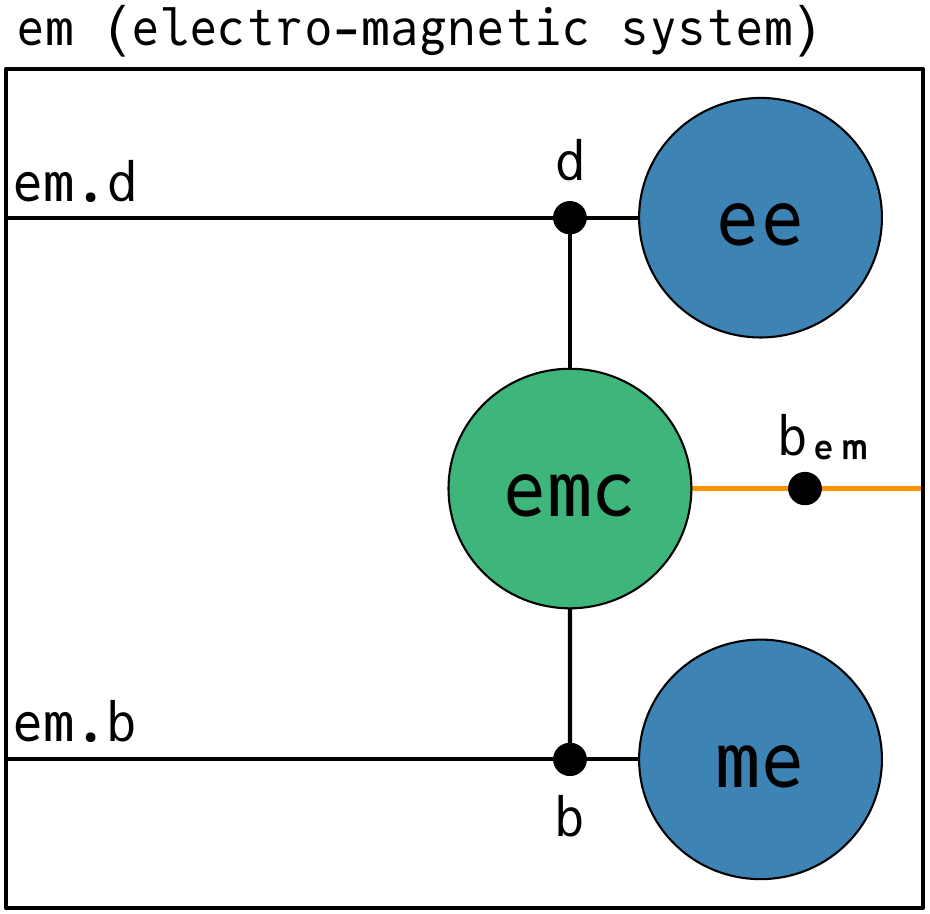}
  \caption{%
    Interconnection pattern of the Maxwell model.
    Boxes $\mathtt{ee}$ and $\mathtt{me}$
    respectively represent
    storage of electric and magnetic energy.
    Box $\mathtt{emc}$
    represents the reversible coupling of
    the electric and the magnetic energy domain,
    and
    the boundary port $\mathtt{b_{em}}$
    accounts for
    exchange of electromagnetic energy
    across $\partial \cZ$.
  }%
	\label{fig:em}
\end{figure}

In this section,
we implement a Maxwell model
describing electromagnetic wave propagation
in a medium with linear polarization and magnetization.
First,
we discuss the three primitive systems
filling the inner boxes of
the interconnection pattern
shown in~\cref{fig:em},
and then
we collect the equations for the composite system.

\subsection{Storage of electric energy}%

The storage component
$(I_\text{ee}, \, E_\text{ee})$
filling box $\mathtt{ee}$
is defined by
its interface
$
I_\text{ee} =
(\{ \mathtt{d} \}, \, \tau_\text{ee})
$
with
$
\tau_\text{ee}(\mathtt{d}) =
((\tilde{\Omega}^2(\cZ), \, \mathtt{electric\_displacement}), \, \mathsf{p})
$
and
its energy function
$
E_\text{ee} \colon \tilde{\Omega}^2(\cZ) \to \bR
$
given by
\begin{equation*}
	E_\text{ee}(\tilde{d})
	\: = \:
	\int_\cZ
	\frac{1}{2 \cdot \epsilon_0 \cdot \epsilon_r} \cdot
	\hodge \tilde{d} \wedge \tilde{d}
  \,,
\end{equation*}
where
$\tilde{d} = \mathtt{d.x}$
is the electric displacement.
The physical constant
$
\epsilon_0 =
8.8541878128 \cdot 10^{-12} \, \mathrm{F} \, \mathrm{m}^{-1}
$
is the vacuum permittivity
and
the parameter
$\epsilon_r \in \bR$
is the relative permittivity
characterizing the linear material response (polarization).

The flow and effort variables
are hence given by
\begin{equation}
  \begin{split}
    \mathtt{d.f}
    \: &= \:
    \dot{\tilde{d}}
    \, \in \, \tilde{\Omega}^2(\cZ)
    \\
    \mathtt{d.e}
    \: &= \:
    \frac{1}{\epsilon_0 \cdot \epsilon_r} \cdot \hodge \tilde{d}
    \: = \:
    e
    \, \in \, \Omega^1(\cZ)
    \,,
  \end{split}
  \label{eq:em_ee}
\end{equation}
where
$e$ is the electric field.

%
%
%

\subsection{Storage of magnetic energy}%

The storage component
$(I_\text{me}, \, E_\text{me})$
filling box $\mathtt{me}$
is defined by
its interface
$
I_\text{me} =
(\{ \mathtt{b} \}, \, \tau_\text{me})
$
with
$
\tau_\text{me}(\mathtt{b}) =
((\Omega^2(\cZ), \, \mathtt{magnetic\_flux}), \, \mathsf{k})
$
and
its energy function
$
E_\text{me} \colon \Omega^2(\cZ) \to \bR
$
given by
\begin{equation*}
	E_\text{me}(b)
	\: = \:
	\int_\cZ
	\frac{1}{2 \cdot \mu_0 \cdot \mu_r} \cdot
	\hodge b \wedge b
  \,,
\end{equation*}
where
$b = \mathtt{b.x}$
is the magnetic flux (density).
Here,
the physical constant
$\mu_0 = 1.25663706212 \cdot 10^{-6} \, \mathrm{N} \, \mathrm{A}^{-2}$
is the vacuum permeability
and
the parameter
$\mu_r \in \bR$
is the relative permeability
characterizing the linear material response (magnetization).

The flow and effort variables are hence given by
\begin{equation}
  \begin{split}
    \mathtt{b.f}
    \: &= \:
    \dot{b}
    \, \in \, \Omega^2(\cZ)
    \\
    \mathtt{b.e}
    \: &= \:
    \frac{1}{\mu_0 \cdot \mu_r} \cdot \hodge b
    \: = \:
    \tilde{h}
    \, \in \, \tilde{\Omega}^1(\cZ)
    \,,
  \end{split}
  \label{eq:em_me}
\end{equation}
where
$\tilde{h}$
is the magnetic field (strength).

%
%
%
%

\subsection{Electro-magnetic coupling}%

The reversible component
$(I_\text{emc}, \, \cD_\text{emc})$
filling box $\mathtt{emc}$
is defined by
its interface
$
I_\text{emc} =
(\{ \mathtt{d}, \, \mathtt{b}, \, \mathtt{b_{em}} \}, \, \tau_\text{emc})
$
with
\begin{equation*}
  \begin{alignedat}{2}
    &\tau_\text{emc}(\mathtt{d})
    \: &&= \:
    ((\tilde{\Omega}^2(\cZ), \, \mathtt{electric\_displacement}), \, \mathsf{p})
    \\
    &\tau_\text{emc}(\mathtt{b})
    \: &&= \:
    ((\Omega^2(\cZ), \, \mathtt{magnetic\_flux}), \, \mathsf{k})
  \end{alignedat}
\end{equation*}
and the Stokes-Dirac structure $\cD_\text{emc}$
given by
\begin{subequations}
  \begin{alignat}{2}
		\left[
			\begin{array}{c}
				\mathtt{d.f} \\
				\mathtt{b.f}
			\end{array}
		\right]
		\: &= \:
		\left[
			\begin{array}{cc}
				0 & -\dd(\_) \\
				\dd(\_) & 0
			\end{array}
		\right]
		\,
    &&\left[
			\begin{array}{c}
				\mathtt{d.e} \\
				\mathtt{b.e}
			\end{array}
		\right]%
		\\
		\left[
			\begin{array}{c}
				\mathtt{b_{em}.f} \\
				\mathtt{b_{em}.e}
			\end{array}
		\right]
		\: &= \:
		\left[
			\begin{array}{cc}
        i^*(\_) & 0 \\
				0 & i^*(\_)
			\end{array}
		\right]
		\,
    &&\left[
			\begin{array}{c}
				\mathtt{d.e} \\
				\mathtt{b.e}
			\end{array}
		\right]
    \,.%
	\end{alignat}%
  \label{eq:em_emc}%
\end{subequations}

This defines a power-preserving relation since
\begin{equation*}
  \begin{split}
    &\int_\cZ \bigl(
      \mathtt{d.e} \wedge \mathtt{d.f}
      \, + \,
      \mathtt{b.e} \wedge \mathtt{b.f}
    \bigr)
    \: = \:
    \\
    &\int_\cZ \dd ( \mathtt{d.e} \wedge \mathtt{b.e} )
    \: = \:
    -\int_{\partial \cZ}
      \mathtt{b_{em}.e} \wedge
      \mathtt{b_{em}.f}
    \,.
  \end{split}
\end{equation*}
We note that
$
-\mathtt{b_{em}.e} \wedge \mathtt{b_{em}.f} =
i^*(e \wedge \tilde{h})
\in {\tilde{\Omega}}^2(\partial \cZ)
$
is the twisted 2-form on $\partial \cZ$
corresponding to the Poynting vector.
The boundary port is defined such that
$\mathtt{b_{em}.f} = 0$
corresponds to
a vanishing electric field
tangential to the boundary
(due to a perfectly conducting wall).

\subsection{Interconnected Maxwell model}%

Combining~\cref{eq:em_ee,eq:em_me,eq:em_emc}
with the equations for the interconnection pattern in~\cref{fig:em}
and eliminating port variables
gives the following system of equations on $\cZ$:
\begin{equation}
  \begin{split}
    \dot{\tilde{d}}
    \: &= \:
    +\dd \tilde{h} + \mathtt{em.d.f}
    \\
    \dot{b}
    \: &= \:
    -\dd e
    \\
    \mathtt{em.d.e}
    \: &= \:
    e
    \,,
  \end{split}
  \label{eq:em}
\end{equation}
where
$
e =
\hodge \tilde{d} / (\epsilon_0 \cdot \epsilon_r)
$
and
$
\tilde{h} =
\hodge b / (\mu_0 \cdot \mu_r)
$.
Moreover,
we have
$\mathtt{b_{em}.f} = i^*(e)$
and
$\mathtt{b_{em}.e} = i^*(\tilde{h})$.

The free charge density $\tilde{q} \in \tilde{\Omega}^3(\cZ)$
is given by
$
\tilde{q} = \dd \tilde{d}
$.
If the current density (source term) $\mathtt{em.d.f}$ vanishes,
charge is a conserved quantity
(Casimir function),
since
\begin{equation*}
  \dot{\tilde{q}}
  \: = \:
  \dd \dot{\tilde{d}}
  \: = \:
  \dd \dd \tilde{h}
  \: = \:
  0
  \,.
\end{equation*}
An initial condition for~\cref{eq:em} has to satisfy
the constraint
$\dd b  = 0$,
called Gauss's law for magnetism.
The constraint is upheld since
$
\dd \dot{b} =
-\dd \dd e = 0
$.

\section{Electro-magneto hydrodynamics model}%
\label{sec:emhd}

\begin{figure}
  \centering
	\includegraphics[width=5.8cm]{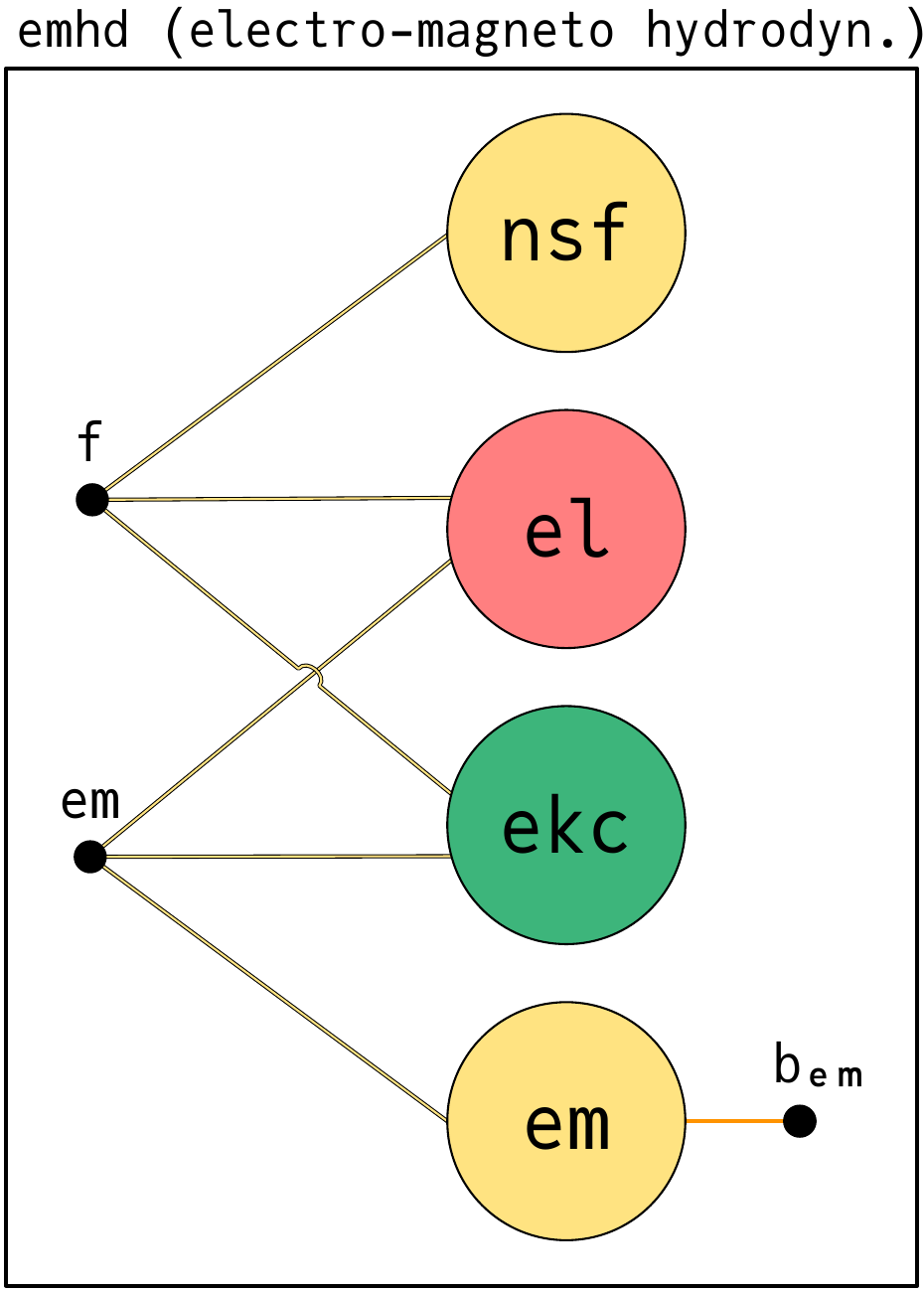}
  \caption{%
    Interconnection pattern of
    the electro-magneto hydrodynamics model.
    Box $\mathtt{nsf}$ represents
    the Navier-Stokes-Fourier system.
    Box $\mathtt{em}$ represents
    the Maxwell system
    describing electromagnetic wave propagation.
    Box $\mathtt{ekc}$ represents
    the electro-kinetic coupling.
    Box $\mathtt{el}$ represents
    electric conduction in the plasma.
    The boundary port
    $\mathtt{b_{em}}$,
    which accounts for
    radiation of electro-magnetic waves
    across the boundary of the spatial domain,
    is not exposed,
    leading to an isolated model.
  }%
  \label{fig:emhd}
\end{figure}

In this section,
we combine the Navier-Stokes-Fourier system
and the Maxwell system
into a model for a charged fluid
interacting with
internally generated
and externally applied
electric and magnetic fields,
see~\cref{fig:emhd}.
As defined next,
this involves two additional systems
describing
the reversible electro-kinetic coupling
and
the irreversible process of electric conduction.
Finally,
we collect the equations defining the EMHD model.

\subsection{Electro-kinetic coupling}%

Box $\mathtt{ekc}$ represents
the reversible coupling of
the plasma's kinetic energy domain
represented by port $\mathtt{f.p}$
and
the electric energy domain
represented by port $\mathtt{em.d}$.
On the one hand,
the motion of charged fluid particles
amounts to a current.
On the other hand,
an electric field
exerts a force on charged fluid particles.
According to special relativity,
the presence of a magnetic flux
leads to a force on moving charges.
The sum of these two forces
is called the Lorentz force.
This coupling is encapsulated by the following component.

The reversible component
$(I_\text{ekc}, \, \cD_\text{ekc})$
filling box $\mathtt{ekc}$
is defined by
its interface
$
I_\text{ekc} =
(\{ \mathtt{f.p}, \, \mathtt{em.d}, \, \mathtt{f.m}, \, \mathtt{em.b} \}, \, \tau_\text{ekc})
$
with
\begin{equation*}
  \begin{alignedat}{2}
    &\tau_\text{ekc}(\mathtt{f.p})
    \: &&= \:
    ((\Omega^1(\cZ), \, \mathtt{momentum}), \, \mathsf{k})
    \\
    &\tau_\text{ekc}(\mathtt{em.d})
    \: &&= \:
    ((\tilde{\Omega}^2(\cZ), \, \mathtt{electric\_displacement}), \, \mathsf{p})
    \\
    &\tau_\text{ekc}(\mathtt{f.m})
    \: &&= \:
    (\tilde{\Omega}^3(\cZ), \, \mathtt{mass})
    \\
    &\tau_\text{ekc}(\mathtt{em.b})
    \: &&= \:
    (\Omega^2(\cZ), \, \mathtt{magnetic\_flux})
  \end{alignedat}
\end{equation*}
and the Dirac structure $\cD_\text{ekc}$
given by
\begin{equation}
	\left[
		\begin{array}{c}
			\mathtt{f.p.f} \\
			\mathtt{em.d.f}
		\end{array}
	\right]
	\: = \:
  c \cdot \hodge \tilde{m} \cdot
	\left[
		\begin{array}{cc}
      \hodge( \hodge(\_) \wedge \hodge b ) & -(\_) \\
      +(\_) &
			0
		\end{array}
	\right]
	\,
	\left[
		\begin{array}{c}
			\mathtt{f.p.e} \\
			\mathtt{em.d.e}
		\end{array}
	\right]
  \,,
	\label{eq:ekc}
\end{equation}
where
$\tilde{m} = \mathtt{f.m.x}$,
$b = \mathtt{em.b.x}$,
and
$c \in \bR$ is a parameter
giving the effective charge per unit of mass.

Simplifying~\cref{eq:ekc} gives
the current density
$
\mathtt{em.d.f} =
c \cdot \hodge \tilde{m} \cdot \hodge \upsilon
$
and
the Lorentz force
$
-\mathtt{f.p.f} =
c \cdot \hodge \tilde{m} \cdot \left(
  \hodge( \upsilon \wedge \hodge b ) + e
\right)
$,
where the term
$
\hodge( \upsilon \wedge \hodge b )
= -\iota_u b
$
corresponds to
the cross product of
the velocity vector field
$u = \upsilon^\sharp$
and
the magnetic flux vector field
$\sharp(\hodge b)$.

\subsection{Electric conduction}%

Electric conduction
is a relaxation process
that counteracts
a non-uniform electric potential
with
an electric current.
As the electric field
is given by the differential of the electric potential,
the former directly reflects
spatial variations of the latter.
According to Ohm's law,
the current density is simply given by
$\kappa_e \cdot \hodge e$,
where $\kappa_e \geq 0$
is the electric conductivity.
This dynamics is encapsulated by the following component.

The irreversible component
$(I_\text{el}, \, \cO_\text{el})$
filling box $\mathtt{el}$
is defined by
its interface
$
I_\text{el} =
(\{ \mathtt{em.d}, \, \mathtt{f.s} \}, \tau_\text{el})
$
with
\begin{equation*}
  \begin{alignedat}{2}
    &\tau_\text{el}(\mathtt{em.d})
    \: &&= \:
    ((\tilde{\Omega}^2(\cZ), \, \mathtt{electric\_displacement}), \, \mathsf{p})
    \\
    &\tau_\text{el}(\mathtt{f.s})
    \: &&= \:
    ((\tilde{\Omega}^3(\cZ), \, \mathtt{entropy}), \, \mathsf{i})
  \end{alignedat}
\end{equation*}
and
the Onsager structure $\cO_\text{el}$
given by
\begin{equation}
	\left[
		\begin{array}{c}
			\mathtt{em.d.f} \\
			\mathtt{f.s.f}
		\end{array}
	\right]
	\: = \:
  \frac{1}{\textcolor{violet}{\theta_0}} \cdot \kappa_e \cdot
	\left[
		\begin{array}{cc}
			\theta \wedge \hodge (\_) &
			-(\_) \wedge \hodge e \\
			-(\_) \wedge \hodge e &
			\frac{1}{\theta} \cdot e \wedge \hodge e \cdot (\_)
		\end{array}
	\right]
	\,
	\left[
		\begin{array}{c}
			\mathtt{em.d.e} \\
			\mathtt{f.s.e}
		\end{array}
	\right]
  \,,%
	\label{eq:el}
\end{equation}
where
$e = \mathtt{em.d.e}$
and
$\theta = \textcolor{violet}{\theta_0} + \mathtt{f.s.e}$.

Simplifying~\cref{eq:el} gives
the electric current density
$
\mathtt{em.d.f} =
\kappa_e \cdot \hodge e
\in \tilde{\Omega}^2(\cZ)
$
and
the entropy production rate
$
-\mathtt{f.s.f} =
\frac{1}{\theta} \cdot e \wedge (\kappa_e \cdot \hodge e)
\in \tilde{\Omega}^3(\cZ)
$.

The total exergy destruction rate
is given by
\begin{equation*}
  \int_\cZ \bigl(
    \mathtt{em.d.e} \wedge \mathtt{em.d.f}
    \, + \,
    \mathtt{f.s.e} \wedge \mathtt{f.s.f}
  \bigr)
  \: = \:
  \textcolor{violet}{\theta_0} \cdot
  \int_\cZ
  \frac{1}{\theta} \cdot e \wedge (\kappa_e \cdot \hodge e)
  \,.
\end{equation*}

\Cref{eq:el} satisfies
the condition for conservation of energy since
$
\theta \wedge \hodge (e)
-(\theta) \wedge \hodge e
= 0
$
and
$
-(e) \wedge \hodge e +
\frac{1}{\theta} \cdot e \wedge \hodge e \cdot (\theta)
= 0
$,
where
$e = \mathtt{em.d.e}$
and
$\theta = \textcolor{violet}{\theta_0} + \mathtt{f.s.e}$.

To improve the model,
one could add a state port
to make the electric conductivity depend on
mass density and temperature.
Once could also construct
an irreversible component for
combined thermal and electric conduction
that takes into account
thermodynamic cross effects.

\subsection{Interconnected EMHD model}%


Combining~\cref{eq:nsf,eq:em,eq:ekc,eq:el}
with the equations for the interconnection pattern in~\cref{fig:emhd}
and eliminating port variables
gives the following system of equations on $\cZ$:
\begin{subequations}
	\begin{align}
		\begin{split}
			\dot{\upsilon}
			\: = \:
      &+ \hodge(\upsilon \wedge \hodge \dd \upsilon)
			-\dd (\hodge (\upsilon \wedge \hodge \upsilon) / 2)
      -\frac{1}{\hodge \tilde{m}} \cdot \dd \pi
      \\
      &+ \frac{1}{\hodge \tilde{m}} \cdot \dd( \mu_v \cdot \hodge \, \dd \, \hodge \, \upsilon ) \\
      &+ \frac{1}{\hodge \tilde{m}} \cdot \hodge_2 \, \dd_\nabla \bigl( \mu_s \cdot \hodge_2(\mathrm{sym}(\nabla \upsilon )) \bigr) \\
      &+ c \cdot \left( \hodge (\upsilon \wedge \hodge b) + e \right)
		\end{split}
		\\
    \begin{split}
      \dot{\tilde{m}}
      \: = \:
      &-\dd (\hodge \tilde{m} \cdot \, \hodge \upsilon)
    \end{split}
		\\
		\begin{split}
      \dot{\tilde{s}}
      \: = \:
      &-\dd ( \hodge \tilde{s} \cdot \, \hodge \upsilon ) \\
			&+ \frac{1}{\theta} \cdot \dd( \kappa_t \cdot \hodge \dd \theta ) \\
			&+ \frac{1}{\theta} \cdot \mu_v \cdot (\hodge \, \dd \, \hodge \, \upsilon) \cdot (\dd \, \hodge \, \upsilon) \\
			&+ \frac{1}{\theta} \cdot \nabla \upsilon^\sharp \dot{\wedge} \bigl( \mu_s \cdot \hodge_2(\mathrm{sym}(\nabla \upsilon )) \bigr) \\
      &+ \frac{1}{\theta} \cdot e \wedge ( \kappa_e \cdot \hodge e )
		\end{split}%
    \\
		\begin{split}
      \dot{\tilde{d}}
      \: = \:
      &+ \dd \tilde{h} \\
      &- c \cdot \hodge \tilde{m} \cdot \hodge \upsilon \\
		  &- \kappa_e \cdot \hodge e
		\end{split}%
    \\
		\begin{split}
      \dot{b}
      \: = \:
      &-\dd e
		\end{split}%
	\end{align}%
	\label{eq:emhd}%
\end{subequations}
The previously stated boundary conditions apply.
Due to the combination of the Maxwell system with the fluid model,
the free charge density $\tilde{q}$
must satisfy the constraint
\begin{equation*}
  \tilde{q}
  \: = \:
  \dd \tilde{d}
  \: = \:
  c \cdot \tilde{m}
  \,,
\end{equation*}
which is upheld since
\begin{equation*}
  \dot{\tilde{q}}
  \: = \:
  \dd \dot{\tilde{d}}
  \: = \:
  \dd \dd \tilde{h} - \dd(c \cdot \hodge \tilde{m} \cdot \hodge \upsilon)
  \: = \:
  - c \cdot \dd(\hodge \tilde{m} \cdot \hodge \upsilon)
  \: = \:
  c \cdot \dot{\tilde{m}}
  \,.
\end{equation*}
Initial conditions
for~\cref{eq:emhd}
must hence satisfy
$
\dd \tilde{d} =
c \cdot \tilde{m}
$
and
$\dd b  = 0$.

We note that
state variables that are
straight differential forms
have negative parity
with respect to time-reversal transformation~%
\cite{2021LohmayerKotyczkaLeyendecker},
whereas
those which are
twisted differential forms
have positive parity.
We expect that this observation plays an important role for
structure-preserving discretization
on staggered space and time grids.

\section{Magneto hydrodynamics model}%
\label{sec:mhd}

\begin{figure}
  \centering
	\includegraphics[width=5.8cm]{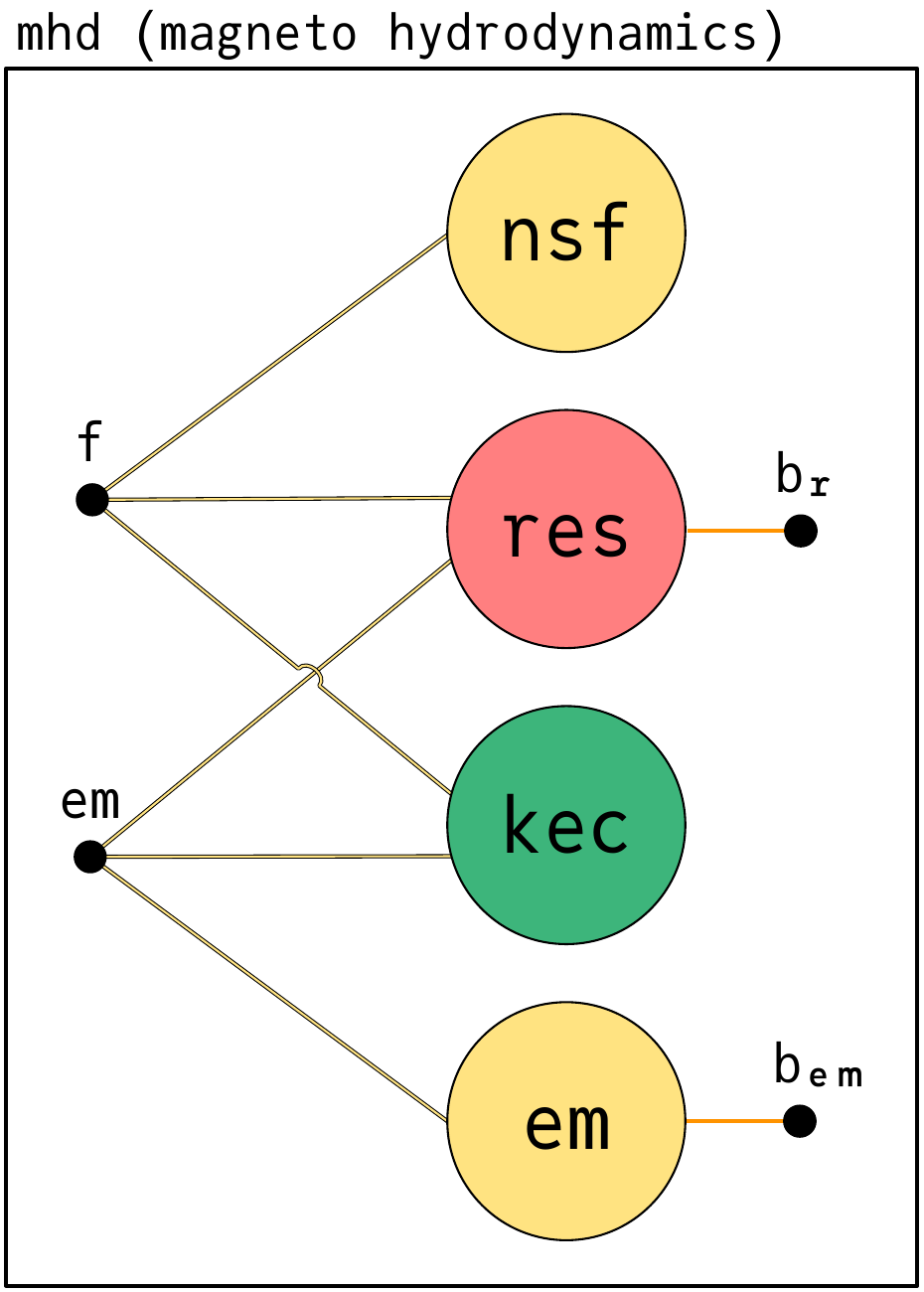}
  \caption{%
    Interconnection pattern of
    the magneto hydrodynamics model.
    Box $\mathtt{nsf}$ represents
    the Navier-Stokes-Fourier system.
    Box $\mathtt{em}$ represents
    the Maxwell system,
    but without a storage component for electric energy.
    Box $\mathtt{kec}$ represents
    the reversible coupling
    of the kinetic and electric energy domains.
    Box $\mathtt{res}$ represents
    the irreversible process due to
    the ohmic resistance of the plasma.
    The boundary port
    $\mathtt{b_r}$
    accounting for
    electric conduction
    across $\partial \cZ$
    is not exposed,
    leading to an isolated model.
  }%
  \label{fig:mhd}
\end{figure}

In this section,
we present a simplification of
the plasma model defined in the previous section.
When the electric field
evolves much faster than
the magnetic field,
the dynamics of the former can be neglected.
Besides the NSF fluid model
and
the Maxwell model
with the storage component for electric energy removed,
the MHD model has two additional subsystems:
Similar to, but distinct from, the EMHD model,
the first system describes
the reversible coupling
of the kinetic and electric energy domains
and
the second system describes
the irreversible process due to
the ohmic resistance of the plasma,
see~\cref{fig:mhd}.

\subsection{Kinetic-electric coupling}%

Box $\mathtt{kec}$ represents
the reversible coupling of
the fluid's kinetic energy domain
represented by port $\mathtt{f.p}$
and
the electric energy domain
represented by port $\mathtt{em.d}$.

The reversible component
$(I_\text{kec}, \, \cD_\text{kec})$
filling box $\mathtt{kec}$
is defined by
its interface
$
I_\text{kec} =
(\{ \mathtt{f.p}, \, \mathtt{em.d}, \, \mathtt{em.b} \}, \, \tau_\text{kec})
$
with
\begin{equation*}
  \begin{alignedat}{2}
    &\tau_\text{kec}(\mathtt{f.p})
    \: &&= \:
    ((\Omega^1(\cZ), \, \mathtt{momentum}), \, \mathsf{k})
    \\
    &\tau_\text{kec}(\mathtt{em.d})
    \: &&= \:
    ((\tilde{\Omega}^2(\cZ), \, \mathtt{electric\_displacement}), \, \mathsf{p})
    \\
    &\tau_\text{kec}(\mathtt{em.b})
    \: &&= \:
    (\Omega^2(\cZ), \, \mathtt{magnetic\_flux})
  \end{alignedat}
\end{equation*}
and the Dirac structure $\cD_\text{kec}$
given by
\begin{equation}
  \begin{bmatrix}
      \mathtt{f.p.f} \\
      \mathtt{em.d.e}
  \end{bmatrix}
  \: = \:
  \begin{bmatrix}
      0 & \hodge( \hodge(\_) \wedge \hodge b ) \\
      -\hodge( \hodge(\_) \wedge \hodge b ) & 0
  \end{bmatrix}
  \,
  \begin{bmatrix}
      \mathtt{f.p.e} \\
      \mathtt{em.d.f}
  \end{bmatrix}
  \,,
	\label{eq:kec}
\end{equation}
where
$b = \mathtt{em.b.x}$.

By simplifying~\cref{eq:kec}, we obtain
the magnetic force
$\mathtt{f.p.f} = \hodge(\tilde{j} \wedge \hodge b)$.
The current density
$\tilde{j} = \dd \tilde{h}$
is determined by the magnetic field through Ampère's law,
which is expressed by the reversible component
filling box $\mathtt{em.emc}$.
Further,~\cref{eq:kec} determines
the electric field
$\mathtt{em.d.e} = -\hodge(\upsilon \wedge \hodge b)$
according to
the ideal MHD condition
$e + \hodge (\upsilon \wedge \hodge b) = 0$.
As the MHD model does not
account for storage of electric energy,
the reversible component
filling box $\mathtt{kec}$
basically transforms port $\mathtt{em.d}$
of the reversible component filling box $\mathtt{em.emc}$
into a port
representing the kinetic energy domain.
In principle,
the two reversible components
filling boxes $\mathtt{kec}$ and $\mathtt{em.emc}$
could be combined into a single component that models
the reversible coupling of
the kinetic and the magnetic energy domain
directly.

\subsection{Ohmic resistance}%

Box $\mathtt{res}$ represents
the irreversible process due to
the ohmic resistance
$\rho_e = 1/\kappa_e \geq 0$ of the plasma.
The model is essentially equivalent to
the electric conduction model
used for the EMHD model,
but expressed on the magnetic side.
The electric conduction model
is defined by Ohm's law
$\mathtt{el.d.f} = \kappa_e \cdot \hodge(\mathtt{el.d.e})$,
which we can write equivalently as
$\mathtt{el.d.e} = \rho_e \cdot \hodge(\mathtt{el.d.f})$.
Combining this with~\cref{eq:em_emc}
for the electro-magnetic coupling
leads to the following component.

The irreversible component
$(I_\text{res}, \, \cO_\text{res})$
filling box $\mathtt{res}$
is defined by
its interface
$
I_\text{res} =
(\{ \mathtt{em.b}, \, \mathtt{f.s}, \, \mathtt{b_r} \}, \tau_\text{res})
$
with
\begin{equation*}
  \begin{alignedat}{2}
    &\tau_\text{res}(\mathtt{em.b})
    \: &&= \:
    ((\Omega^2(\cZ), \, \mathtt{magnetic\_flux}), \, \mathsf{k})
    \\
    &\tau_\text{res}(\mathtt{f.s})
    \: &&= \:
    ((\tilde{\Omega}^3(\cZ), \, \mathtt{entropy}), \, \mathsf{i})
  \end{alignedat}
\end{equation*}
and
the Onsager structure $\cO_\text{res}$
given by
\begin{subequations}
	\begin{align}
    \begin{split}
      \left[
        \begin{array}{c}
          \mathtt{em.b.f} \\
          \mathtt{f.s.f}
        \end{array}
      \right]
      \: &= \:
      \frac{1}{\textcolor{violet}{\theta_0}} \cdot
      \left[
        \begin{array}{cc}
          A(\_) & B(\_) \\
          C(\_) & D(\_)
        \end{array}
      \right]
      \,
      \left[
        \begin{array}{c}
          \mathtt{em.b.e} \\
          \mathtt{f.s.e}
        \end{array}
      \right]
      \\
      A(\_)
      \: &= \:
      \dd \bigl( (\rho_e \cdot \hodge \dd(\_) ) \wedge \theta \bigr) \\
      B(\_)
      \: &= \:
      -\dd \bigl( (\rho_e \cdot \hodge \dd \tilde{h} ) \wedge (\_) \bigr) \\
      C(\_)
      \: &= \:
      -\dd \tilde{h} \wedge \bigl( \rho_e \cdot \hodge \dd(\_) \bigr) \\
      D(\_)
      \: &= \:
      \frac{1}{\theta} \cdot \dd \tilde{h} \wedge ( \rho_e \cdot \hodge \dd \tilde{h} ) \wedge (\_)
    \end{split}
		\label{eq:res_domain}
		\raisetag{1.1em}
    \\
		\begin{split}
      \mathtt{b_r.f}
      \: &= \:
      i^*\biggl(
        \frac{1}{\textcolor{violet}{\theta_0}} \, \biggl(
          \bigl( \rho_e \cdot \hodge \dd (\mathtt{em.b.e}) \bigr) \wedge \theta
          \\ & \qquad \qquad
          -(\rho_e \cdot \hodge \dd \tilde{h}) \wedge \mathtt{f.s.e}
        \biggr)
      \biggr)
      \\
      \mathtt{b_r.e}
      \: &= \:
      i^*(\mathtt{em.b.e})
      \,,%
    \end{split}
		\raisetag{1.1em}
	\end{align}%
	\label{eq:res}%
\end{subequations}
where
$\tilde{h} = \mathtt{em.b.e}$
and
$\theta = \textcolor{violet}{\theta_0} + \mathtt{f.s.e}$.

Simplification of~\cref{eq:res} gives
the rate
$
\mathtt{em.b.f} =
\dd( \rho_e \cdot \hodge \dd \tilde{h} )
\in \Omega^2(\cZ)
$
at which the magnetic flux changes
due to the ohmic resistance of the plasma.
Moreover, we obtain
the entropy production rate
$
-\mathtt{f.s.f} =
\frac{1}{\theta} \cdot \dd \tilde{h} \wedge (\rho_e \cdot \hodge \dd \tilde{h})
\in \tilde{\Omega}^3(\cZ)
$.
For the boundary port, we have
$
\mathtt{b_r.f} =
i^*(\rho_e \cdot \hodge \dd \tilde{h})
\in \Omega^1(\partial \cZ)
$
and
$
\mathtt{b_r.e} =
i^*(\tilde{h})
\in \tilde{\Omega}^1(\partial \cZ)
$.

To show that~\cref{eq:res} defines
a non-negative definite symmetric operator,
we again
apply integration by parts
to obtain a weak form
with implicit boundary conditions.
Letting the test functions
be equal to
the effort variables $\mathtt{em.b.e}$ and $\mathtt{f.s.e}$
gives the following power balance equation:
\begin{equation*}
	\begin{split}
		\int_\cZ \bigl(
      \mathtt{em.b.e} &\wedge \mathtt{em.b.f}
      +
      \mathtt{f.s.e} \wedge \mathtt{f.s.f}
    \bigr)
		+
		\int_{\partial \cZ}
      \mathtt{b_r.e} \wedge \mathtt{b_r.f}
		\: = \:
		\\
		\frac{1}{\textcolor{violet}{\theta_0}} \cdot \int_\cZ \Bigl[
      &+\dd(\mathtt{em.b.e}) \wedge \bigl( \rho_e \cdot \hodge \dd(\mathtt{em.b.e}) \bigr) \wedge \theta
			\\
      &-\dd(\mathtt{em.b.e}) \wedge (\rho_e \cdot \hodge \dd \tilde{h}) \wedge \mathtt{f.s.e}
			\\
      &-\mathtt{f.s.e} \wedge \dd \tilde{h} \wedge \bigl( \rho_e \cdot \hodge \dd(\mathtt{em.b.e}) \bigr)
			\\
      &+\mathtt{f.s.e} \wedge \biggl( \frac{1}{\theta} \cdot \dd \tilde{h} \wedge (\rho_e \cdot \hodge \dd \tilde{h}) \biggr) \wedge \mathtt{f.s.e}
		\Bigr]
		\: = \:
    \\
    \textcolor{violet}{\theta_0} \cdot \int_\cZ
    &\frac{1}{\theta} \cdot \dd \tilde{h} \wedge (\rho_e \cdot \hodge \dd \tilde{h})
		\: \geq \: 0
    \,.
	\end{split}
\end{equation*}
Again,
we see that the operator is symmetric
and the second law is satisfied.

\Cref{eq:res_domain} satisfies the condition
for conservation of energy
because
$A(\tilde{h}) + B(\theta) = 0$
and
$B(\tilde{h}) + D(\theta) = 0$,
where
$\tilde{h} = \mathtt{em.b.e}$
and
$\theta = \textcolor{violet}{\theta_0} + \mathtt{f.s.e}$.

\subsection{Interconnected MHD model}%

Without the storage component for electric energy,
\cref{eq:em} for the Maxwell system reduces to
the Maxwell-Faraday equation
$
\dot{b} =
-\dd(\mathtt{em.d.e})
$
and Ampère's law
$
-\mathtt{em.d.f} =
\dd \tilde{h}
$,
where
$
\tilde{h} =
\hodge b / (\mu_0 \cdot \mu_r)
$.
Combining this with~\cref{eq:nsf,eq:kec,eq:res}
and the equations for the interconnection pattern in~\cref{fig:mhd}
and eliminating port variables
gives the following system of equations on $\cZ$:
\begin{subequations}
	\begin{align}
		\begin{split}
			\dot{\upsilon}
			\: = \:
      &+ \hodge(\upsilon \wedge \hodge \dd \upsilon)
			-\dd (\hodge (\upsilon \wedge \hodge \upsilon) / 2)
      -\frac{1}{\hodge \tilde{m}} \cdot \dd \pi
      \\
      &+ \frac{1}{\hodge \tilde{m}} \cdot \dd( \mu_v \cdot \hodge \, \dd \, \hodge \, \upsilon ) \\
      &+ \frac{1}{\hodge \tilde{m}} \cdot \hodge_2 \, \dd_\nabla \bigl( \mu_s \cdot \hodge_2(\mathrm{sym}(\nabla \upsilon )) \bigr) \\
      &+ \frac{1}{\hodge \tilde{m}} \cdot \hodge \bigl( \hodge \dd \tilde{h} \wedge \hodge b \bigr)
		\end{split}
		\\
    \begin{split}
      \dot{\tilde{m}}
      \: = \:
      &-\dd (\hodge \tilde{m} \cdot \, \hodge \upsilon)
    \end{split}
		\\
		\begin{split}
      \dot{\tilde{s}}
      \: = \:
      &-\dd ( \hodge \tilde{s} \cdot \, \hodge \upsilon ) \\
			&+ \frac{1}{\theta} \cdot \dd( \kappa_t \cdot \hodge \dd \theta ) \\
			&+ \frac{1}{\theta} \cdot \mu_v \cdot (\hodge \, \dd \, \hodge \, \upsilon) \cdot (\dd \, \hodge \, \upsilon) \\
			&+ \frac{1}{\theta} \cdot \nabla \upsilon^\sharp \dot{\wedge} \bigl( \mu_s \cdot \hodge_2(\mathrm{sym}(\nabla \upsilon )) \bigr) \\
      &+ \frac{1}{\theta} \cdot \dd \tilde{h} \wedge ( \rho_e \cdot \hodge \dd \tilde{h} )
		\end{split}%
    \\
		\begin{split}
      \dot{b}
      \: = \:
      &+\dd \hodge (\upsilon \wedge \hodge b) \\
      &-\dd \bigl( \rho_e \cdot \hodge \dd \tilde{h} \bigr)
		\end{split}%
	\end{align}%
	\label{eq:mhd}%
\end{subequations}
The boundary conditions stated previously
for the respective subsystems apply.
Concerning the boundary port of
the Maxwell system (without electric energy storage),
we have
$
\mathtt{b_{em}.f} =
-i^*(\upsilon \wedge \hodge b)
\in \Omega^1(\partial \cZ)
$
and
$
\mathtt{b_{em}.e} =
-i^*(\tilde{h})
\in \tilde{\Omega}^1(\partial \cZ)
$.
Initial conditions
for~\cref{eq:mhd}
must satisfy Gauss's law
$\dd b  = 0$.

\section{Conclusion}%
\label{sec:conclusion}

We presented two plasma models
using the Exergetic Port-Hamiltonian Systems (EPHS) modeling language.
The compositional graphical syntax allowed us
to gradually build up the models from simpler parts.
As subsystems,
we implemented
an ideal fluid model,
a Navier-Stokes-Fourier fluid model
that reuses the ideal fluid model as a subsystem,
and a Maxwell electromagnetism model.
The hierarchical organization of increasingly complex models
as well as the energy-based perspective
simplify understanding and communication.
In particular,
this allows people who are not domain experts
to understand the architecture and meaning of relatively complex physical models.
Moreover,
models and their parts can be easily reused, removed, or replaced,
as shown in this work.
This can save a lot of time
when models have to be modified or extended.
The three kinds of
primitive systems at the bottom of the hierarchical model specifications
represent
storage
as well as
reversible and irreversible exchange of energy.
Based on their structure,
all models are guaranteed to respect
the first and second law of thermodynamics.

Besides shedding light on
the energetic, thermodynamic, and modular structure,
we have expressed the considered fluid models
using exterior calculus
as a coordinate-invariant language.
Although it is unfortunately absent from most curricula,
the geometric approach has a very long, if not the longest, tradition in science.
In recent times,
it again becomes increasingly clear that
a deeper knowledge of the geometric character of physical quantities
significantly enhances understanding.

Leveraging the structured and coordinate-invariant expression
of the considered fluid models,
future work should investigate
their spatial and temporal discretization.
Making use of compositionality,
the involved primitive subsystems
should be transformed into discrete analogues,
which can then be interconnected
just as their continuous counterparts.
It could be instrumental to build upon prior work
that derives numerical schemes
based on
discrete analogues of exterior calculus
as well as
variational methods from mechanics,
see e.g.~\onlinecite{2011PavlovMullenTongKansoMarsdenDesbrun,2022CoueraudGay}.
If such methods could be understood as natural transformations
applied to EPHS models,
it could become much easier for users to readily apply them,
especially in the context to networks of interconnected systems.
Regarding the goal of structure-preserving discretization,
it could also be helpful to study
the involved differential complexes more deeply,
see e.g.~\onlinecite{2021ArnoldHu}.
This might also clarify how
to express
the isotropic-deviatoric splitting
of the strain rate tensor
in the coordinate-invariant language,
in order to fully disentangle
the volume and shear viscosity models.
Finally,
the topic of interconnecting models
through their boundary ports
should receive more attention,
both in the continuous and the discrete setting.

\section*{Conflict of interest statement}%

The authors report no conflict of interest.

\section*{Author contribution statement}%

\textbf{Markus Lohmayer}:
  Conceptualization,
  Investigation,
  Writing -- Original Draft,
  Writing -- Review \& Editing,
  Visualization;
\textbf{Michael Kraus}:
  Investigation,
  Review \& Editing;
\textbf{Sigrid Leyendecker}:
  Supervision,
  Review \& Editing,
  Funding

\section*{Data availability}%

Data sharing is not applicable to this article
as no new data were created or analyzed in this study.

\appendix
\section{Geometric foundation}%
\label{sec:geometry}

Here,
we want to
introduce the geometric concepts
that are
used in the main part of the paper.
More details can be found e.g.~in
the lectures by~\onlinecite{2021Crane}
or the textbooks~\onlinecite{2012Lee,1988AbrahamMarsdenRatiu}.

\subsection{Linear duality}%

Given a finite-dimensional vector space $V$,
its \textbf{dual space} $V^*$
is the vector space of all linear functions
from $V$ to $\bR$.
Vector addition on $V^*$ is defined by
$
(\alpha_1 + \alpha_2)(v) =
\alpha_1(v) + \alpha_2(v)
$
for
any two dual vectors (or covectors)
$\alpha_1, \, \alpha_2 \in V^*$
and
any vector $v \in V$.
Scalar multiplication on $V^*$
is also inherited from $\bR$, i.e.~%
$
(c \cdot \alpha)(v) =
c \cdot \alpha(v)
$
for
any scalar $c \in \bR$,
any covector $\alpha \in V^*$,
and
any vector $v \in V$.

The \textbf{duality pairing}
$
\langle \_ \mid \_ \rangle \colon
V^* \times V \to \bR
$
is simply defined by
$\langle \alpha \mid v \rangle = \alpha(v)$
for
any covector $\alpha \in V^*$
and
any vector $v \in V$.
A basis
$(e_1, \, \ldots, \, e_n)$ for $V$
determines
the corresponding dual basis
$(e^1, \, \ldots, \, e^n)$ for $V^*$
by requiring
$\langle e^i \mid e_j \rangle = \delta^i_j$
for all $i, \, j = 1, \, \ldots, \, n$,
where
$n = \dim(V) = \dim(V^*)$
and
$\delta^i_j = 1$ if $i=j$ and $\delta^i_j = 0$ otherwise.
It hence holds that $V^{**} = V$.

Given a linear map $f \colon V \to W$
between two vector spaces,
the \textbf{dual map} (or linear adjoint)
$f^* \colon W^* \to V^*$
is defined by
$
\langle f^*(\alpha) \mid v \rangle =
\langle \alpha \mid f(v) \rangle
$
for any
$\alpha \in W^*$
and
$v \in V$.
Assuming a choice of basis for both $V$ and $W$,
linear maps $V \to W$
can be represented as matrices.
The matrix for $f^*$ then simply is
the transpose of the matrix for $f$.

\subsection{Tensor algebra}%

A $(p, q)$-\textbf{tensor} on $V$
(\emph{contravariant} of order $p$
and \emph{covariant} of order $q$)
can be seen as
a multilinear map
\begin{equation*}
  t \colon
  \underbrace{V^* \times \ldots \times V^*}_\text{$p$ copies} \times
  \underbrace{V \times \ldots \times V}_\text{$q$ copies} \to \bR
  \,.
\end{equation*}
Multilinear maps are linear in each argument
when the other arguments are held fixed.
The vector space of \mbox{$(p, q)$-tensors} is denoted by
$V^{\otimes p} \otimes (V^*)^{\otimes q}$
and
$V^{\otimes p} = V \otimes \ldots \otimes V$ ($p$ copies)
is called the $p$-th \textbf{tensor power} of $V$.
Vector addition and scalar multiplication are inherited from $\bR$.
For instance,
let $t_1$ and $t_2$ be two \mbox{$(0, 2)$-tensors} on $V$.
Their sum $t = t_1 + t_2$ is defined by
$
t(v_1, \, v_2) =
t_1(v_1, \, v_2) +
t_2(v_1, \, v_2)
$
for all $v_1, \, v_2 \in V$.
The \mbox{$(0, 0)$-tensors} are scalars
and
$V^{\otimes 0} \cong \bR$ is
the unit for the tensor product (of vector spaces).
Considering again a \mbox{$2$-covariant} tensor as example,
$t \colon V^* \otimes V^*$
may be seen as
a map from the unit
$t \colon \bR \to V^* \otimes V^*$
(using scalar multiplication)
and by duality
it gives a map
$t \colon V \to V^*$
or
$t \colon V \otimes V \to \bR$.
The \textbf{tensor product} (of tensors)
is defined by multiplication of the resulting scalars.
E.g.~the tensor product
$t = t_1 \otimes t_2$ of
a \mbox{$(k,0)$-tensor} $t_1$ and
a \mbox{$(l,0)$-tensor} $t_2$ is
the \mbox{$(k+l,0)$-tensor} $t$
defined by
$
t(\alpha_1, \, \ldots, \, \alpha_k, \, \alpha_{k+1}, \, \ldots, \, \alpha_{k+l}) =
t_1(\alpha_1, \, \ldots, \, \alpha_k) \cdot
t_2(\alpha_{k+1}, \, \ldots, \, \alpha_{k+l})
$
for all $\alpha_1, \, \ldots, \, \alpha_{k+l} \in V^*$.
Equipped with the tensor product,
the formal sum
$\bigoplus_{k=0}^\infty V^{\otimes k}$
becomes a graded algebra
called the \textbf{tensor algebra} on $V$.

\subsection{Exterior algebra}%

The \textbf{exterior algebra} $\Lambda(V)$
on a vector space $V$
is the graded subalgebra of its tensor algebra
that includes only antisymmetric tensors.
A tensor is called antisymmetric if
swapping two arguments changes the sign of the resulting scalar.
Tensors in
the $k$-th \textbf{exterior power}
$\Lambda^k(V) \subset V^{\otimes k}$
are called \mbox{$k$-vectors}.
The \textbf{exterior product} (or wedge product)
$\wedge \colon \Lambda^k(V) \times \Lambda^l(V) \to \Lambda^{k+l}(V)$
is the antisymmetrized tensor product
defined by
\begin{equation*}
  \begin{split}
    &(v_1 \wedge v_2)(\alpha_1, \, \ldots, \, \alpha_{k+l})
    \: = \:
    \\
    &\frac{1}{k! \cdot l!} \cdot
    \sum_{\sigma \in S_{k+l}} \mathrm{sgn}(\sigma) \cdot
    (v_1 \otimes v_2)
    \bigl( \alpha_{\sigma(1)}, \, \ldots, \, \alpha_{\sigma(k+l)} \bigr)
    \,,
  \end{split}
\end{equation*}
where
$S_{k+l}$ is the set of all permutations
of the indices $1, \, \ldots, \, k+l$
and
$\mathrm{sgn} \colon S_{k+l} \to \{ -1, \, +1 \}$
yields their sign.
Due to the anti-symmetry,
we have
\begin{equation}
  v_1 \wedge v_2
  \: = \:
  (-1)^{k \cdot l} \, \, v_2 \wedge v_1
  \label{eq:wedge_anticommutativity}
\end{equation}
for all
$v_1 \in \Lambda^k(V)$
and
$v_2 \in \Lambda^l(V)$.
\mbox{$1$-vectors} encode a directed length
and
\mbox{$2$-vectors} encode an oriented area.
According to~\cref{eq:wedge_anticommutativity},
the orientation of a \mbox{$2$-vector} is reversed
if the two \mbox{$1$-vectors} spanning the area
(thought of as a  parallelogram)
are swapped.
If they are linearly dependent,
the resulting \mbox{$2$-vector} is
the $2^\text{nd}$ grade zero vector
(as the area of the parallelogram is zero).
\mbox{$0$-vectors} are scalars
and
\mbox{$3$-vectors} encode an oriented volume.
Assuming $\dim(V) = 3$,
we have
$\dim(\Lambda^0(V)) = \dim(\Lambda^3(V)) = 1$
and
$\dim(\Lambda^1(V)) = \dim(\Lambda^2(V)) = 3$.

\subsection{Hodge duality}%

An inner product on an $n$-dimensional vector space $V$
induces a linear isomorphism
$\star \colon \Lambda^k(V) \to \Lambda^{n-k}(V)$
called the Hodge star.
For $V = \bR^3$
with
an orthonormal basis $(e_1, \, e_2, \, e_3)$,
we have
$\star 1 = e_1 \wedge e_2 \wedge e_3$,
$\star e_1 = e_2 \wedge e_3$,
$\star e_2 = e_3 \wedge e_1$,
$\star e_3 = e_1 \wedge e_2$, etc.
The Hodge dual $\star v$ of a \mbox{$1$-vector} $v$
is oriented in the plane that is orthogonal to $v$.
Hence,
the Hodge star converts between scalars and volume elements
and between orthogonal line and surface elements.
Up to a sign,
it is its own inverse,
as we have
$\star \star v = (-1)^{k \cdot (n-k)} \, v$
for any $v \in \Lambda^k(V)$.
Due to the symmetry of the inner product,
we have
$v_1 \wedge \star v_2 = v_2 \wedge \star v_1$
for any \mbox{$k$-vectors} $v_1$ and $v_2$.

\subsection{Smooth manifolds}%

A \textbf{smooth manifold} $M$
is a topological space,
meaning that it has
a notion of neighborhoods around points.
For some neighborhood (open set) $U \subset M$,
a \textbf{coordinate chart} on $U$ is
a smooth isomorphism
$x \colon U \to x(U)$
that takes any point in $U$
to its coordinate representation
in $x(U) \subseteq \mathbb{R}^n$,
where $n = \dim(M)$.
While a single chart suffices for a flat space
such as $\bR^3$,
in general it takes multiple overlapping charts
to cover a manifold.
For each overlap of two charts,
there is a smooth isomorphism
called a chart transition map.

Given that it is not empty,
the boundary $\partial M$ of $M$ is seen as
a $(n-1)$-dimensional submanifold
with \emph{inclusion} map
$i \colon \partial M \hookrightarrow M$.
At the boundary,
$M$ is locally isomorphic to a half-space of $\bR^n$.

\subsection{The derivative and the tangent bundle}%

Let
$f: M \to N$
be a smooth function
between two smooth manifolds.
The \textbf{derivative} of $f$
evaluated at point $p \in M$
is the linear function
\begin{equation*}
  \begin{split}
    \mathrm{T}_p f \colon
    \mathrm{T}_p M &\to \mathrm{T}_{f(p)} N
    \\
    v &\mapsto
    \dv{t} \, \Big\vert_{t=0} \: \: f \bigl( c(t) \bigr)
    \,,
  \end{split}
\end{equation*}
where
$c : \mathbb{I} \to M$
with $\mathbb{R} \supseteq \mathbb{I} \ni 0$
is an arbitrary smooth curve on $M$
such that
$c(0) = p$ and $\dot{c}(0) = v$.
Hence,
the derivative 
of a smooth curve $c$
at some point (here $0$)
is a vector $v \in \mathrm{T}_{c(0)} M$,
which is seen to be
tangent to the curve at that point.
The \textbf{tangent space}
of $M$
over point $p$,
denoted by $\mathrm{T}_p M$,
is simply the vector space of
all possible tangent vectors at $p$,
when considering all possible curves passing through $p$.
Finally,
the derivative $\mathrm{T} f$
of any smooth function $f$
is the linear function
that propagates tangent vectors along $f$.
%
The disjoint union of all tangent spaces
$\mathrm{T} M = \sqcup_{p \in M} \mathrm{T}_p M$
forms again a smooth manifold,
called the \textbf{tangent bundle} over $M$.
We have
$\dim(\mathrm{T} M) = 2 \cdot \dim(M)$,
since for every point $p \in M$,
there are
$\dim(\mathrm{T}_p M) = \dim(M)$
directions for change.
So,
$\mathrm{T}$ sends
a manifold $M$ to its tangent bundle $\mathrm{T} M$
and it sends
a smooth map $f : M \to N$ between manifolds to
its derivative $\mathrm{T} f : \mathrm{T} M \to \mathrm{T} N$.
For any composite function $f = f_2 \circ f_1$,
$\mathrm{T}$ satisfies the chain rule
(functor property)
$\mathrm{T} f = \mathrm{T} f_2 \circ \mathrm{T} f_1$.

Let $\mathrm{pr} \colon \mathrm{T} M \to M$
denote the bundle projection,
which maps a tangent vector at point $p$ to
the point $p$ itself.
A \textbf{section}
$s$ of the bundle $\mathrm{T} M$
is a smooth map $s \colon M \to \mathrm{T} M$
such that $\mathrm{pr} \circ s = \mathrm{id}_M$.
A section of $\mathrm{T} M$
is called a \textbf{vector field}.
We write $\Gamma(\mathrm{T} M)$ for
the infinite-dimensional vector space of such sections.
An \textbf{integral curve}
of a vector field $X \in \Gamma(\mathrm{T} M)$
is a smooth curve
$c : \mathbb{I} \to M$,
with $\mathbb{I} \subseteq \mathbb{R}$,
such that
for each $t \in \mathbb{I}$,
the tangent vector to $c$ at $c(t)$
is $X \vert_{c(t)}$.

\subsection{The cotangent bundle and the differential}%

Given a manifold $M$,
we can define
its \textbf{cotangent bundle}
$\mathrm{T}^* M = \sqcup_{p \in M} \mathrm{T}^*_p M$,
where
the cotangent space
$\mathrm{T}^*_p M$
is the dual space of
the tangent space
$\mathrm{T}_p M$.
A section of the cotangent bundle is called
a \textbf{covector field}.

We recall that
for a function $f \colon M \to N$
between two smooth manifolds,
the derivative is a map
$\mathrm{T} f \colon \mathrm{T} M \to \mathrm{T} N$
that sends any pair
$(p, \, v)$
with $p \in M$ and $v \in \mathrm{T}_p M$
to the pair
$(q, \, w)$
with $q = f(p)$ and $w \in \mathrm{T}_{q} N$.
Here,
$w$ is the local change of $f(p)$,
when the local change of $p$ is $v$.
To make composition work (chain rule),
one has to propagate also the points
$q$ and $p$
and not only their local changes
$v$ and $w$.
In contrast to the derivative,
the \textbf{differential}
only applies to
functions $f : M \to \bR$
\emph{on} manifolds.
The differential $\dd f \in \Gamma(\mathrm{T}^* M)$
is a covector field.
At point $p$,
$\dd f \vert_p \in \mathrm{T}^*_p M$,
is a linear function
that sends
a vector $v \in \mathrm{T}_p M$
to the corresponding infinitesimal change
$w \in \mathrm{T}_{f(p)} \bR \cong \bR$.
Hence,
$w = \langle \dd f \vert_p \mid v \rangle$.
The definition of the gradient
used in VC
is based on an inner product.
The differential instead uses linear duality,
making it independent of such extra structure.

\subsection{Differential forms and the exterior derivative}%

Generalizing vector and covector fields,
we can define $(p,q)$-\textbf{tensor fields}
as sections of
${(\mathrm{T} M)}^{\otimes p} \otimes {(\mathrm{T}^* M)}^{\otimes q}$
with
the tensor product extending to fields in a pointwise manner.

\textbf{Differential forms} are tensor fields
in the exterior algebra of the cotangent bundle.
We write
$\Omega^k(M) \coloneqq \Gamma(\Lambda^k(\mathrm{T}^* M))$
for the infinite-dimensional vector space
of \mbox{$k$-forms} on $M$.
\mbox{$0$-forms} are smooth functions on $M$, i.e.~%
$\Omega^0(M) \cong C^\infty(M)$,
and
\mbox{$1$-forms} are covector fields.
A \mbox{$2$-form} $\alpha \in \Omega^2(M)$ gives
at every point $p \in M$
a bilinear map
$\alpha \vert_p \colon \mathrm{T}_p M \times \mathrm{T}_p M \to \bR$
that satisfies
$\alpha(v_1, \, v_2) = -\alpha(v_2, \, v_1)$
for all $v_1, \, v_2 \in \mathrm{T}_p M$.
The exterior product extends
in a pointwise manner
to a map
$\wedge \colon \Omega^k(M) \times \Omega^l(M) \to \Omega^{k+l}(M)$,
which satisfies
the graded anticommutativity property
in~\cref{eq:wedge_anticommutativity}.

For measuring the local change of differential forms,
the differential
$\dd \colon C^\infty(M) \to \Gamma(\mathrm{T}^* M)$
extends to a map
$\dd \colon \Omega^k(M) \to \Omega^{k+1}(M)$,
called the \textbf{exterior derivative}.
This is uniquely determined by requiring that
the product rule
\begin{equation}
  \dd(\alpha \wedge \beta)
  \: = \:
  \dd \alpha \wedge \beta + (-1)^k \cdot \alpha \wedge \dd \beta
\end{equation}
holds
for all
$\alpha \in \Omega^k(M)$,
$\beta \in \Omega^l(M)$
and that
$\dd \dd \alpha = 0$
for any differential form $\alpha$.
The nilpotency corresponds to the fact that
a Hessian matrix
has no antisymmetric part.
Applied to \mbox{$0$-forms},
the exterior derivative is similar to the gradient in VC.
On \mbox{$1$-forms},
it is similar to the curl
and
on \mbox{$2$-forms}
it is similar to the divergence.
The nilpotency hence corresponds also to the fact that
the curl of a gradient field
and
the divergence of a curl field
vanish.

Given a smooth map
$f \colon M \to N$
between manifolds
and a differential form
$\alpha \in \Omega^k(N)$,
the \textbf{pullback} of $\alpha$ along $f$ is
denoted by $f^*(\alpha) \in \Omega^k(M)$
and it is defined by
$
f^*(\alpha) \vert_p(v_1, \ldots, v_k) =
\alpha \vert_{f(p)}((\mathrm{T}_p f)(v_1), \ldots, (\mathrm{T}_p f)(v_k))
$
for all
$p \in M$
and
$v_1, \ldots, v_k \in \mathrm{T}_p M$.
The pullback
distributes over the exterior product, i.e.
$f^*(\alpha \wedge \beta) = f^*(\alpha) \wedge f^*(\beta)$
and it
commutes with the exterior derivative, i.e.
$f^*(\dd \alpha) = \dd(f^*(\alpha))$.

\subsection{Integration and Stokes theorem}%

Let $M$ be a manifold with
$\dim(M) = n$
and
let $N$ be a submanifold of $M$ with
$\dim(N) = k$.
In particular,
we may have $N = M$
implying $k = n$.

A differential \mbox{$k$-form} on $M$
can be integrated over $N$.
At each point $p \in M$,
a \mbox{$k$-form} gives a linear map that
sends a \mbox{$k$-vector} on $\mathrm{T}_p M$ to a real number.
We hence think of \mbox{$k$-forms} as
somehow measuring oriented $k$-dimensional volumes.
For instance,
a \mbox{$1$-form} provides a `scale' to measure
a \emph{signed} length-like quantity associated to curves,
as it assigns to each tangent vector a number
that can be interpreted as an infinitesimal `length'.
Summing these up
for all tangent vectors along the curve
gives the integral of the \mbox{$1$-form} along the curve,
thought of as its total `length'.

The integral theorems of VC are unified into
the generalized Stokes (or Stokes-Cartan) theorem,
which states that
\begin{equation}
  \int_{N} \dd \alpha
  \: = \:
  \int_{\partial N} i^* ( \alpha )
	\label{eq:stokes}
\end{equation}
for all
$\alpha \in \Omega^{k-1}(N)$.
Here,
$i^*(\alpha)$ denotes
the pullback of $\alpha$
along the inclusion
$i \colon \partial N \hookrightarrow N$,
thought of as
restriction to the boundary.
The orientation of $\partial N$
is induced by
the orientation of $N$
via an outwards pointing vector field.
For $k = \dim(N) = 1$,
the right hand side looks as in~\cref{eq:stokes_1d_rhs}.
If the boundary $\partial N$ is empty,
the right hand side is zero.
Applying Stokes theorem twice
with integrand $\dd \dd \alpha$ shows that
the nilpotency of the exterior derivative
also reflects the topological fact that
the boundary of any boundary is empty.
Combining Stokes theorem with
the product rule for the exterior derivative
gives the following integration by parts formula:
For all
$\alpha \in \Omega^k(N)$
and
$\beta  \in \Omega^{(n-k-1)}(N)$,
we have
\begin{equation*}
	\begin{split}
    \int_{N} \dd ( \alpha \wedge \beta )
		\: &= \:
		\int_{\partial N} i^* ( \alpha \wedge \beta )
		\: = \:
		\int_{\partial N} i^* \alpha \wedge i^* \beta  \\
		\: &= \:
		\int_{N} \dd \alpha \wedge \beta
		\: + \:
		{(-1)}^k \cdot
		\int_{N} \alpha \wedge \dd \beta
    \,.
	\end{split}%
\end{equation*}

\subsection{Straight vs twisted differential forms}%
\label{ssec:straight_twisted}

We make the physically meaningful distinction between
straight and twisted \mbox{$k$-forms},
see e.g.~\onlinecite{1985Burke}.


The integral of
a (straight) \mbox{$k$-form} on $M$
over $N$
changes sign if
the orientation of $N$ changes.
For instance,
this applies to a $1$-from
that represents an electric field,
as integrating it over a curve with reversed orientation
results in a voltage with opposite sign.
%
The voltage is measured between
the two endpoints of the curve.

The integral of
a twisted \mbox{$k$-form} (or $k$-pseudoform) on $M$
over $N$
changes sign if
the transverse orientation of $N$ changes.
The transverse orientation
(or pseudoorientation)
of $N$ (in $M$)
is the orientation of
the $(n-k)$-dimensional space around $N$
and
it is determined by
the orientation of $N$
and
the orientation of the ambient space $M$.
For instance,
this applies to a twisted $1$-from
representing a magnetic field strength,
as its integral
over a closed curve
changes sign if
the orientation of the curve is reversed
or
the handedness of the $3$-dimensional ambient space changes.
The latter determines whether
the left or right hand screw rule is used to
determine the transverse orientation.
%
%
The integral is equal to
the current that passes through
a surface whose boundary is the closed curve.
%

A twisted \mbox{$n$-form} on $M$
is also called a density
and
it can be integrated over
($n$-dimensional submanifolds of) $M$
without the need for an orientation.
For instance,
this applies to extensive quantities
such as mass or entropy,
which are given by twisted \mbox{$3$-forms}.

The vector space of twisted \mbox{$k$-forms} on $M$
is denoted by $\tilde{\Omega}^k(M)$.
The exterior derivative takes
straight forms to straight forms
and
twisted forms to twisted forms.
The exterior product of
a straight \mbox{$k$-form} and
a twisted \mbox{$l$-form} gives
a twisted \mbox{$(k+l)$-form}.
Stokes theorem equally applies to twisted forms.


%
%

%
%

%
%

\subsection{Riemannian metric and induced structure}%

A \textbf{Riemannian metric} on a manifold $M$
is a symmetric positive-definite \mbox{$(0, 2)$-tensor}
$g \in \Gamma(\mathrm{T}^* M \otimes \mathrm{T}^* M)$,
which gives an inner product on every tangent space of $M$.

On a Riemannian manifold $(M, \, g)$,
the so-called \textbf{musical isomorphisms}
denoted by
$\flat \colon \mathrm{T} M \to \mathrm{T}^* M$
and
$\sharp \colon \mathrm{T}^* M \to \mathrm{T} M$
map between
vectors and covectors
or between
vector fields and \mbox{$1$-forms}.
For some vector $v \in \mathrm{T}_p M$,
the flat map is defined by
$v^\flat = g(v, \cdot) \in \mathrm{T}^*_p M$.
The sharp map is defined similarly
using the inverse of $g$.

The inner product on each tangent space
induces a Hodge star isomorphism
$\star \colon \Omega^k(M) \to \tilde{\Omega}^{n-k}(M)$,
see~\onlinecite{2005Ramanan}.
%
We again have
$\star \star \alpha = {(-1)}^{k \cdot (n-k)} \, \alpha$
for all
$\alpha \in \Omega^k(M)$.
For the case $n = 3$,
this simplifies to $\star \star \alpha = \alpha$.
Due to the symmetry of the inner product,
we have
$\alpha \wedge \star \beta = \beta \wedge \star \alpha$
for all $\alpha, \beta \in \Omega^k(M)$.
The same applies for the other direction
$\star \colon \tilde{\Omega}^k(M) \to \Omega^{n-k}(M)$.

The \textbf{Riemannian volume form} (or density)
on $(M, \, g)$
is given by
$
\star 1 \in \tilde{\Omega}^n(M)
$.

\subsection{Interior product and Lie derivative}%

The \textbf{interior product}
$\iota_X \colon \Omega^k(M) \to \Omega^{k-1}(M)$
fixes a vector field $X \in \Gamma(\mathrm{T} M)$
as the first argument of a (twisted) \mbox{$k$-form}.
As shown in~\onlinecite{2003Hirani},
it can be expressed
using the Hodge star:
\begin{equation}
	\begin{split}
		\raisetag{3.3em}
		\iota_{{(\_)}^\sharp} (\_) \colon
		\Omega^1(M) \times
		\Omega^k(M)
		&\to
		\Omega^{k-1}(M) \\
		\left( \upsilon, \, \alpha \right)
		&\mapsto
		\iota_{\upsilon^\sharp} \alpha
		= {(-1)}^{(k+1)n} \,
		\star (\upsilon \wedge \star \alpha)
    \,.
	\end{split}
	\label{eq:interior_product}
\end{equation}

The \textbf{Lie derivative} $\cL$
does not depend on the metric
and measures the change of a tensor
along the (local) flow of a vector field.
The Lie derivative of a (twisted) \mbox{$k$-form} $\alpha$
can be computed with
Cartan's magic formula
\begin{equation*}
  \cL_X \alpha
  \: = \:
  \dd ( \iota_X \alpha )
  \, + \,
  \iota_X ( \dd \alpha)
  \,.
\end{equation*}

\subsection{Covariant derivative and bundle-valued forms}%

A \textbf{covariant derivative}
is a directional derivative
for tensor fields.
In contrast to
the exterior derivative and Lie derivative,
there is no unique covariant derivative
defined on a smooth manifold.
However,
for a Riemannian manifold,
there is a canonical choice,
called the Levi-Civita connection.
For the relevant case of a vector field
$u \in \Gamma(\mathrm{T} M)$,
the covariant derivative of $u$
along a tangent vector
$v \in \mathrm{T} M$
is denoted by
$\nabla_v \, u \in \Gamma(\mathrm{T} M)$.
Since $\nabla$ is linear
with respect to the direction,
we write
$\nabla u \in \Gamma(\mathrm{T} M) \otimes \Omega^1(M)$.
Pairing the second leg (or form part)
of the $\mathrm{T} M$-valued $1$-form $\nabla u$
with some tangent vector $v$ again gives $\nabla_v \, u$.

To define a duality pairing for \textbf{bundle-valued forms},
we use the binary operation
\begin{equation}
  \dot{\wedge} \colon
  ( \Gamma(\mathrm{T} M) \otimes \Omega^k(\cZ) )
  \times
  ( \Gamma(\mathrm{T}^* M) \otimes \tilde{\Omega}^{n-k}(M) )
  \to \tilde{\Omega}^n(M)
  \label{eq:wedge_dot_pairing}
\end{equation}
defined by
$
(u \otimes \beta) \dot{\wedge} (\alpha \otimes \gamma) =
\langle \alpha \mid u \rangle \cdot \beta \wedge \gamma
$.
For $k = 0$,
we implicitly identify
$\Gamma(\mathrm{T} M) \cong \Gamma(\mathrm{T} M) \otimes \Omega^0(M)$
and
$\Gamma(\mathrm{T}^* M) \cong \Gamma(\mathrm{T}^* M) \otimes \Omega^0(M)$.
%
We may then write
$
u \dot{\wedge} (\alpha \otimes (\hodge 1)) =
(\iota_u \alpha) \, \hodge 1 =
\hodge (\iota_u \alpha)
$.

The exterior covariant derivative $\dd_\nabla$
is the formal adjoint of $-\nabla$.
For the relevant case of
a $\Gamma(\mathrm{T}^* M)$-valued twisted $2$-form $T$,
it gives a map
$
\dd_\nabla \colon
\Gamma(\mathrm{T}^* M) \otimes \tilde{\Omega}^2(M) \to
\Gamma(\mathrm{T}^* M) \otimes \tilde{\Omega}^3(M)
$
that is determined by
the integration by parts formula~%
\begin{equation*}
  \int_M
  \dd ( u \dot{\wedge} T )
  \: = \:
  \int_{\partial M}
  i^*(u \dot{\wedge} T)
  \: = \:
  \int_M u \dot{\wedge} \dd_\nabla T
  \: + \:
  \int_M \nabla u \dot{\wedge} T
  \,,
\end{equation*}
see~\onlinecite{2007KansoArroyoTongYavariMarsdenDesbrun}.

\section{Relationship with the metriplectic or GENERIC formalism}%
\label{sec:metriplectic}

Here,
we want to clarify how the EPHS language
is related to
the metriplectic formalism,
which ensures the thermodynamically consistent combination of
reversible (Hamiltonian) and irreversible (gradient) dynamics,
see e.g.~\onlinecite{1986Morrison}.
Although there are subtle differences,
what we say here also applies to
the GENERIC formalism~\cite{1997GrmelaOettinger},
when assuming a quadratic dissipation potential.

To focus on the core idea,
we consider a metriplectic system,
whose state space $\cX$ is
a finite-dimensional vector space.
A metriplectic dynamics
is then determined by
\begin{equation*}
  \dot{x}
  \: = \:
  L(x) \, \dd E(x)
  \, + \,
  \tilde{M}(x) \, \dd S(x)
  \,.
\end{equation*}
Here,
$x \in \cX$ is
the combined state variable.
The function
$E \colon \cX \to \bR$ yields the total energy.
Similarly,
functions $S$ and $N$
yield the total entropy and mass,
respectively.
For every $x \in \cX$,
the skew-symmetric matrix
$L(x) = -{(L(x))}^\mathrm{T}$
satisfies
the two conditions
$L(x) \, \dd S(x) = L(x) \, \dd N(x) = 0$.
Similarly,
the symmetric, non-negative definite matrix
$\tilde{M}(x) = {(\tilde{M}(x))}^\mathrm{T}$
satisfies
$
\tilde{M}(x) \, \dd E(x) =
\tilde{M}(x) \, \dd N(x) = 0
$.

The metriplectic structure ensures that
energy is conserved:
\begin{equation*}
  \begin{split}
    \dot{E}(x(t))
    \: &= \:
    \langle \dd E(x) \mid \dot{x} \rangle
    \\
    \: &= \:
    \langle \dd E(x) \mid L(x) \, \dd E(x) \rangle
    \, + \,
    \langle \dd E(x) \mid \tilde{M}(x) \, \dd S(x) \rangle
    \\
    \: &= \:
    0 + \langle \tilde{M}(x) \, \dd E(x) \mid \dd S(x) \rangle
    \: = \:
    0
  \end{split}
\end{equation*}
The reversible part conserves energy
due to skew-symmetry of $L(x)$
and
the irreversible part conserves energy
due to symmetry of $\tilde{M}(x)$ and
the degeneracy condition
$\tilde{M}(x) \, \dd E(x) = 0$.
Moreover,
entropy grows monotonically:
\begin{equation*}
  \begin{split}
    \dot{S}(x(t))
    \: &= \:
    \langle \dd S(x) \mid L(x) \, \dd E(x) \rangle
    \, + \,
    \langle \dd S(x) \mid \tilde{M}(x) \, \dd S(x) \rangle
    \\
    \: &= \:
    \langle \dd S(x) \mid \tilde{M}(x) \, \dd S(x) \rangle
    \: \geq \:
    0
  \end{split}
\end{equation*}
The reversible part conserves entropy
due to skew-symmetry of $L(x)$ and
the degeneracy condition
$L(x) \, \dd S(x) = 0$.
The irreversible part has
non-negative entropy production,
since $\tilde{M}(x)$ is symmetric and non-negative definite.
Further quantities,
such as the total mass $N$,
are conserved,
given that both $L$ and $\tilde{M}$ satisfy
the relevant degeneracy conditions.

The metriplectic evolution equation
can equivalently be written as
\begin{equation*}
  \dot{x}
  \: = \:
  \biggl(
    L(x)
    \, - \,
    \frac{1}{\textcolor{violet}{\theta_0}} \cdot \tilde{M}(x)
  \biggr) \, \dd H(x)
  \,,
\end{equation*}
where
$H \colon \cX \to \bR$
is the exergy storage function
defined by
\begin{equation*}
  H(x)
  \: = \:
  E(x)
  \, - \,
  \textcolor{violet}{\theta_0} \cdot S(x)
  \, - \,
  \textcolor{violet}{\mu_0} \cdot N(x)
  \, + \,
  \mathrm{const.}
\end{equation*}
In principle,
$
\textcolor{violet}{\theta_0},
\textcolor{violet}{\mu_0}
\in \bR
$
are arbitrary constants.
More details can be found in~%
\onlinecite{2021LohmayerKotyczkaLeyendecker}.

Without a fundamental loss of generality,
we now \textit{assume} that
none of the individual state variables
in the combined state $x$
have the quantity energy.
Consequently,
$S$ and $N$ are simply given by
a sum of state variables
and
the components of $\dd S(x)$ and $\dd N(x)$ are,
independently of $x \in \cX$,
either $0$ or $1$,
indicating which state variables have the quantity entropy or mass.
We can thus write
\begin{equation*}
  \dd H(x)
  \: = \:
  \dd E(x)
  \, - \,
  \textcolor{violet}{\theta_0} \cdot \dd S
  \, - \,
  \textcolor{violet}{\mu_0} \cdot \dd N
  \,.
\end{equation*}
Components of $\dd H$ and $\dd E$
are equal
if the respective quantity
is neither entropy nor mass.
Otherwise,
the component of $\dd H$ additionally has a shift by either
$-\textcolor{violet}{\theta_0}$ or
$-\textcolor{violet}{\mu_0}$.
It follows that
$\dd H$ is fully determined by $\dd E$,
and vice versa,
provided that
the physical quantities of
the individual state variables are known.

\begin{figure}
  \centering
	\includegraphics[width=5.8cm]{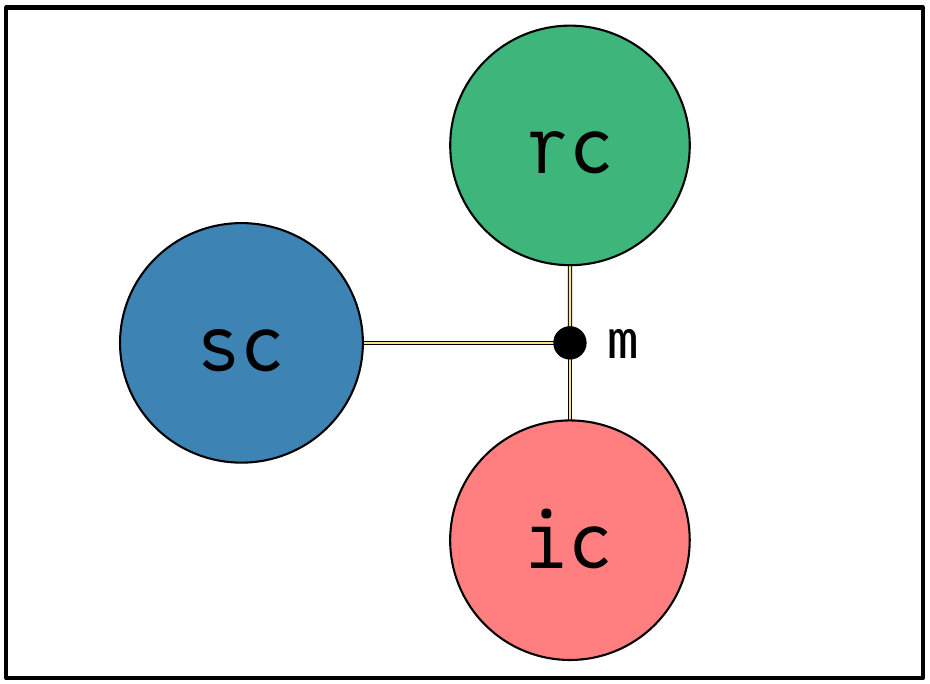}
  \caption{%
    Interconnection pattern
    for a generic, isolated EPHS model.
    Boxes $\mathtt{sc}$, $\mathtt{rc}$ and $\mathtt{ic}$
    are filled by
    a storage component, reversible component, and irreversible component,
    respectively.
    Each component has the same interface
    with a multiport $\mathtt{m}$.
  }%
  \label{fig:generic}
\end{figure}

To satisfy the degeneracy condition
$\tilde{M}(x) \, \dd E(x) = 0$,
$\tilde{M}$ must depend on $x$  through $\dd E$
or equivalently through $\dd H$.
This means we can \textit{assume} that
there exists a $M$
such that
\begin{equation}
  \begin{split}
    \tilde{M}(x) \, \dd S(x)
    \: &= \:
    M \bigl( \dd H(x) \bigr) \, \dd S(x)
    \\
    \: &= \:
    -\frac{1}{\textcolor{violet}{\theta_0}} \cdot
    M \bigl( \dd H(x) \bigr) \, \dd H(x)
    \,.
  \end{split}
  \label{eq:irreversible}
\end{equation}
The degeneracy conditions on $M$ are
\begin{subequations}
\begin{equation}
  \begin{split}
    &M \bigl( \dd H(x) \bigr) \, \dd E(x)
    \: = \:
    \\
    &M \bigl( \dd H(x) \bigr) \, \bigl(
      \dd H(x)
      \, + \,
      \textcolor{violet}{\theta_0} \cdot \dd S
      \, + \,
      \textcolor{violet}{\mu_0} \cdot \dd N
    \bigr)
    \: = \:
    0
  \end{split}
\end{equation}
and
\begin{equation}
  M \bigl( \dd H(x) \bigr) \, \dd N
  \: = \:
  0
  \,.
\end{equation}%
\label{eq:irreversible_degeneracy}%
\end{subequations}
This allows us to recast
the metriplectic system as an EPHS model.
We first define an interface $I$
that reflects the structure and physical meaning of $\cX$.
If each component of $x \in \cX$ represents
a separate physical quantity,
$I$ has a port for each component.
More generally,
$\cX$ is a Cartesian product of spaces
such that
each factor or subspace corresponds to a physical quantity
and a port of $I$.
To draw the interconnection pattern
of the EPHS model in a simple and general manner,
all ports are combined into one multiport $\mathtt{m}$,
see~\cref{fig:generic}.
Box $\mathtt{sc}$ is filled by
the storage component $(I, \, E)$.
Its port variables hence satisfy
$\mathtt{sc.m.x} = x$,
$\mathtt{sc.m.f} = \dot{x}$, and
$\mathtt{sc.m.e} = \dd H(x)$.
Here,
$\mathtt{sc.m.x}$ stands for
the concatenation of
the state variables
$\mathtt{sc.m.} p \mathtt{.x}$,
with $p$ ranging over the individual ports of
the multiport $\mathtt{sc.m}$.
The meaning of
$\mathtt{sc.m.f}$ and
$\mathtt{sc.m.e}$
follows analogously.
Box $\mathtt{rc}$ is filled by
the reversible component $(I, \, \cD)$
with the Dirac structure $\cD$ given by
$\mathtt{rc.m.f} = -L(\mathtt{rc.m.x}) \, \mathtt{rc.m.e}$.
Box $\mathtt{ic}$ is filled by
the irreversible component $(I, \, \cO)$
with the Onsager structure $\cO$ given by
$
\mathtt{ic.m.f} =
\frac{1}{\textcolor{violet}{\theta_0}} \cdot
M(\mathtt{ic.m.e}) \, \mathtt{ic.m.e}
$.

An idiomatic EPHS model is obtained by
splitting the three components into
various small, reusable components,
which are organized into
(a hierarchy of) reusable subsystems
that are relatively easy to understand.
To guarantee thermodynamic consistency
for any composite system,
it suffices to
check the degeneracy conditions
once for each reusable component.
As a precise statement
of the conditions
requires some definitions,
we refer to~\onlinecite{2024LohmayerLynchLeyendecker}.
In essence,
$\dd S$ and $\dd N$ are determined by
the physical quantities associated to each port
and
$\dd H(x)$ corresponds to the effort variable.
For a reversible component we then have
$L(x) \, \dd S = L(x) \, \dd N = 0$
and
for an irreversible component we have~%
\cref{eq:irreversible_degeneracy}.

Concerning
spatially-distributed systems,
it is important to note that
the degeneracy conditions on $L$
do not apply at the boundary of the spatial domain,
since entropy and mass can enter of leave the system
due to a reversible process (advection).
Similarly,
the degeneracy conditions on $M$
do not apply at the boundary,
since energy and mass can enter or leave the system
due to an irreversible processes.
Finally,
the symmetry and non-negative definiteness of $M$
become apparent when
the operator equation
required to hold within the spatial domain
is written in a weak form,
where boundary ports are included implicitly.
Regarding this, we also refer to~\onlinecite{2006Oettinger}.

\section*{References}%

\bibliography{literature.bib}

\end{document}